# The European AI Liability Directives

—

## Critique of a Half-Hearted Approach and Lessons for the Future

Philipp Hacker[*]

**Abstract:**

The optimal liability framework for AI systems remains an unsolved problem across the globe. With ChatGPT and other large generative models taking the technology to the next level, solutions are urgently needed. In a much-anticipated move, the European Commission advanced two proposals outlining the European approach to AI liability in September 2022: a novel AI Liability Directive (AILD) and a revision of the Product Liability Directive (PLD). They constitute the final cornerstone of AI regulation in the EU. Crucially, the liability proposals and the AI Act are inherently intertwined: the latter does not contain any individual rights of affected persons, and the former lack specific, substantive rules on AI development and deployment. Taken together, these acts may well trigger a "Brussels effect" in AI regulation, with significant consequences for the US and other countries.

Against this background, this paper makes three novel contributions. First, it examines in detail the Commission proposals and shows that, while making steps in the right direction, they ultimately represent a half-hearted approach: if enacted as foreseen, AI liability in the EU will primarily rest on disclosure of evidence mechanisms and a set of narrowly defined presumptions concerning fault, defectiveness and causality.

Hence, second, the article suggests amendments to the proposed AI liability framework. They are collected in a concise Annex at the end of the paper. I argue, inter alia, that the dichotomy between the fault-based AILD Proposal and the supposedly strict liability PLD Proposal is fictional and should be abandoned; that an EU framework for AI liability should comprise one fully harmonizing regulation instead of two insufficiently coordinated directives; and that the current proposals unjustifiably collapse fundamental distinctions between social and individual risk by equating high-risk AI systems in the AI Act with those under the liability framework.

Third, based on an analysis of the key risks AI poses, the final part of the paper maps out a road for the future of AI liability and regulation, in the EU and beyond. More specifically, I make four key proposals. Effective compensation should be ensured by combining truly strict liability for certain high-risk AI systems with general presumptions of defectiveness, fault and causality in cases involving SMEs or non-high-risk AI systems. The paper introduces a novel distinction between illegitimate- and legitimate-harm models to delineate strict liability's scope. Truly strict liability should be reserved for high-risk AI systems that, from a social perspective, should not cause harm (illegitimate-harm models, e.g., autonomous vehicles or medical AI). Models meant to cause some unavoidable harm by ranking and rejecting individuals (legitimate-harm models, e.g., credit scoring or insurance scoring) may only face rebuttable presumptions of defectiveness and causality. General-purpose AI systems and Foundation Models should only be subjected to high-risk regulation, including liability for high-risk AI systems, in specific high-risk use cases for which they are deployed. Consumers, in turn, ought to be liable based on regular fault, in general.

---

[*] Professor Dr. Philipp Hacker, LL.M. (Yale), Chair for Law and Ethics of the Digital Society, European New School of Digital Studies, European University Viadrina. I am grateful for comments by Maximilian Eber, Andreas Engel, Rasmus Rothe, and Sandra Wachter. My team provided excellent research assistance, particularly Sarah Großheim and Marco Mauer.

Furthermore, innovation and legal certainty should be fostered through a comprehensive regime of safe harbours, defined quantitatively to the best extent possible. Moreover, trustworthy AI remains an important goal for AI regulation. Hence, the liability framework must specifically extend to non-discrimination cases and provide for clear rules concerning explainability (XAI).

Finally, awareness for the climate effects of AI, and digital technology more broadly, is rapidly growing in computer science. In diametrical opposition to this shift in discourse and understanding, however, EU legislators have long neglected environmental sustainability in both the draft AI Act and the proposed liability regime. To counter this, I propose to jump-start sustainable AI regulation via sustainability impact assessments in the AI Act and sustainable design defects in the liability regime. In this way, the law may help spur not only fair AI and XAI, but also sustainable AI (SAI).

## Table of Contents

# Overview of Tables

**Table 1 (p. 6):**

| | HRAIS | Non-HRAIS | Manufacturer | Provider | Comm. User | Consumer |
|---|---|---|---|---|---|---|
| Disclosure of evidence (PLD) | 8(1) | 8(1) | 8(1) | - | - | - |
| Disclosure of evidence (AILD) | 3(1) | - | - | 3(1) | 3(1) | 3(1) |
| Presumption (defectiveness: PLD) | 9(2) & 9(4) | 9(2) & 9(4) | 9(2) & 9(4) | - | - | - |
| Presumption (fault: AILD) | 3(5) | - | - | 3(5) | 3(5) | 3(5) |
| Presumption (causal link: defect → output, PLD) | 9(3) & 9(4) | 9(3) & 9(4) | 9(3) & 9(4) | - | - | - |
| Presumption (causal link: fault → output, AILD) | 4(2), 4(3) & 4(5) | 4(5) & 4(6) | - | 4(2) & 4(5) | 4(3) & 4(5) | 4(6) |

*Table 1: Overview of the applicability of the AILD and PLD proposals to: HRAIS (High-Risk AI Systems); manufacturers (= developers of AI or AI components); providers (= developers who place system on the market under own name); commercial users; and consumers. Note that there is significant overlap between the terms of "manufacturer" and "provider". Furthermore, the PLD applies to other economic operators besides manufacturers (e.g., related service providers).*

**Table 2 (p. 9):**

| PLD Proposal | AILD Proposal |
|---|---|
| Claim rooted in EU law | Claim rooted in Member State law |
| Material and procedural aspects of product liability | Procedural aspects of liability for AI systems |
| Applicable to physical products and software, including AI systems | Applicable to AI systems only |
| Supposedly strict liability | Fault-based liability |
| Claims against manufacturers and other entities in the supply chain | Claims against manufacturers, professional users and consumers |
| Eligible damage: property, death or personal injury, and data loss | Eligible damage: potentially also fundamental rights and primary financial loss |
| Full harmonization | Minimum harmonization |

*Table 2: Overview of core differences between the PLD and the AILD Proposal*

**Table 3 (p. 33):**

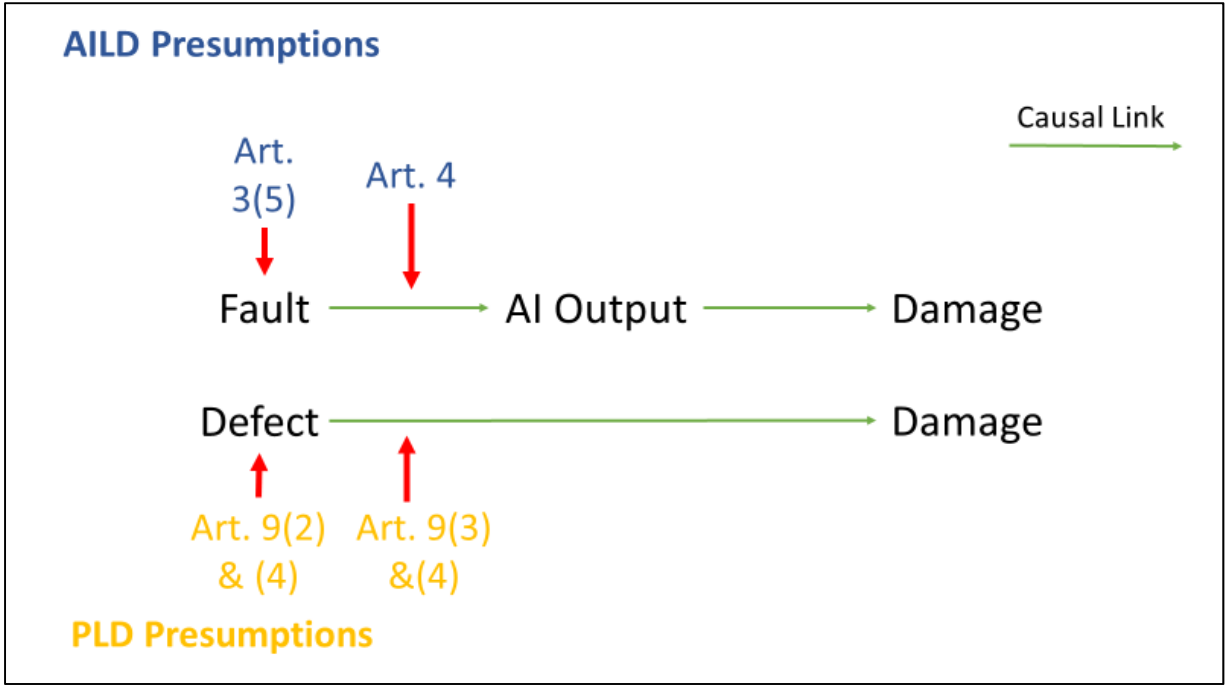

*Table 3: Overview of the presumptions of fault, defectiveness, and causality under the proposed AI liability framework*

# I. Introduction

We are currently witnessing the next quantum leap in artificial intelligence (AI) technology, with new large generative AI models launched on almost a weekly basis, transforming the way humans work, create, and communicate.[1] These models are not restricted to a particular task anymore, but operate on a wide range of problems and may produce human-level text (e.g., ChatGPT, GPT-4), images (e.g., DALL·E 2), videos (e.g., Synthesia), or art (e.g., Stable Diffusion). The accelerating rollout of this new generation of AI systems raises a host of regulatory issues,[2] and endows with new urgency the question of who should, in a legal sense, be responsible if AI systems make mistakes.

With the publication of two proposals concerning directives for AI liability,[3] the EU Commission unveiled, in September 2022, the final cornerstone of its approach to AI regulation. While many instruments of EU law are formulated in a technology neutral manner and will therefore generally also apply to AI – such as the non-discrimination framework,[4] unfair commercial practices law,[5] the GDPR,[6] general contract and tort law[7] – the last years have increasingly seen proposals directed explicitly, or implicitly, toward AI systems. Most prominently, the proposal for an AI Act, published in April 2021,[8] is currently being negotiated in the EU trilogue. The Council of the European Union adopted a formal position on December 6, 2022;[9] the European Parliament followed on June 14, 2023, notably with specific rules on Foundation Models and generative AI.[10] The final AI Act will probably be finalized at the end of 2023 or early 2024. Moreover, instruments ostensibly addressed towards online platforms, such as the recently enacted Digital Markets Act (DMA) and the Digital Services Act (DSA), also contain crucial constraints on and provisions for AI models.[11]

---

[1] See, e.g., Rishi Bommasani and others, 'On the opportunities and risks of foundation models' (2021) arXiv preprint arXiv:210807258; Roberto Gozalo-Brizuela and Eduardo Garrido-Merchan, 'ChatGPT is not all you need. A State of the Art Review of Large Generative AI Models' (2023) Working Paper, arXiv preprint arXiv:230104655; Deep Ganguli and others, 'Predictability and surprise in large generative models' (2022) ACM Conference on Fairness, Accountability, and Transparency 1747; Masahiro Suzuki and Yutaka Matsuo, 'A survey of multimodal deep generative models' (2022) 36 Advanced Robotics 261.

[2] See, e.g., Philipp Hacker, Andreas Engel and Marco Mauer, 'Regulating ChatGPT and other Large Generative AI Models' (2023) ACM Conference on Fairness, Accountability, and Transparency (FAccT '23) 1112; Natali Helberger and Nicholas Diakopoulos, 'ChatGPT and the AI Act' (2023) 12 Internet Policy Review; Lilian Edwards, 'Regulating AI in Europe: four problems and four solutions' (2022).

[3] European Commission, Proposal for a Directive of the European Parliament and of the Council on Liability for Defective Products (2022) [PLD Proposal]; European Commission, Proposal for a Directive of the European Parliament and of the Council on adapting non-contractual civil liability rules to artificial intelligence (2022) [AILD Proposal].

[4] See, e.g., Sandra Wachter, 'The Theory of Artificial Immutability: Protecting Algorithmic Groups Under Anti-Discrimination Law' (2022) arXiv preprint arXiv:220501166; Jeremias Adams-Prassl, Reuben Binns and Aislinn Kelly-Lyth, 'Directly Discriminatory Algorithms' (2022) The Modern Law Review; Sandra Wachter, Brent Mittelstadt and Chris Russell, 'Why fairness cannot be automated: Bridging the gap between EU non-discrimination law and AI' (2021) 41 Computer Law & Security Review 105567; Mireille Hildebrandt, 'The issue of bias. The framing powers of machine learning' in Marcello Pelillo and Teresa Scantamburlo (eds), Machines We Trust Perspectives on Dependable AI (MIT Press 2021); Sandra Wachter, Brent Mittelstadt and Chris Russell, 'Bias preservation in machine learning: the legality of fairness metrics under EU non-discrimination law' (2020) 123 W Va L Rev 735; Frederik J Zuiderveen Borgesius, 'Strengthening legal protection against discrimination by algorithms and artificial intelligence' (2020) 24 The International Journal of Human Rights 1572; Meike Zehlike, Philipp Hacker and Emil Wiedemann, 'Matching code and law: achieving algorithmic fairness with optimal transport' (2020) 34 Data Mining and Knowledge Discovery 163; Frederik Zuiderveen Borgesius, 'Discrimination, artificial intelligence, and algorithmic decision-making' (2018); Philipp Hacker, 'Teaching fairness to artificial intelligence: existing and novel strategies against algorithmic discrimination under EU law' (2018) 55 Common Market Law Review 1143.

What had been missing so far to complete the picture was the adaptation of the civil liability framework to AI. Margot Kaminski has recently criticized the narrow and specific risk-regulation approach underlying the AI Act proposal and encouraged lawmakers to consider other perspectives, like civil liability, as alternative modes of regulation[12] With the two proposals published on September 28, 2022, the Commission now finally does venture into the field of liability, with a dual – and controversial – methodological approach.[13] On the one hand, the AILD Proposal seeks to harmonize procedural questions, such as disclosure of evidence and burden of proof, across Member States' national liability regimes for the purposes of AI liability, while largely tying these instruments to violations of the AI Act. On the other hand, and independently of the AI Act, the PLD Proposal suggests a general update of classical product liability, with a specific view, however, toward digital products more generally and AI more specifically. Indeed, as Herbert Zech has pointed out, administrative frameworks like the AI Act and liability regimes constitute two complementary approaches to regulating AI, directly (via specific regulation in the AI Act) and indirectly (via incentives generated by the liability framework).[14] The proposal for an AI Act outlines a regulatory and oversight framework for AI systems, particularly those considered as high-risk. Building on it, the AILD Proposal and PLD Proposal now seek to integrate the AI Act into civil (product) liability while aligning this field with the new risks and realities of the digital economy.[15] Crucially, the liability proposals and the AI Act complement one another insofar as the latter does not contain any individual rights of affected persons,[16] and the former lack specific, substantive rules on AI development and deployment.

---

[5] See, e.g., Natali Helberger and Agustin Reyna, 'The perfect match? A closer look at the relationship between EU consumer law and data protection law' (2017) 54 Common Market Law Review; Natali Helberger and others, 'EU Consumer Protection 2.0' (2021); Philipp Hacker, 'Manipulation by algorithms. Exploring the triangle of unfair commercial practice, data protection, and privacy law' (2021) European Law Journal, https://doi.org/10.1111/eulj.12389.

[6] See, e.g., Ronan Hamon and others, 'Bridging the gap between AI and explainability in the GDPR: towards trustworthiness-by-design in automated decision-making' (2022) 17 IEEE Computational Intelligence Magazine 72; Brent Mittelstadt, Chris Russell and Sandra Wachter, Explaining explanations in AI (2019); Sandra Wachter, Brent Mittelstadt and Chris Russell, 'Counterfactual explanations without opening the black box: Automated decisions and the GDPR' (2017) 31 Harv JL & Tech 841; Philipp Hacker, 'A legal framework for AI training data—from first principles to the Artificial Intelligence Act' (2021) 13 Law, Innovation and Technology 257; Joshua Gacutan and Niloufer Selvadurai, 'A statutory right to explanation for decisions generated using artificial intelligence' (2020) 28 International journal of law and information technology 193.

[7] See, e.g., Andrea Bertolini, Artificial intelligence and civil liability (2020); Herbert Zech, Liability for AI: public policy considerations (Springer 2021); Gerhard Wagner, 'Robot Liability' in S.Lohsse/R.Schulze/D.Staudenmayer (ed), Liability for Artificial Intelligence and the Internet of Things (Nomos 2019); Stefan Grundmann and Philipp Hacker, 'Digital technology as a challenge to European Contract Law' (2017) 13 European Review of Contract Law 255; Paulius Čerka, Jurgita Grigienė and Gintarė Sirbikytė, 'Liability for damages caused by artificial intelligence' (2015) 31 Computer Law & Security Review 376; Christiane Wendehorst, 'Strict liability for AI and other emerging technologies' (2020) 11 Journal of European Tort Law 150; Christiane Wendehorst, The Proposal for an Artificial Intelligence Act COM (2021) 206 from a Consumer Policy Perspective (2021); Gerald Spindler, 'User liability and strict liability in the Internet of Things and for robots' in Sebastian Lohsse, Reiner Schulze and Dirk Staudenmayer (eds), Liability for artificial intelligence and the internet of things (Nomos 2019); Emiliano Marchisio, 'In support of "no-fault" civil liability rules for artificial intelligence' (2021) 1 SN Social Sciences 1; Francesco Paolo Patti, 'The European road to autonomous vehicles' (2019) 43 Fordham Int'l LJ 125; Philipp Hacker and others, 'Explainable AI under contract and tort law: legal incentives and technical challenges' (2020) 28 Artificial Intelligence and Law 415.

[8] See, e.g. Michael Veale and Frederik Zuiderveen Borgesius, 'Demystifying the Draft EU Artificial Intelligence Act—Analysing the good, the bad, and the unclear elements of the proposed approach' (2021) 22 Computer Law Review International 97.

[9] Unless otherwise noted, all references to the AI Act are to the following document (called Council Version for disambiguation when necessary): Council of the EU, Interinstitutional File: 2021/0106(COD), General approach of Nov. 25, 2022, Doc. No. 14954/22, https://data.consilium.europa.eu/doc/document/ST-14954-2022-INIT/en/pdf.

[10] DRAFT Compromise Amendments on the Draft Report, Proposal for a regulation of the European Parliament and of the Council, Brando Benifei & Ioan-Dragoş Tudorache (May 9, 2023), https://www.europarl.europa.eu/meetdocs/2014_2019/plmrep/COMMITTEES/CJ40/DV/2023/05-11/ConsolidatedCA_IMCOLIBE_AI_ACT_EN.pdf (= AI Act EP Version).

[11] See, e.g., Meike Zehlike and others, 'Beyond incompatibility: Trade-offs between mutually exclusive algorithmic fairness criteria in machine learning and law' (2022) Working Paper, http://ssrncom/abstract=4279866; Philipp Hacker, Johann Cordes and Janina Rochon, 'Regulating Gatekeeper AI and Data: Transparency, Access, and Fairness under the DMA, the GDPR, and beyond' (2022) Working Paper, https://arxivorg/abs/221204997; cf. also Rupprecht Podszun and Philipp Bongartz, 'The digital markets act: moving from competition law to regulation for large gatekeepers' (2021) 10 Journal of European Consumer and Market Law; Martin Eifert and others, 'Taming the giants: The DMA/DSA package' (2021) 58 Common Market Law Review 987.

Taken together, the EU AI package may well trigger a "Brussels effect[17]" in AI regulation.[18] The US government, for example, is commenting on and following the developments closely. Many rules of the AI Act, and the liability package, will apply to AI developed in the US, and in other places on the planet, if deployed or used in the EU.[19] The EU's regulatory approach to AI will therefore have global repercussions, and may be emulated to various degrees in other jurisdictions. This highlights again the importance of "getting things right" regarding the EU's AI regulation in this highly innovative, globally competitive environment.

The Commission proposals were a long time in the making and constitute a much-anticipated, final puzzle piece of the regulation of AI in Europe. In 2018, the Commission published an evaluation of the PLD,[20] which was followed by specific reports of expert groups the following year.[21] Numerous scholars have weighed in on the issue,[22] and the debate about the published proposals will likely be vivid as well. Particularly in high-tech sectors such as AI development, liability rules need to strike an intricate balance between effective compensation of injured persons and incentives and affordances for effective innovation and deployment of AI systems. It should not be forgotten, however, that society overall does not have an interest, and should not facilitate, the development of AI systems which lead to unnecessary harm for end users.

---

[12] Margot E Kaminski, 'Regulating the Risks of AI' (2023) 103 Boston University Law Review (forthcoming).
[13] See also Gerald Spindler, 'Die Vorschläge der EU-Kommission zu einer neuen Produkthaftung und zur Haftung von Herstellern und Betreibern Künstlicher Intelligenz' (2022) Computer und Recht 689.
[14] Herbert Zech, 'Liability for AI: public policy considerations' (2021) 22 ERA Forum 147, 150.
[15] PLD Proposal, 2.
[16] See, e.g., Christiane Wendehorst, AI liability in Europe: anticipating the EU AI Liability Directive (Ada Lovelace Institute 2022), 7.
[17] Anu Bradford, 'The brussels effect' (2012) 107 Nw UL Rev 1.
[18] Mireille Hildebrandt, 'Global competition and convergence of AI Law' in Jan. M. Smits and others (eds), (Edward Elgar, forthcoming), 12-13; Wendehorst, AI liability in Europe: anticipating the EU AI Liability Directive, 2022, 9.
[19] See Article 2(1) AI Act.
[20] European Commission, Report from the Commission to the European Parliament, the Council and the European Economic and Social Committee on the Application of the Council Directive on the approximation of the laws, regulations and administrative provisions of the Member States concerning liability for defective products (85/374/EEC) (2018).
[21] Expert Group on Liability and New Technologies, Report on Liability for Artificial Intelligence and Other Emerging Digital Technologies (2019).
[22] See Fn. 7.

This article analyzes to what extent this balance has been meaningfully struck, suggests specific amendments, and maps out paths for the future development of AI liability in the EU and beyond. To this end, Part II very briefly recaps the main risks of AI. Based on this, Part III provides an overview of the two proposed liability directives and highlights several core differences. Part IV engages in a detailed comparison, systematization and critique of the most relevant provisions of both proposals against the background of current law. Drawing on this assessment, Part V suggests strategies for the future development of AI liability regarding effective compensation, the safeguarding of innovation, incentives for trustworthy AI, and the achievement of sustainable AI regulation.

## II. Risks of AI

The risks of AI have been analyzed in numerous scholarly contributions.[23] Hence, this section limits itself to recalling the major risks which AI regulation should address. First, AI may exhibit technical autonomy, defined here as independence from human intervention coupled with a quasi-cognitive learning and/or adaptation capacity.[24] This feature may lead to output unforeseeable by human observers, including even the developers themselves. This concerns not only probabilistic, but just as much deterministic[25] AI models (see also Part V.3.a)).[26]

Second, AI systems exhibit different levels of opacity and complexity (see also Part V.3.b)).[27] Deep learning systems in particular are hard to interpret and hence technically opaque. Hence, it may be onerous or even close to impossible to provide an explanation for the model's output, in a specific case or more generally.[28] Moreover, institutional secrecy may hamper access to the model and its underlying data.[29] Both dimensions of opacity – technical and institutional – erect significant, and sometimes prohibitive, impediments to proving fault, defectiveness and causality for harm in the context of AI liability.[30]

---

[23] See, e.g., David Lorge Parnas, 'The real risks of artificial intelligence' (2017) 60 Communications of the ACM 27; Herbert Zech, 'Risiken Digitaler Systeme' (2020) Weizenbaum Series 1; Kaminski, 'Regulating the Risks of AI'.

[24] See, e.g., Catherine Tessier, 'Robots autonomy: Some technical issues', Autonomy and Artificial Intelligence: A Threat or Savior? (Springer 2017) 180; Stuart J. Russell and Peter Norvig, Artificial Intelligence: A Modern Approach (3rd Global ed. edn, Pearson Education, Inc. 2016) 39.

[25] These are models which, when analysing the same input, always deliver the same output. Probabilistic systems operate with random variables and thus deliver unforeseeable outcomes by definition, but are rarer; see David L Poole and Alan K Mackworth, Artificial Intelligence: Foundations of Computational Agents (2nd ed. edn, Cambridge University Press 2017), 29; Andreas Lindholm and others, Machine Learning (Cambridge University Press 2022), 127.

[26] Andrew D Selbst, 'Negligence and AI's human users' (2020) 100 BUL Rev 1315, 1331 et seq.; Roman V Yampolskiy, 'Unpredictability of AI: On the impossibility of accurately predicting all actions of a smarter agent' (2020) 7 Journal of Artificial Intelligence and Consciousness 109.

[27] Jenna Burrell, 'How the machine 'thinks': Understanding opacity in machine learning algorithms' (2016) 3 Big Data & Society 2053951715622512; Zachary C Lipton, 'The mythos of model interpretability: In machine learning, the concept of interpretability is both important and slippery' (2018) 16 Queue 31.

[28] Alejandro Barredo Arrieta and others, 'Explainable Artificial Intelligence (XAI): Concepts, taxonomies, opportunities and challenges toward responsible AI' (2020) 58 Information Fusion 82.

[29] Burrell, 'How the machine 'thinks': Understanding opacity in machine learning algorithms'.

[30] See, e.g., Zech, Liability for AI: public policy considerations.

Third, AI systems may perpetuate or exacerbate illegal discrimination (see also Part V.3.c)),[31] for example by using biased training data sets or incorrect target variables correlated with protected attributes.[32] Large, inclusive data sets are necessary, but difficult to establish and expensive to buy on the market.[33] Fourth, privacy and data protection are challenged by the need for large training data sets, particularly for deep learning purposes.[34] These risks, however, are covered by the GDPR will not be the focus of this paper. Fifth, the use of large data sets entails significant challenges for IT and cyber security.[35] Centrally stored data sets and models are prone to attacks seeking to perturb their functioning (e.g., adversarial attacks[36]) or to siphon off data,[37] with specific secure AI tools seeking to rise to the challenge.[38]

In the past, computer science and legal scholarship has mostly, to varying degrees, highlighted these five risks of AI.[39] Significantly, all of these risks are potentially compounded, and aggravated, by the growing interconnection between devices, data sets and applications, exacerbating difficulties of proving fault, defectiveness or causality.[40]

In recent years, however, yet another, crucial risk has emerged: the ecological dimension of AI, and technology more generally.[41] The non-negligible contribution of information and computing technology (ICT) in general and of machine learning in particular is only starting to be appreciated in the computer science and sustainability literature.[42] As the final part of this paper will show, it remains largely uncharted territory in the AI regulation landscape – a state that should change swiftly lest the window of opportunity offered by the current regulatory debate surrounding AI closes. As I shall argue, sustainable AI and technology regulation can, and should, contribute to the mitigation of climate change.

---

[31] See, e.g., Andrea Romei and Salvatore Ruggieri, 'A multidisciplinary survey on discrimination analysis' (2014) 29 The Knowledge Engineering Review 582; Eirini Ntoutsi and others, 'Bias in data-driven artificial intelligence systems—An introductory survey' (2020) 10 Wiley Interdisciplinary Reviews: Data Mining and Knowledge Discovery e1356.
[32] See, e.g., Ziad Obermeyer and others, 'Dissecting racial bias in an algorithm used to manage the health of populations' (2019) 366 Science 447.
[33] Ben Dickson, 'Why data remains the greatest challenge for machine learning projects' VentureBeat <https://venturebeat.com/ai/why-data-remains-the-greatest-challenge-for-machine-learning-projects/> accessed November 14, 2022.
[34] Cf. Ian Goodfellow, Yoshua Bengio and Aaron Courville, Deep Learning (MIT Press 2016) 19 et seqq.
[35] See, e.g., Jian-hua Li, 'Cyber security meets artificial intelligence: a survey' (2018) 19 Frontiers of Information Technology & Electronic Engineering 1462; Yuri Demchenko and others, Big security for big data: Addressing security challenges for the big data infrastructure (Springer 2013).
[36] Naveed Akhtar and Ajmal Mian, 'Threat of adversarial attacks on deep learning in computer vision: A survey' (2018) 6 IEEE Access 14410.
[37] See, e.g., Li, 'Cyber security meets artificial intelligence: a survey'; Colin Tankard, 'Big data security' (2012) 2012 Network Security 5.
[38] Tim Keary, 'MLsec could be the answer to adversarial AI and machine learning attacks' VentureBeat <https://venturebeat.com/security/mlsec-ai-machine-learning/> accessed November 14, 2022.
[39] See Fn. 23.
[40] See, e.g., Zech, Liability for AI: public policy considerations.
[41] See, e.g., Andrew A Chien, 'Computing's grand challenge for sustainability' (2022) 65 Communications of the ACM 5; Eng. Nat. Acad. of Sciences, Medicine, Fostering Responsible Computing Research: Foundations and Practices (2022), 54 et seq..
[42] See, e.g., Charlotte Freitag and others, 'The real climate and transformative impact of ICT: A critique of estimates, trends, and regulations' (2021) 2 Patterns 100340; ACM Tech. Policy Council, ACM TechBrief: Computing and Climate Change (2021); Josh Cowls and others, 'The AI gambit: leveraging artificial intelligence to combat climate change—opportunities, challenges, and recommendations' (2021) AI & Society 1.

## III. Overview of the new regime

The AI-specific update of the PLD and the new rules of the AILD Proposal are comparatively modest in reach and ambition compared to the much more far-reaching proposal the European Parliament submitted in October 2020. While the latter included a truly strict liability system for high-risk AI and a general reversal of the burden of proving fault for non-high-risk systems, the Commission proposals chiefly consist of disclosure regimes and case-specific alleviations of the burden of proof. Nonetheless, they continue to differentiate between high-risk and non-high-risk systems, as per the AI Act, setting up an intricate system of obligations for these different model uses, and for different types of addressees (see *Table 1*).

|  | HRAIS | Non-HRAIS | Manufacturer | Provider | Comm. User | Consumer |
|---|---|---|---|---|---|---|
| Disclosure of evidence (PLD) | 8(1) | 8(1) | 8(1) | - | - | - |
| Disclosure of evidence (AILD) | 3(1) | - | - | 3(1) | 3(1) | 3(1) |
| Presumption (defectiveness: PLD) | 9(2) & 9(4) | 9(2) & 9(4) | 9(2) & 9(4) | - | - | - |
| Presumption (fault: AILD) | 3(5) | - | - | 3(5) | 3(5) | 3(5) |
| Presumption (causal link: defect → output, PLD) | 9(3) & 9(4) | 9(3) & 9(4) | 9(3) & 9(4) | - | - | - |
| Presumption (causal link: fault → output, AILD) | 4(2), 4(3) & 4(5) | 4(5) & 4(6) | - | 4(2) & 4(5) | 4(3) & 4(5) | 4(6) |

*Table 1: Overview of the applicability of the AILD and PLD proposals to: HRAIS (High-Risk AI Systems); manufacturers (= developers of AI or AI components); providers (= developers who place system on the market under own name); commercial users; and consumers. Note that there is significant overlap between the terms of "manufacturer" and "provider". Furthermore, the PLD applies to other economic operators besides manufacturers (e.g., related service providers).*

### 1. The contentious choice of *two* directives

To start with, the regulation of AI liability is based on Article 114 TFEU. This is indeed justified to prevent fragmentation of the digital single market, prohibitive legal costs for companies providing AI across the EU, and obstacles for injured persons aiming to sue providers in cross-border settings.[43]

Currently, tort liability for AI systems rests on a dual-track regime: largely harmonized liability according to the PLD (to the extent that it currently applies to software[44]); and largely unharmonized liability according to the rules of national tort law (covered by the AILD Proposal). The very reasons advanced for the use of Article 114 TFEU, however, also clearly speak for the establishment of only *one* harmonized European regime of tort liability for products, including AI systems; under such a regime, Member States would be precluded from applying national liability systems with vastly divergent rules to product liability. This would indeed be a crucial element for the digital single market and facilitate the EU-wide rollout of AI models, and software more generally. At the moment, however, such a demarche is politically unlikely as most Member States continue to cherish their national tort regimes.

---
[43] Cf. Bertolini, Artificial intelligence and civil liability, 11.
[44] See, e.g., Geraint Howells, Christian Twigg-Flesner and Chris Willett, 'Product liability and digital products' in Tatiana-Eleni Synodinou and others (eds), EU Internet Law (Springer 2017).

It is precisely this dual-track system of European and Member State law which has likely prompted the publication of two separate proposals, by two different DG's, for AI liability. Even under the assumption that Member State tort liability for products remains operative, the choice of *two* directives with diverging content does not seem without alternatives. At a minimum, however, the same disclosure mechanisms, reversals of the burden of proof and other rules should apply to tort claims involving AI systems irrespective of whether the claim is based on Member State or EU product liability law.

The current choice of two directives, one addressed to AI specifically and one targeting product liability including AI, gives rise to significant problems. Importantly, this concerns the delimitation of the PLD's and the AILD Proposal's scope, which are supposed to be non-overlapping.[45] The AILD Proposal only applies to AI, the revised PLD more broadly to software. Some companies, however, claim to use AI when they really only harness standard software or slightly more complex algorithms, but not AI *stricto sensu*.[46] This could lead victims to choose the wrong compensation regime.

Hence, any instrument specifically tailored to AI needs to ensure that whoever in public statements *claims* to be using AI may also be *sued* under the AI liability regime. In a similar vein, the scope of contractual obligations in contracts for the sale of goods or digital content or services is chiefly influenced by the wording of advertisements.[47] Such a provision would reduce friction between the two regimes, and reduce legal uncertainty for potential plaintiffs.

### 2. General aim and thrust of the two directives

While the PLD Proposal contains amendments for material as well as procedural product liability law, the AILD Proposal follows a purely procedural approach: it only harmonises the rules for the disclosure and burden of proof when fault-based liability involving AI systems is at stake under Member State law. The two proposals hence follow different doctrinal approaches, but they do share several objectives. Most prominently, first, both proposals seek to ensure effective compensation of injured persons and prevent liability gaps arising from the peculiarities of AI technology.[48] Essentially, and convincingly, victims of AI systems should enjoy the same level of protection as injured persons in scenarios not involving AI,[49] including the ease of compensation.

Second, the new regime is supposed to foster legal certainty in an area of law plagued by hitherto unforeseeable ad-hoc adaptation to AI systems.[50] This, in turn, should facilitate the insurability of liability risks for companies and spur AI uptake and innovation, particularly with SMEs.[51] The Commission notes that in its survey of 2020, liability constituted the top external barrier to AI adoption by companies in the EU (43%).[52] Third, the incentivisation of trustworthy AI has been a long-standing goal of the EU initiative to regulate AI systems.[53] Therefore, it is not surprising that both proposals are meant to facilitate the rollout and uptake of trustworthy AI.[54] The PLD Proposal, together with several other EU initiatives, also aims to "minimise" the risks of digital products.[55] This requirement will be taken up in greater detail in the analysis of the concept of defectiveness and of the breach of a duty of care.

---

[45] PLD Proposal, 4.
[46] Cf. Giovanni Bruner, 'No, Artificial Intelligence doesn't exist (yet)' Towards Data Science <https://towardsdatascience.com/no-artificial-intelligence-doesnt-exist-yet-3318d83fdfe8> accessed November 14, 2022.
[47] See Art. 7(1)(d) of the Revised Sale of Goods Directive 2019/771; Art. 8(1)(b) of the Digital Content and Services Directive 2019/770.
[48] Rec. 3-4 AILD Proposal; Rec. 3 PLD Proposal.
[49] Rec. 29 AILD Proposal.
[50] Rec. 6 AILD Proposal; Rec. 4 PLD Proposal.

### 3. Relationship and key differences between the two directives

Since the Commission chose to operate with two directives, the relationship between them must be clarified. Both apply in parallel, but cover different and, according to the Commission,[56] mutually exclusive aspects.[57] Injured persons therefore may choose if they would like to proceed under the PLD, the AILD Proposal framework, or both. This is relevant because the covered damage positions, addressees, and substantive liability requirements differ between the frameworks (see *Table 2*).

| **PLD Proposal** | **AILD Proposal** |
|---|---|
| Claim rooted in EU law | Claim rooted in Member State law |
| Material and procedural aspects of product liability | Procedural aspects of liability for AI systems |
| Applicable to physical products and software, including AI systems | Applicable to AI systems only |
| Supposedly strict liability | Fault-based liability |
| Claims against manufacturers and other entities in the supply chain | Claims against manufacturers, professional users and consumers |

---

[51] AILD Proposal, 2.
[52] Ipsos, European enterprise survey on the use of technologies based on AI (2020) 58.
[53] HLEG AI, Ethics Guidelines for Trustworthy AI (European Commission 2019); Recital 5 AI Act; Johann Laux, Sandra Wachter and Brent Mittelstadt, 'Trustworthy Artificial Intelligence and the European Union AI Act: On the Conflation of Trustworthiness and the Acceptability of Risk' (2022) Working Paper, https://ssrncom/abstract=4230294; see also Mariarosaria Taddeo, 'Modelling trust in artificial agents, a first step toward the analysis of e-trust' (2010) 20 Minds and Machines 243; Wolter Pieters, 'Explanation and trust: what to tell the user in security and AI?' (2011) 13 Ethics and Information Technology 53; Frances S Grodzinsky, Keith W Miller and Marty J Wolf, 'Developing artificial agents worthy of trust:"Would you buy a used car from this artificial agent?"' (2011) 13 Ethics and Information Technology 17, 21; Andrea Ferrario, Michele Loi and Eleonora Viganò, 'In AI we trust incrementally: A multi-layer model of trust to analyze human-artificial intelligence interactions' (2020) 33 Philosophy & Technology 523.
[54] AILD Proposal p. 2; see also Rec. 4 PLD Proposal; PLD Proposal, 4-5.
[55] PLD Proposal, 4; more precisely, the aim should be, under an economic standard: to minimize the sum of 1) the cost of precaution and 2) the cost of expected damage, see Guido Calabresi, The cost of accidents: a legal and economic analysis (Yale University Press 1970) 31; Steven Shavell, Foundations of economic analysis of law (Harvard University Press 2004) 178.
[56] PLD Proposal, 4.
[57] See European Commission, Questions and answers on the revision of the Product Liability Directive (2022), under 9.

| | |
|---|---|
| Eligible damage: property, death or personal injury, and data loss | Eligible damage: potentially also fundamental rights and primary financial loss |
| Full harmonization | Minimum harmonization |

*Table 2: Overview of core differences between the PLD and the AILD Proposal*

The PLD Proposal applies to physical products as well as software, including AI systems. It harmonizes supposedly strict liability of manufacturers, and of other entities in the supply chain under certain conditions, based on product liability rooted in EU law. In contrast, the AILD Proposal restricts its scope to AI systems. It harmonizes procedural aspects of fault-based liability under Member State law. These rules not only cover claims against manufacturers but also against professional and non-professional users (= consumers).[58] Furthermore, eligible damage positions under the PLD are restricted to life, health, property, and loss of data, while Member State law is generally more open and may include the infringement of fundamental rights or primary financial loss.[59]

The two proposals finally also differ regarding their degree of harmonization. The PLD Proposal provides for full harmonization (Article 3 PLD Proposal). This article codifies the case law of the CJEU concerning the existing PLD,[60] in which the Court has repeatedly maintained that the PLD must be interpreted as being fully harmonizing within its scope of application.[61]

In contrast, the AILD Proposal only strives for minimum harmonisation.[62] Hence, a complete reversal of the burden of proof concerning fault remains possible. In some Member States, for example, this applies to drivers of motor vehicles, irrespective of the involvement of AI (as long as drivers can still be said to be 'driving').[63] Germany, for instance, has even introduced specific (and contested) duties for drivers of highly automated vehicles.[64] Similarly, Member States are free to introduce or maintain (truly) strict liability for AI, which is foreseen in some Members States for keepers of autonomous vehicles (AVs),[65] as long as the trigger of liability is not the defectiveness of the product. However, minimum harmonization by the AILD Proposal carries the real danger of fragmented liability system for AI across Member States, which could be a significant impediment to the EU-wide deployment of AI, particularly by SMEs.

---

[58] Ibid, under 9.
[59] Ibid, under 9.
[60] CJEU, Case C-183/00, María Victoria González Sánchez v. Medicina Asturiana ECLI:EU:C:2002:255 paras. 23 et seq.; CJEU, Case C-52/00: Commission of the European Communities v French Republic ECLI:EU:C:2002:252 paras. 13 et seq.; CJEU, Case C-154/00, Commission v. Greece, ECLI:EU:C:2002:254 paras. 9 et seq.
[61] Thomas Verheyen, 'Full Harmonization, Consumer Protection and Products Liability: A Fresh Reading of the Case Law of the ECJ' (2018) 26 European Review of Private Law 119, 121.
[62] Rec. 14 AILD Proposal.
[63] UK: fault-based; Germany presumption of driver's fault, see Melinda Florina Lohmann, 'Liability issues concerning self-driving vehicles' (2016) 7 European Journal of Risk Regulation 335, 336; cf. also Patti, 'The European road to autonomous vehicles' 129, 134.
[64] §§ 1a et seqq. StVG, see, e.g., for black box access: Wagner, 'Robot Liability' 27, 46.
[65] Patti, 'The European road to autonomous vehicles' 129, 134; Lohmann, 'Liability issues concerning self-driving vehicles' 335, 336: strict liability, e.g., in German, Swiss and French tort law.

### 4. The contentious choice of two *directives*

Even fully harmonizing directives always create the risk of diverging transpositions. The Data Protection Directive offers a prime example of this. The fragmented data protection landscape across Member States was one of the key reasons for the choice of a regulation as the instrument for the GDPR.[66] Hence, to prevent fragmentation and diverging transpositions between the Member States in this highly sensitive area for innovation, AI liability should equally be spelled out in *one* comprehensive *regulation*, not two directives.[67] This draws on the proposal of the European Parliament for AI liability, which also recommended the use of a regulation.[68]

A unified, coherent framework not only benefits the AI industry;[69] transpositions diverging between Member States are also prone to exacerbate difficulties for consumers to know what regime applies and how to sue providers located in different Member States.[70] Overall, the traditional competence of Member States concerning liability regimes must, at least in the non-contractual sphere, give way to the realities of a unified digital single market. Hence, the competence in Article 114 TFEU should be harnessed to provide an EU-wide, comprehensive framework for AI liability and software more generally.

## IV. A Critique of the two AI liability proposals

A unified framework for AI liability laid down in one single regulation remains an ideal that will, given political realities in Member States, probably not be realized in the scope of the current legislative initiative.[71] Therefore, the following section reviews and critiques in detail the various elements of the two proposals, highlighting differences between the two directives and/or the current legal framework, and suggesting amendments along the way. This concerns particularly the requirements for the applicability of the directives; the substantial trigger of liability; the new disclosure of evidence mechanism; the new burden of proof framework; the potential for representative actions; and some remaining elements exclusive to the PLD Proposal.

### 1. Applicability

The applicability requirements of the two directives differ substantially both regarding the material and the personal scope of application.

#### a) AILD Proposal: transplants from the AI Act

To delineate its scope, the AILD Proposal heavily relies on references to the AI Act. This not only implies that its applicability cannot yet be finally gauged, as the AI Act continues to be revised at the level of the Council and the European Parliament and will likely face a prolonged trialogue procedure. More importantly, however, the wisdom of transplanting concepts from the AI Act to the AI liability framework must be questioned in the first place.

---

[66] Rec. 9 GDPR.
[67] Cf. Bertolini, Artificial intelligence and civil liability, 67.
[68] European Parliament resolution of 20 October 2020 with recommendations to the Commission on a civil liability regime for artificial intelligence, (2020/2014(INL)).
[69] See, e.g., Wendehorst, AI liability in Europe: anticipating the EU AI Liability Directive, 8.
[70] See the remarks by MEP Axel Voss in: Bertuzzi, The new liability rules for AI.
[71] Wendehorst, AI liability in Europe: anticipating the EU AI Liability Directive, 4.

### i. Material applicability

The material applicability of the AILD Proposal hinges chiefly on the definition of AI and the classification of AI systems as high-risk (Articles 1 and 2(1) and (2) AILD Proposal).

### (1) Definition of AI

The procedural novelties of the AILD Proposal – disclosure of evidence and reversal of the burden of proof under certain circumstances – only apply in cases involving AI (Articles 3 and 4 AILD Proposal). Hence, the definition of this term is equally crucial to the AILD Proposal framework as for the AI Act. Understandably, the AILD Proposal imports the definition of AI, and of high-risk AI systems, from the AI Act (Article 2(1) and (2) AILD Proposal). Ironically, it is precisely this definition which continues to haunt lawmakers engaged in finalizing the AI Act.[72]

Indeed, they face a daunting challenge. Even in computer science, scholars do not even closely agree on a definition for AI.[73] Typically, concepts of AI distinguish between two main subtypes, machine learning on the one and logic-or expert-based approaches on the other hand.[74] This difference is now embodied in the new Recitals 6a and 6b specifying the more general definition of AI contained in Article 3(1) AI Act Council Version.[75] The core problem remains to distinguish AI from complex, but more traditional software, particularly when it uses statistical methods or hand-crafted rules. In the final version of the AI Act's Council compromise text, statistical methods are relegated to Recital 6a.[76] Hence, the definition of AI will likely comprise rather simple software tools, such as linear regression,[77] which have been used in a variety of industries, research and education for decades. Moreover, Recital 6b AI Act concerns logic- and knowledge-based approaches, which include expert systems and other rule-based models. Unless the criterion of "autonomy" in Article 3(1) AI Act is endowed with a meaning warranting a certain learning capacity or complexity threshold,[78] Recital 6b risks to bring relatively simple hand-crafted inference systems under the ambit of the AI Act.[79]

---

[72] Luca Bertuzzi, 'Artificial Intelligence definition, governance on MEPs' menu' EURACTIV <https://www.euractiv.com/section/digital/news/artificial-intelligence-definition-governance-on-meps-menu> accessed November 9, 2022.
[73] Overview in: Russell and Norvig, Artificial Intelligence: A Modern Approach-5; Poole and Mackworth, Artificial Intelligence: Foundations of Computational Agents 3-7; Matt O'Shaughnessy, 'One of the Biggest Problems in Regulating AI Is Agreeing on a Definition' Carnegie Endowment <https://carnegieendowment.org/2022/10/06/one-of-biggest-problems-in-regulating-ai-is-agreeing-on-definition-pub-88100> accessed October 25, 2022.
[74] Goodfellow, Bengio and Courville, Deep Learning; Poole and Mackworth, Artificial Intelligence: Foundations of Computational Agents 645 et seqq.
[75] It might, in general, have been wiser to focus on advanced machine learning rather than "AI" for the moment, see Philipp Hacker, 'Europäische und nationale Regulierung von Künstlicher Intelligenz' (2020) Neue Juristische Wochenzeitschrift 2142, 2142 et seq.; but see Hildebrandt, 'Global competition and convergence of AI Law', 2-3.
[76] "Machine learning approaches include for instance supervised, unsupervised and reinforcement learning, using a variety of methods including deep learning with neural networks, statistical techniques […]."
[77] Linear regression is a simple but very popular machine learning method for solving numerical problems, see Lindholm and others, Machine Learning.
[78] See Fn. 24; this problem remains under the EP Version.
[79] Cf. also the critique in the US Non-Paper on the AI Act, summarized in Luca Bertuzzi, 'The US unofficial position on upcoming EU Artificial Intelligence rules' EURACTIV (October 26, 2022)

With such a broad scope of application, the AI Act, and by implication the AILD Proposal, must be considered "advanced software regulation" rather than specific "AI regulation". This, perhaps inadvertently, significantly approximates the AILD Proposal to the PLD Proposal. The latter applies to software in general, with some burden of proof reversals reserved, however, to particularly complex software systems. This would indeed be an important consideration for the AI Act (and the AILD Proposal) as well: if a software surpasses a certain threshold of complexity and its outcomes hence become difficult to foresee, control or explain (without further analysis tools), it should not matter from a risk-based perspective if the underlying model qualifies as AI or not under some – rather arbitrary – definition. Crucially, the very risks of unforeseeability and opacity often ascribed to AI can easily be triggered by non-AI models containing thousands of features, too.[80]

In my view, such potentially unforeseeable and opaque "conventional" yet complex software should follow the same rules as potentially unforeseeable and opaque machine learning models. An even more radical approach would completely abandon a definition of AI in favor of full-blown risk regulation, covering any software triggering specific risks such as opacity, discrimination, or unforeseeability, for example. Under such a framework, which can only be sketched here, distinct provisions would be connected to each specific risk, ensuring a tailored regulatory environment for software exhibiting certain risks irrespective of the – often contested – AI label. While such a shifted focus of the AI Act to cover complex software may not occur at this stage of the legislative process anymore, it nevertheless constitutes the most defensible option under a risk-based approach in my view for reasons of specificity and proportionality.

Under either option, a significant grey zone will appear at the borders of the definition (of AI or complex software). Here, technical standards (Art. 40 and 41 AI Act) and an expansive list of concrete examples for software not falling under the relevant definition are urgently needed to mitigate uncertainty about the applicability of the AI Act and large parts of the AILD Proposal.[81]

### (2) Risk categories

The second consequential transplant from the AI Act to the AILD Proposal concerns the risk classification of AI systems (Article 2(2) AILD Proposal). The disclosure of evidence rules, and part of the alleviation of the burden of proof, only apply to high-risk AI systems (Articles 3 and 4(2) to (4) AILD Proposal). While an alignment with the AI Act concerning the definition of AI (or software) is paramount, it is submitted that the AILD Proposal must be decoupled from the AI Act concerning risk classification: the high-risk category of the AI Act is both over- and underinclusive, particularly when directly transplanted to an AI liability framework.

### (a) Under-inclusiveness: missing high-risk use cases

In the context of the digital economy, the AI Act will (likely) apply only to systems used for employment, credit scoring, and products falling under the New Legislative Framework of product safety, such as medical devices.[82] At least under the Commission proposal and the

---

<https://www.euractiv.com/section/digital/news/the-us-unofficial-position-on-upcoming-eu-artificial-intelligence-rules/> accessed November 7, 2022.
[80] Cf. Lipton, 'The mythos of model interpretability: In machine learning, the concept of interpretability is both important and slippery'; I am grateful for discussions with Maximilian Eber on this subject.
[81] I am grateful for discussions with Rasmus Rothe on this subject.
[82] Cf. also Veale and Borgesius, 'Demystifying the Draft EU Artificial Intelligence Act—Analysing the good, the bad, and the unclear elements of the proposed approach'.

version adopted by the Council on December 6, 2022, this excludes certain AI applications, such as AVs,[83] emotion recognition systems[84] or insurance pricing models (the latter are currently only covered regarding life and health insurance[85]). Importantly, these excluded models are just as much prone to significantly damage affected persons materially or immaterially as those considered high-risk under the AI Act. AVs are the epitome of products threatening life and limb of affected persons, such as passengers and other traffic participants. Emotion recognition systems pointedly intrude on the privacy, potentially even the intimacy, of affected persons,[86] which is why the European Parliament added them to the list of high-risk systems in its position adopted on June 14, 2023.[87] Concerning insurance, it bears noting that any risk for which a (rational) person takes out an insurance will, by implication, be significant.

Hence, the wisdom of simply transplanting risk categories from the AI Act must be questioned, even though it does ensure coherence between the instruments. Coherence, however, is not a goal in itself, particularly if there are good reasons for differentiating. While the AI Act does not contain subjective rights of affected persons and therefore implicitly adopts a more abstract vision of risk geared toward society at large, a regime for damages necessarily must consider individual specificities and differences as well.

The materialization of certain risks may be very unequally distributed across individuals. If, in a given society of one million citizens, only five individuals stand to be significantly damaged by a certain AI application, this may not be enough to categorize the application as high-risk under the AI Act. However, such an individually pronounced risk must nevertheless be addressed by an effective system of liability and compensation. Hence, in the case of low total expected damage yet significant variance of the risk between affected individuals, the individual and the social risk classification may diverge. This is a phenomenon well-known from the theory of social choice in general.[88]

This theoretical difference undermines the proposed equation of risk categories from the AI Act with those of the AILD Proposal. To do so conflates social and individual risk and negates the right to an effective remedy for those affected by low-probability AI damage. To the very least, legislators would have to acknowledge that such significant risks may arise for affected individuals concerning AVs, emotion recognition systems and insurance pricing models, even if they continue to maintain that the risk level of those systems is not high enough on aggregate to merit their qualification as high-risk under the AI Act (although, in my view, such a qualification would be well justified). Hence, such applications must at least qualify as high-risk under the AILD Proposal.[89]

---

[83] Technically, autonomous vehicles will be considered high-risk (Article 6(1) and (2) AI Act), but are exempt from all of the core obligations of the AI Act (Articles 2(2) and 84 and Annex II Section B No. 2, 3, 6 and 7 AI Act), hence rendering the relevant references in Articles 3 and 4 AILD Proposal inapplicable to them.
[84] See Article 52(2a) AI Act.
[85] See Annex III (5)(e) AI Act, and the deleted predecessor under (d).
[86] Andrew McStay, 'Empathic media and advertising: Industry, policy, legal and citizen perspectives (the case for intimacy)' (2016) 3 Big data & Society 2053951716666868, 6; Philipp Hacker, 'Manipulation by algorithms. Exploring the triangle of unfair commercial practice, data protection, and privacy law' (2021) European Law Journal, 32.
[87] Annex III (1)(aa) AI Act EP Version.
[88] Amartya Sen, 'Social Choice Theory' (1986) 3 Handbook of Mathematical Economics 1073 et seq.; Matthew D. Adler, 'The Social Welfare Function: A New Tool for Regulatory Policy Analysis' in Stefan Grundmann and Philipp Hacker (eds), Theories of Choice (OUP 2021), 155.
[89] While specific liability rules exist in all EU countries concerning AVs targeting owners and drivers, it still seems desirable to provide for product liability to shift liability from car insurance companies to manufacturers, who control AV development and risks.

### (b) Over-inclusiveness: ChatGPT et al.

Conversely, it should be asked how the providers of broadly applicable models, such as ChatGPT, GPT-4, DALL·E 2, and Stable Diffusion, may best the treated. The Council version of the AI Act, adopted in December 2022, denominated them general-purpose AI systems (GPAIS). While the definition of and provisions on GPAIS in the AI Act are hotly debated as of the writing of this manuscript,[90] they are usually called "Foundation Models" or "large generative models" in computer science.[91] These terms cover the latest generation of machine learning models, such as ChatGPT, DALL·E 2, and Stable Diffusion. Based on this, the European Parliament introduced Art. 28b in its position on the AI Act in June 2023 to tackle the specific challenges of Foundation Models and generative AI.

The Council text generally subjects providers of GPAIS to the full range of obligations for high-risk AI if the GPAIS may be used as a high-risk system or as a component thereof (Article 4b(1) AI Act Council Version), unless a limited exception applies (Article 4c AI Act Council Version). As a result, almost all GPAIS would have to comply with the full range of the AI Acts high-risk obligations and, by extension, be subject to the stricter liability rules under the AILD (and the PLD) Proposal. This concerns many powerful models (such as speech or image recognition or text-to-speech/speech-to-text conversion tools, but also generic ML model building platforms offering pre-trained models). Many of these models are currently offered as open-source software and used extensively by developers (e.g., TensorFlow;[92] cf. Article 3(1b) AI Act Council Version).

The Parliament, in turn, opted for a much preferable staggered approach containing three levels of obligations for providers of Foundation Models,[93] which had also been suggested in the literature.[94] First, all providers have to comply with minimum requirements concerning, for example, data governance, cybersecurity, technical documentation, as well as risk assessment, mitigation and management (Level 1, Art. 28b(2) AI Act EP Version). Second, if used in specific high-risk applications, the full range of obligations for high-risk AI systems applies (Level 2, Art. 6 et seqq. AI Act EP Version). Third, duties for sharing of information among members of the AI value chain (Foundation Model provider; entities providing fine-tuning; professional users) are foreseen (Art. 28(2)-(2b) AI Act EP Version).

Notably, the minimum requirements in Level 1 might eventually be restricted to particularly large and "systemically relevant" Foundation Models, similar to the approach in the DSA.[95] Irrespective of these limitations, the final version of AI Act will very likely feature specific rules on generative AI, too, along the lines proposed by the European Parliament in Art. 28b(4)

---

[90] See, e.g., Carlos Ignacio Gutierrez and others, 'A Proposal for a Definition of General Purpose Artificial Intelligence Systems' (2022) Working Paper 2022, https://ssrncom/abstract=4238951.

[91] Mark Chen and others, 'Evaluating large language models trained on code' (2021) arXiv preprint arXiv:210703374; Rishi Bommasani and others, 'On the opportunities and risks of foundation models' (2021) arXiv preprint arXiv:210807258.

[92] https://www.tensorflow.org/learn?hl=en.

[93] Foundation Models are defined as "an AI system model [sic] that is trained on broad data at scale, is designed for generality of output, and can be adapted to a wide range of distinctive tasks", Art. 3(1c) AI Act AP Version; see also Rishi Bommasani and others, 'On the opportunities and risks of foundation models' (2021) arXiv preprint arXiv:210807258, 2.

[94] Philipp Hacker, Andreas Engel and Marco Mauer, 'Regulating ChatGPT and other Large Generative AI Models' (2023) ACM Conference on Fairness, Accountability, and Transparency (FAccT '23) 1112; see also Natali Helberger and Nicholas Diakopoulos, 'ChatGPT and the AI Act' (2023) 12 Internet Policy Review.

[95] Kai Zenner, 'A law for foundation models: the EU AI Act can improve regulation for fairer competition' OECD AI Policy Observatory Blog (July 20, 2023), <https://oecd.ai/en/wonk/foundation-models-eu-ai-act-fairer-competition>.

AI Act. According to these amendments, three basic, additional rules apply with respect to generative AI: any interaction of generative AI with humans must be rendered transparent; illegal output; and copyrighted training data must be disclosed.

These new rules, while presenting steps into the right direction, raise a number of concerns. Most importantly, the comprehensive risk assessment and management should not be located at Level 1.[96] This requirement implies that Foundation Model providers need to map out, meticulously described, and established integrative monitoring and mitigation systems for thousands of potential high-risk use cases of their general-purpose systems – even though only a fraction of these might eventually be implemented in practice. Such a system would have to include any possible high-risk use case of the model – hence, almost every imaginable high-risk scenario on the planet. Such an obligation does not seem useful, nor quite meaningfully fulfillable given the very large number of purely hypothetical scenarios involved.[97] Ultimately, this would amount to an inefficient compliance exercise which would better be shifted to Level 2.

It would be difficult for any company developing Foundation Models, but particularly for non-commercial open-source providers, to meet the obligations of the AI Act for *all possible* high-risk use cases (Council Version) or to map out all the risks (EP Version). Already, a recent study has concluded, Europe will likely struggle to develop competitive Foundation Models.[98] Such obligations, driving up compliance costs, will add another competitive disadvantage to the EU landscape for Foundation Model development. Hence, the restriction of the most onerous rules of Level 1 to systematically relevant Foundation Models seems sensible from a competitive perspective. Turning to liability again, the combination of the AI Act rules on Foundation Models and generative AI with liability may be over-deterrent for open-source providers. Hence, to the very least, the open-source exception contained in Recital 13 PLD Proposal should be extended to the AILD Proposal (see below, Part IV.1.c)i.(2)).

### ii. Personal applicability

Regarding personal applicability, the AILD Proposal covers providers and users, both as defined in the AI Act (Article 2(3) and (4) AILD Proposal). Providers are developers who also put the product onto the market (Article 3(2) AI Act). The reference is again important for the coordination with the AI Act (Rec. 15 AILD Proposal), but leaves the personal scope of applicability open, to a certain extent, as the definitions of provider and user in the AI Act were modified during the legislative process and may still be changed until the final version of the AI Act. Significantly, users now also include consumers, as per the final version of the AI Act's compromise text (Article 3(4) AI Act), even though they are exempted from most obligations (Article 2(8) AI Act). Hence, the AILD Proposal generally applies to consumers using AI systems as well, as will be discussed in detail below (IV.4.d) Case 4, Part IV.4.e) and Part V.1.f)).

---

[96] Hacker, Engel and Mauer, 'Regulating ChatGPT and other Large Generative AI Models', 1115; see there also for a critique of the other rules in Art. 28b AI Act EP Version.
[97] See also the US Non-Paper on the AI Act, summarized in Bertuzzi, 'The US unofficial position on upcoming EU Artificial Intelligence rules'.
[98] Future of Life Institute, Emerging Non-European Monopolies in the Global AI Market (2022).

### b) PLD Proposal: a new software regulation

In contrast to the AILD Proposal, the PLD Proposal draws much less on the AI Act to delineate its perimeter of application. Rather, in a small revolution for the PLD regime, it now explicitly covers software of all types.

### i. Material applicability

The PLD Proposal finally contains the long-awaited extension of the product liability framework to the digital economy. It now includes, in its definition of "product" in Article 4(1), digital manufacturing files[99] and software.[100] This is crucial as the CJEU had recently excepted information (contained in a print newspaper) from the concept of a product in the *Krone* case[101] – which may have been read as an implicit exclusion of standalone software.[102]

#### (1) Software and AI

Importantly, software in the sense of the revised PLD includes AI[103] (which, after all, is only a specific software which includes a more or less complex mathematical model[104]). The PLD therefore finally expands its material scope of applicability to generally include non-high-risk AI systems (as per the AI Act), and any type of software beyond AI.[105] It reaches much further, in this sense, than the AILD Proposal and the AI Act itself.

#### (2) Open-source software

To facilitate innovation and access to software, however, the revised PLD excludes free and open-source software that is developed or supplied outside the course of a commercial activity (Rec. 13 PLD). This exception should be welcomed as contributors to open-source software often work on the code in their free time and could be significantly discouraged from contributing if they were subject to PLD liability. To make this exception binding, though, it should be included in the definition of a product in Article 4(1) PLD.[106]

In fact, many widely used commercial software tools do rely, in parts, on open-source software.[107] This immediately raises the question of the applicability of the PLD if an open-source component of a commercially developed and sold software contains a bug. On the one hand, one could argue that any liability of the developer of the commercial software would threaten to undermine the liability privilege of open-source contributors: the commercial provider could turn around and seek indemnification from the open-source programmers. However, since there are often no economically viable alternatives to open-source software, open-source programmers should be in a strong bargaining position to exclude indemnification in their Terms of Service. Hence, it seems more important to incentivize commercial providers to thoroughly vet open-source software they include into their own software products. In the past, bugs or security deficiencies in open-source software have triggered some of the most aggressive and damaging hacking campaigns.[108] Hence, incentives for code vetting by commercial providers must be reinforced, and they should therefore be liable under the PLD (and the AILD Proposal to the extent that an open-source exception is added) for bugs in open-source software integrated into their own products.

---

[99] This denotes the digital version or digital template of a movable (Article 4(2) PLD Proposal).
[100] Currently, it is debated whether software constitutes a product, see, e.g., Howells, Twigg-Flesner and Willett, 'Product liability and digital products'; Wagner, 'Robot Liability' 27, 41.
[101] CJEU, Case C-65/10, Krone, para. 42.
[102] See Gerhard Wagner, 'Software as a product' in Sebastian Lohsse, Reiner Schulze and Dirk Staudenmayer (eds), *Smart Products* (Nomos 2022), 157, 171 et seqq.; Gerhard Wagner, 'Liability Rules for the Digital Age - Aiming for the Brussels Effect' (2023) European Journal of Tort Law (forthcoming), 13-14.
[103] Commission, Questions and answers on the revision of the Product Liability Directive, under 8.: "The revised PLD […] makes clear that software, including AI systems, is a product".
[104] See e.g., Lindholm and others, Machine Learning.
[105] Spindler, 'Die Vorschläge der EU-Kommission zu einer neuen Produkthaftung und zur Haftung von Herstellern und Betreibern Künstlicher Intelligenz', 690.
[106] See now also Wagner, 'Liability Rules for the Digital Age - Aiming for the Brussels Effect', 15.
[107] See. e.g. application softwares Google Chrome, Spotify as mentioned in: Elmar Geese, 'Kommerzielle Open-Source-Software in Unternehmen nutzen' 2021) <https://www.computerweekly.com/de/meinung/Kommerzielle-Open-Source-Software-in-Unternehmen-nutzen> accessed 3 November 2022; Mislav Streicher, 'What is Commercial Open Source Software' 2020) <https://www.webiny.com/blog/what-is-commercial-open-source> accessed 4 November 2022; on commercial open source firms: Dirk Riehle, 'The Commercial Open Source Business Model' (5th Americas Conference on Information Systems, AMCIS 2009) 18, 19.

## ii. Personal applicability

The PLD Proposal suggests applying the liability regime to "economic operators" (Articles 1 and 7 PLD Proposal), the new generic term replacing the concept of "producers" used in the current version of the PLD (Article 1 PLD). The latter comprises manufacturers of the finished product or of components, producers of raw materials, importers, and persons representing themselves as producers, e.g. by affixing a trademark (Article 3(1) and (2) PLD). Subsidiary liability of suppliers is established in Article 3(3) PLD.

### (1) Economic operators

Article 4(15) PLD Proposal now expands the scope by defining the term "economic operator". It includes any of the entities covered under the PLD, but adds the provider of a related service, the authorized representative, and the fulfilment service provider (the supplier now being called the distributor).[109] The importer and the authorized representative of the manufacturer, as well as the fulfilment service provider in a subsidiary manner, can be held liable if the manufacturer itself is established outside the EU (Article 7(2) and (spell three) PLD Proposal). If neither of them can be identified, distributors (= entities along the supply chain, Article 4(15) PLD Proposal) can be sued. This recognizes the importance of logistics channels and hybrid products. Furthermore, as an acknowledgment of the importance of e-commerce, online platforms can be held liable under the same conditions as distributors if they are not shielded by the hosting exception under the Digital Services Act.[110] In a doctrinally intriguing extension, the PLD Proposal takes up the recent CJEU judgement in *Wathelet*, whereby contractual liability under sales law covers intermediaries not clearly acting as agents of a represented entity.[111] Furthermore, the PLD does not, in contrast to the AILD Proposal, apply to users as defined in the AI Act.

In practice, AI will often be developed by an independent company and integrated into a software product developed by a second company. In this case, the PLD Proposal follows the approach of the current PLD framework: if the component is defective, the manufacturer (= developer) of the component and the manufacturer of the end product are jointly and severally liable (Articles 5(1), 7(1)(2), 11 and 4(16) and Recitals 26 and 40 PLD Proposal).

---

[108] Zakir Durumeric and others, The matter of heartbleed (2014); Haytham M Mohamed and Omar El-Gayar, 'Security Vulnerability Impact on Open Source: A Social Media Exploration' (2022).
[109] Each of the entities is defined under Article 4(11) to (15) PLD Proposal.
[110] Online platforms need to fall under the exception to the hosting liability exception, contained in Article 5(3) of the original Digital Services Act proposal. This means that the average consumer needs to believe that the platform itself, or an entity under its control, provides the object of the transaction.
[111] CJEU, Case C-149/15, Wathelet, ECLI:EU:C:2016:840.

Should the injured person receive compensation from the end product manufacturer, that entity can then usually turn around to seek reimbursement from the component manufacturer according to the Member State rules on joint and several liability.[112]

### (2) Embedded software and related services

Significantly, the PLD now holds liable entities providing related services, being defined in Article 4(4) PLD Proposal as digital services "integrated into, or inter-connected with, the product in such a way that its absence will prevent the program from performing one or more of its functions". Doctrinally, this is an important paradigm shift as the PLD does not, generally, apply to services, but only to products.

The Commission rightly acknowledges the fact that devices increasingly contain digital components critical for their functioning, for example Internet of Things devices, and that product liability must not hinge on formal categorization as a service or manufacturing process. Such embedded products are now rightly covered by the PLD Proposal as well, and the developers of the digital component held responsible. This is fully justified as it cannot make a difference, from the point of view of compensation as well as incentivization, if a software is sold separately or as a bundle with a physical product. For example, if a microphone is equipped with an AI-based natural language processing (NLP) software to automatically transform speech into text, the NLP system would be considered a related service as it is essential to function of that microphone. The developers of the NLP software are subject to PLD liability as a consequence. Similarly, and importantly, those providing training data for an AI system offer an essential related service. Hence, the developers of the training data – of crucial importance for machine learning[113] – are now subject to product liability as well.[114] Art. 10 AI Act will have to serve as a reference point for the standard of defectiveness (see also below, Part V.2.a)).

Note that non-essential related-service software generally constitutes a component of the final product (Article 4(3) PLD Proposal), and its manufacturer is therefore covered by the PLD as well (Article 4(11) PLD Proposal). Recital 15 PLD Proposal specifies the limits of related service provider responsibility: related services "should be considered as components of the product to which they are inter-connected, when they are within the control of the manufacturer of that product, in the sense that they are supplied by the manufacturer itself or that the manufacturer recommends them or otherwise influences their supply by a third party." Hence, there are two ways in which entities supplying related services can be liable under the PLD Proposal: as related services providers if the service is essential to at least one product function (Article 4(4) and 16 PLD Proposal); or as component manufacturers if they relate to the product manufacturer by way of recommendation or influence of the latter (Article 4(16) in conjunction with Recital 15 PLD Proposal).

### 2. Substantial trigger of liability under the directives

Given the diverging doctrinal points of departure of both proposals, the substantial criteria triggering liability differ significantly, at least at first glance. While the claim under the AILD Proposal presupposes fault, the PLD requires the defectiveness of the product in question.

---

[112] See, e.g., § 426 BGB for German law.
[113] Hacker, 'A legal framework for AI training data—from first principles to the Artificial Intelligence Act'.
[114] Spindler, 'Die Vorschläge der EU-Kommission zu einer neuen Produkthaftung und zur Haftung von Herstellern und Betreibern Künstlicher Intelligenz', 691.

### a) AILD Proposal: Fault

Fault is the key trigger of liability for cases brought under the AILD Proposal. Traditionally, fault implies intent or negligence on the part of the tortfeasor.[115] While the proposal refrains from defining the term to grant leeway to Member States continuing their traditional tort law regimes, the disclosure of evidence and reversal of the burden of proof, in most cases, requires a substantive violation of the AI Act.[116] In some cases, the violation of an AI-specific duty of care outside of the AI Act suffices to trigger the obligations of the AILD Proposal.[117] Such a link between the violation of substantial AI regulation and procedural alleviations had long been suggested by scholars.[118]

Recital 15 AILD Proposal contains a noteworthy restriction of the types of fault covered by the directive. Succinctly put, the alleviations (disclosure; burden reversal) are not supposed to apply if a human agent intercedes between the AI output and the damage. This particularly concerns the use of AI as recommender or decision-support systems. Alleviations are only provided if decisions are based on human fault, but fully automated from the moment of AI output on. The idea behind this restriction is that, if a human agent intervenes, causality (and non-compliance) are just as difficult to prove as in cases not involving AI (Recital 15 AILD Proposal).

In my view, there are two problems with this restriction. First, limiting the applicability of a provision to fully automated decision making is well known from Article 22(1) GDPR.[119] In this context, some scholars have long warned that such a wording invites circumvention of the rule by having AI output "rubberstamped" by human actors blindly following machine advice.[120] The Article 29 Working Party,[121] and recent court decisions,[122] rightly point to a broader understanding of automated decision making, though, which will only be invalidated by substantive human intervention.

Even if not installed to circumvent the provision on purpose, humans monitoring and potentially correcting AI decisions may be prone to automation bias.[123] While the empirical strength of this claim remains disputed,[124] it is nevertheless clear that the limitation significantly restricts the scope of application of the AILD Proposal. In practice, many consequential decisions are passed through human review, particularly in case of AI output with a low confidence interval. Similarly, the very invocation of data subject rights under Article 22(3) GDPR, such as a human intervention, would paradoxically remove the very decision from the scrutiny of the AILD.

---

[115] See, e.g., Simon Deakin and Zoe Adams, Markesinis and Deakin's Tort Law (8th edn, Oxford University Press 2019), 87; Wendehorst, AI liability in Europe: anticipating the EU AI Liability Directive, 4.
[116] Articles 3 and 4(2) to (4) AILD Proposal.
[117] Article 4(5) and (6) AILD Proposal.
[118] Mario Martini, 'Algorithmen als Herausforderung für die Rechtsordnung' (2017) 72 Juristen Zeitung 1017, 1023-1024; Hacker, 'Europäische und nationale Regulierung von Künstlicher Intelligenz'; Technologies, Report on Liability for Artificial Intelligence and Other Emerging Digital Technologies 49 et. seqq.; Wendehorst, 'Strict liability for AI and other emerging technologies' 150, 159-160; Zech, 'Liability for AI: public policy considerations' 147, 152-153.
[119] Spindler, 'Die Vorschläge der EU-Kommission zu einer neuen Produkthaftung und zur Haftung von Herstellern und Betreibern Künstlicher Intelligenz', 700.
[120] See, e.g., Sandra Wachter, Brent Mittelstadt and Luciano Floridi, 'Why a right to explanation of automated decision-making does not exist in the general data protection regulation' (2017) 7 International Data Privacy Law 76, 92; Lea Katharina Kumkar and David Roth-Isigkeit, 'A Criterion-Based Approach to GDPR's Explanation Requirements for Automated Individual Decision-Making' (2021) 12 J Intell Prop Info Tech & Elec Com L 289, 292.
[121] Article 29 Working Party, Guidelines on Automated individual decision-making and Profiling for the purposes of Regulation 2016/679, WP251rev.01, 2018, 21.
[122] See, on the Amsterdam Uber robo-firing case decided by the Amsterdam Cort of Appeals, Jakob Turner, 'Amsterdam Court Upholds Appeal in Algorithmic Decision-Making Test Case: Drivers v Uber and Ola' Fountain Court Blog (June 4, 2023), https://www.fountaincourt.co.uk/2023/04/amsterdam-court-upholds-appeal-in-algorithmic-decision-making-test-case-drivers-v-uber-and-ola/.
[123] Saar Alon Barkat and Madalina Busuioc, 'Human-AI Interactions in Public Sector Decision-Making:'Automation Bias' and 'Selective Adherence'to Algorithmic Advice' (2022) XX Journal of Public

Moreover, Article 14 AI Act would even compel providers to enable human oversight of high-risk AI systems, and Art. 29(1), (1a) and (4) AI Act require professional users to implement such oversight, rendering much of the AILD Proposal potentially inapplicable and obsolete. The latest version of the AI Act has therefore rightly deleted the reference to automated decision-making in its requirements for high-risk AI systems.[125] Instead, Article 6(3) and Recital 32 AI Act now spell out a carve-out for purely accessory use of AI systems, such as in translation or document management regarding high-risk activities. This is convincing from the systematic perspective of the AI Act. However, even errors in translation or document managing may result in significant individual harm to, and concomitant difficulty of proof for, injured persons. Hence, the carve-out for accessory use should not be transplanted to the AILD Proposal; and the ill-fated reference to automated decision-making in Recital 15 AILD Proposal should be deleted.

Second, it must be questioned whether it is really equally difficult to prove causality and non-compliance in cases of human intervention with and without AI involvement. In the former case, claimants have to prove that the human agent violated a duty of care concerning the monitoring and override of the AI system.[126] While such a demonstration may be equally difficult in the case of oversight of complex non-AI machinery, it is clearly more challenging than the demonstration of a breach of the duty of care in non-technical or non-scientific environments. Hence, the presumption of non-compliance contained in Article 3(5) AILD Proposal should apply equally if a human agent took, or failed to take, the final decision leading to the damage caused by the AI output.

### b) PLD Proposal: defectiveness of the product

The key requirement triggering liability under the PLD Proposal is, as under the current PLD, the defectiveness of the product (Article 6 PLD Proposal). From an economic and a technical perspective, defectiveness is the crucial element as economic operators, such as manufacturers/developers, are best positioned to avoid defects leading to harm.[127] They usually control the development and production risks, incur the least social costs for mitigating them (least-cost avoider argument), and hence should be incentivized to do precisely that.[128] If rightly

---

Administration Research and Theory, 3, 7; Stefan Strauß, 'Deep Automation Bias: How to Tackle a Wicked Problem of AI?' (2021) 5 Big Data and Cognitive Computing 18.

[124] See, e.g., Nina Grgić-Hlača, Christoph Engel and Krishna P Gummadi, 'Human decision making with machine assistance: An experiment on bailing and jailing' (2019) 3 Proceedings of the ACM on Human-Computer Interaction 1; Alon Barkat and Busuioc, 'Human-AI Interactions in Public Sector Decision-Making:'Automation Bias' and 'Selective Adherence'to Algorithmic Advice', 12.

[125] See the now deleted Article 6(3)(a) AI Act.

[126] See, e.g., A Michael Froomkin, Ian Kerr and Joelle Pineau, 'When AIs outperform doctors: confronting the challenges of a tort-induced over-reliance on machine learning' (2019) 61 Ariz L Rev 33, 97; Hacker and others, 'Explainable AI under contract and tort law: legal incentives and technical challenges'; David Schneeberger, Karl Stöger and Andreas Holzinger, The European legal framework for medical AI (Springer 2020) 219.

[127] Cf. Shavell, Foundations of economic analysis of law.

[128] Wagner, 'Liability Rules for the Digital Age - Aiming for the Brussels Effect', 9; Wagner, 'Robot Liability' 27, 57; Herbert Zech, 'Liability for autonomous systems: tackling specific risks of modern IT' in

configured, the liability framework may then theoretically minimize the sum of precaution and accident costs.

### i. The consumer expectations test

The new provision on defectiveness only obliquely mirrors these economic and technical considerations. Rather, it echoes Article 6 PLD by recurring on the consumer expectation perspective. Again, a comprehensive assessment is needed ("taking all circumstances into account"). As has repeatedly been noted, the consumer expectation standard is vague at best and bordering on a concept devoid of any content at worst.[129] In fact, US product liability law therefore uses the so-called risk-utility test,[130] which is also gaining traction in EU law (see below).[131] Under this test, the product is considered effective if the safety gain from an alternative design is greater than the utility reduction (including increased cost) of that alternative design.[132]

The PLD, in turn, attempts to add contours to the consumer expectation standard by listing several criteria that need to be considered. It currently mentions three criteria: the presentation of the product; its expected use; at the time of putting it into circulation (Article 6(1) PLD). This list is bound to be expanded by the PLD Proposal, which contains several aspects important for AI.

### (1) Learning after deployment

One impediment to finding defects in AI systems is the potential ability to learn after being put onto the market, the time relevant according to Article 6(c) PLD at the moment. Defects arising after that moment are hence generally excluded from liability (later-defect defence, see also below, Part IV.6.b)). The PLD Proposal now includes into the comprehensive assessment any potential ability to learn after deployment (Article 6(1)(c) PLD Proposal).

This is clearly a meaningful step forward as the developers are the least-cost avoiders regarding defects occurring due to continued learning. For example, if an ML steering component of an autonomous vehicle picks up aggressive driving styles from other, human drivers after deployment, the economic operators must ensure that the adaptive driving feature does not lead to aggressive driving beyond a certain risk tolerance threshold (at which point, for the sake of example, the defect would arise). Similarly, if a chatbot or a language translation tool learns from suggestions by users, the economic operators must ensure, via curation mechanisms or manual supervision, that the system does not deteriorate due to erroneous suggestions by or discriminatory interactions with users, as happened in the case of Microsoft's Tay bot.[133]

However, it must be mentioned that only some AI systems do continue to learn after being deployed (so-called online learning).[134] The learning process can be, and often is, stopped after

---

S.Lohsse/R.Schulze/D.Staudenmayer (ed), Liability for Artificial Intelligence and the Internet of Things (Nomos 2019) 187, 196-197; Zech, 'Liability for AI: public policy considerations' 147, 152-153.
[129] Wagner, 'Robot Liability' 27, 35-36; Gerhard Wagner, '§ 823' in C. H. Beck (ed), Münchener Kommentar zum BGB (2017), para. 731-733; Simon Whittaker, Liability for Products (Oxford University Press 2005) 481-494.
[130] David G Owen, 'Risk-Utility Balancing in Design Defect Cases' (1997) 30 U Mich JL Reform 239.
[131] Wagner, 'Liability Rules for the Digital Age - Aiming for the Brussels Effect', 18.
[132] Ibid.
[133] Elle Hunt, 'Tay, Microsoft's AI chatbot, gets a crash course in racism from Twitter', The Guardian (March 24, 2016), <https://www.theguardian.com/technology/2016/mar/24/tay-microsofts-ai-chatbot-gets-a-crash-course-in-racism-from-twitter>, last accessed November 28, 2022.
[134] Ekaba Bisong, 'Batch vs. online learning', Building Machine Learning and Deep Learning Models on Google Cloud Platform (Springer 2019), 199.

the model was trained by the developers. Hence, the PLD correctly identifies the economic operators as the least-cost avoiders for emergent damage.

### (2) Control after deployment

Another item on the list targets situations in which the manufacturer retains control of the product after placing it on the market or putting it in service (irrespective of whether the system continues to learn or not). Again, the economic operator, by exercising control, significantly influences the risks exhibited by the system and hence must continue to be liable for defects until that control is relinquished (Article 6(1)(e) PLD Proposal). This could be a situation where a machine is (partially) remote-controlled by the manufacturer or developers during its operation.

### (3) Product safety and cybersecurity

The PLD Proposal embeds product safety requirements into the updated feature list, expressly linking the civil liability framework with the administrative product safety regime (Article 6(1)(f) PLD Proposal). This is all the more important as the latter regime is being overhauled, with a view to AI software, in parallel to the PLD.[135] Significantly, the provision explicitly mentions safety-relevant cybersecurity requirements. These are not only contained in Article 32 GDPR but also in the proposed Cyber Resilience Act[136] and in Article 15 AI Act.[137] The PLD hence cross-references, and thereby integrates, important IT-related requirements from existing and future EU IT law.

### ii. Defects in AI systems

The PLD fails to address one fundamental aspect concerning the defectiveness of AI systems, however. As is well known, AI models make mistakes that humans would not have made. Conversely, they may avoid mistakes humans would have made. The Tesla crash which occurred in May 2016 epitomizes this tendency.[138] The autopilot mistook a large truck crossing on an intersection for a distant bridge and steered the vehicle right into the truck, killing the Tesla passenger. Clearly, a human driver in any normal state would not have mistaken the truck for a bridge. This raises the question whether this bug amounts to a defect under product liability law.

As is well known, under EU product liability law, three different types of defect may be found: manufacturing, design and instruction (= warning) defects.[139] In a construction defect case, the concrete product in question negatively deviates from the design blueprint.[140] An instruction or warning defect occurs if the instructions for installation, use or maintenance contain an error or a relevant omission (cf. Article 6(1)(a) PLD Proposal).[141] Both types of defect do not present any major challenges regarding AI-based product.

---

[135] European Commission, Proposal for a Regulation of the European Parliament and of the Council on machinery products, COM/2021/202 final.
[136] European Commission, Proposal for a Regulation on cybersecurity requirements for products with digital elements - Cyber Resilience Act (2022)
[137] Spindler, 'Die Vorschläge der EU-Kommission zu einer neuen Produkthaftung und zur Haftung von Herstellern und Betreibern Künstlicher Intelligenz', 693.
[138] Danny Yadron and Dan Tynan, 'Tesla driver dies in first fatal crash while using autopilot mode' The Guardian <https://www.theguardian.com/technology/2016/jun/30/tesla-autopilot-death-self-driving-car-elon-musk> accessed November 17, 2022.
[139] Mathias Reimann, 'Product liability', Comparative Tort Law (Edward Elgar Publishing 2021) 243.
[140] Bertolini, Artificial intelligence and civil liability, 52; Gerhard Wagner, 'Produkthaftung für autonome Systeme' (2017) 217 Archiv für die civilistische Praxis 707, 725; BGH Case VI ZR 107/08, Airbag, BGHZ 181, 253 para. 15.

Most cases concerning AI liability will, however, turn on design defects. While it is possible, of course, that a specific copy of an AI model in a specific device suffers from a unique bug – for example, because the download was interrupted and part of the model not installed (= construction defect) – or that the instructions of use contain a mistake (= instruction defect), most cases will likely involve errors that occur in all the equivalent models deployed on all comparable devices. For instance, if an image recognition model fails to recognize certain structures (truck vs. bridge), or a scoring system does not function well for specific levels of income, or a face recognition program does not correctly identify persons above a certain age, this is an error that applies to all versions of the model in the same way. All of these examples are therefore instances of potential design errors. The crux, hence, lies in the definition and finding of a design error in AI contexts.

### (1) Anthropomorphic approach

In the legal literature, an anthropomorphic and a system-oriented approach to design defects in AI systems offer competing conclusions. Under the former, a judge would have to find a defect if an average human being would not have made the mistake (and consumers would have expected the human not to make this mistake).[142] This affords the advantage of incentivizing operators to proactively find and eradicate truck-as-a-bridge mistakes and similar edge cases. On the other hand, it could be the case that an AI system such as the autopilot avoids all possible mistakes, including all made by humans, but that it still continues to confuse tracks and bridges under some very rare and unpredictable conditions (apparently, in the Tesla case, the sun was setting behind the crossroads). Imagine that it would be costly, but not impossible to eliminate this error. Overall, the AI system even with this error reduces the overall accident rate by 90% compared to human drivers.

In such a situation, the anthropomorphic standard would indeed find a defect in a truck-as-a-bridge case. This may lead to over-deterrence. Moreover, determining whether an error of the AI system would have been committed by a human is not only highly difficult to operationalize, but also quite arbitrary. From a technical point of view, and from the point of view of the injured person, it does not make any difference if the error would have been made by a human replacement or not.

### (2) AI-specific approach

By contrast, the system-oriented approach would abstract from the individual case and evaluate the risk level of the entire AI system rolled out across all applicable devices (in this case: cars). Only if a malfunctioning exists at the systemic level, a defect is found.[143] In my view, this is the correct approach. Technically, a model is usually centrally trained and then deployed in many different devices or scenarios. The model may be adapted during or after deployment, a situation contemplated by Article 7(4) PLD Proposal.[144] Generally, however, the design is determined at the modelling stage, not at the stage of the individually deployed model. This points to the centrality of directing incentives toward that stage. An AI-specific perspective would then take into account technical standards, but also needs to consider situations in which these are still lacking.

---

[141] Bertolini, Artificial intelligence and civil liability, 52; Wagner, 'Produkthaftung für autonome Systeme', 748; BGH Case VI ZR 107/08, Airbag, BGHZ 181, 253 para. 29.
[142] Georg Borges, 'Haftung für selbstfahrende Autos' (2016) 32 Computer und Recht 272, 275-276; Christian Gomille, 'Herstellerhaftung für automatisierte Fahrzeuge' (2015) JuristenZeitung 76, 77.
[143] For outlines and limits of such a system-related concept of defect, see Mark A. Geistfeld, 'A roadmap for autonomous vehicles: State tort liability, automobile insurance, and federal safety regulation' (2017) 105 Calif L Rev 1611, 1645-1647; see also Wagner, 'Robot Liability' 27, 44-45.
[144] In this case, the entity substantially modifying the model is considered a manufacturer of the new product.

### (a) Technical standards

As scholars have suggested, defectiveness should generally be defined by comparison with technical standards governing specific industry branches.[145] The main advantage is that this would align PLD liability with the AI Act and thus provide a certain amount of cohesion and legal certainty. Indeed, the AI Act encourages the establishment of technical standards (see Articles 40 and 41 AI Act). If such standards do exist, they should exhaustively define defectiveness in their scope of application, or at least define safe harbours for developers. This would be conducive to innovation while providing potential claimants with clear guidelines as to when a defect has occurred (given that they have access to the necessary data, see below). Often, however, such standards do not (yet) exist.[146] In that case, different reference points are needed.

Regarding autonomous driving, one could compare the error rate of the entire AI system (e.g., of one specific Tesla type which runs on the same AI model or at least comparable versions of the model) to that of the average human driver, based on, for example, accidents per kilometre. Here, one could argue that any error rate of the system under scrutiny that is lower than the average human error rate excludes a defect, as the reduction of accidents is a key aim of traffic regulation.[147] However, equating defectiveness with infra-human performance eliminates all incentives from product liability law to improve the system any further than the breakeven point with human performance. Conversely, even AI models performing less well than humans on average might be socially highly advantageous, and ultimately not defective: they may make solutions, such as medical treatment, available where otherwise there would simply be none. Even imperfect solutions might be preferable to the status quo.

Hence, in my view, an AI-specific approach cannot be restricted to a comparison with human performance. While the perspective must be systemic because AI models continue to be developed and deployed beyond individual instances, any standard of defectiveness for AI products must fulfil two objectives: allow the rollout even of AI models with infrahuman performance if these clearly improve upon the status quo; and include incentives for improving the system even in the domain of suprahuman performance.

### (b) Infrahuman performance

If the model does not, on average, perform as well as humans in a given task,[148] this should trigger a rebuttable presumption of defectiveness. On the basis of a risk-utility framework, the economic operators would have to make two comparisons to rebut the presumption: with the status quo without the AI model; and with reasonable alternative AI model designs, taking the list and Article 6(1) PLD Proposal into account.

---

[145] Geistfeld, 'A roadmap for autonomous vehicles: State tort liability, automobile insurance, and federal safety regulation'; Spindler, 'User liability and strict liability in the Internet of Things and for robots', 129, 143; see also Gerald Spindler, 'Interaction between product liability and regulation at the European level' in Fabrizio Cafaggi and Horatia Muir Watt (eds), The Regulatory Function of European Private Law (Edward Elgar Publishing 2009).
[146] Spindler, 'User liability and strict liability in the Internet of Things and for robots', 137.
[147] For German law cf. BGH, Case VI ZR 155/14, Freeway service station, NJW 2015, 1174 (BGH) para. 12; BVerfG, Case 1 BvR 118/71, Promotional Trips, BVerfGE 40, 371 (BVerfG), para. 33.
[148] This may be measured by appropriate performance measures, see Lindholm and others, Machine Learning, 88; if time is considered depends on the task at hand.

- To avoid a defect, a model, first, needs to improve upon the status quo without the model in terms of safety. For models with infrahuman performance, this suggests that they cannot, by definition, replace human decision makers. They may improve the risk profile for affected persons, however, if they are integrated into a smart decision-making framework under which the model makes a recommendation which is reviewed by humans. Operators need to show, however, for example by way of Randomized Controlled Trials, that this set up improves safety for affected persons vis-à-vis a purely human decision-making process. Another setting in which an improvement can be demonstrated is the expanded reach of decisions. For example, models detecting cancer hold the potential to bring medical diagnostics to persons who would otherwise remain entirely untreated.[149] Here, again, operators will have to demonstrate that affected persons, in this case patients, are on average better off with AI facilitated diagnostics than without it. In the case of expanding medical care to areas where there was none before, there are strong arguments in favour of such a conclusion, even though they are not uncontested.[150]
- Second, as in general product liability law, a defect still arises if the reasonable alternative design exists which was not implemented by the operators. In other words, a risk-utility analysis must show that the elimination of the error would be inefficient, i.e., that it would be costlier to implement an equally performing alternative design risk which avoids the error than to pay for expected damages. The application of this test is no longer specifically restricted to the United States.[151] The German Federal Private Law Court (BGH) famously ruled in 2009 in the Airbag case that such a risk utility analysis is an important part of the EU test for design defects.[152] Importantly, this implies, for example, that public or private healthcare providers cannot blindly deploy mediocre AI to substitute for investment in human resources if additional personnel, realistically available for hire, would improve the situation for patients even more than the contemplated AI system. In other words, the financial neglect of a healthcare system cannot serve as a pretext to justify a liability shield for mediocre AI.
- Third, if errors cannot be efficiently avoided, operators must effectively warn consumers of them.[153]

### (c) Suprahuman performance

Suprahuman performance, in turn, should not be considered a *carte blanche* for economic operators to assume non-defectiveness. Rather, incentives for innovation and improvement need to be set in this domain as well. This cannot mean, however, that only the industry leader is considered free of defects and any competitors not quite meeting its performance rate are deemed defective. Such an interpretation would lead to a dangerous winner-takes-all logic,[154] which has proven deleterious to competition in markets governed by network effects. Article 6(2) PLD Proposal equally rules out such an interpretation by noting that the introduction of a better product in the market cannot be the sole reason for finding a defect. Rather, to incentivize continued innovation and performance in the suprahuman domain, economic operators should be able to avoid the finding of the design defect under the following three conditions:

- First, a risk-utility analysis as in the case of infrahuman performance must show the absence of a reasonable alternative design.
- Second, the entire product needs to exhibit positive social utility.[155] This criterion is generally not met, for example, by malicious AI developed to deceive and manipulate.[156]
- Third, operators must again warn of errors which cannot be remedied by a reasonable alternative design.[157]

---

[149] See, e.g, Eric J Topol, 'High-performance medicine: the convergence of human and artificial intelligence' (2019) 25 Nature medicine 44, 50-51; Giorgio Quer and others, 'Wearable sensor data and self-reported symptoms for COVID-19 detection' (2021) 27 Nature Medicine 73-77; Eric R.Krishnan/P.Rajpurkar/E.J.Topol, 'Self-supervised learning in medicine and healthcare' (2022) online first Nature Biomedical Engineering 1-7.
[150] Froomkin, Kerr and Pineau, 'When AIs outperform doctors: confronting the challenges of a tort-induced over-reliance on machine learning'; see also Schneeberger, Stöger and Holzinger, The European legal framework for medical AI; Hacker and others, 'Explainable AI under contract and tort law: legal incentives and technical challenges'.
[151] See also Deakin and Adams, Markesinis and Deakin's Tort Law 91.
[152] BGH Case VI ZR 107/08, Airbag, BGHZ 181, 253 para. 17.
[153] Cf. Howard Latin, 'Good warnings, bad products, and cognitive limitations' (1993) 41 UCLA L Rev 1193.

These criteria, in my view, balance the burden for economic operators with incentives to improve in the suprahuman performance space and to effectively protect affected persons, by alternative designs or, failing that, by (necessarily less effective) warnings.

### (d) Summary

Overall, the defectiveness of products should therefore be linked to technical standards. The AI Act encourages their development and endows them with binding effect. Where they do not exist, however, different solutions need to be found. Infrahuman AI performance should trigger a presumption of defectiveness which may be rebutted by, inter alia, demonstrating an improvement upon the status quo without the model. In the suprahuman AI performance domain, in turn, manufacturers still need to implement reasonable alternative designs to avoid design defects, setting incentives to further improve AI systems.

### 3. Disclosure of evidence and access to information

Under both proposals as they currently stand, a central concern remains: the key points triggering liability (fault; defectiveness) have to be proven, at least generally, without alleviation of the burden of proof. In general product liability cases, however, a pronounced asymmetry of information exists between the manufacturer and the injured person concerning the potentially defective product.[158] This asymmetry is even exacerbated regarding AI systems, which tend to be technically complex.[159] Both proposals, hence, foresee disclosure of evidence mechanisms.

---

[154] Wagner, 'Robot Liability' 27, 45.
[155] Hans-Bernd Schäfer and Claus Ott, Lehrbuch der ökonomischen Analyse des Zivilrechts (5th edn, Springer 2012) 157; see also Owen, 'Risk-Utility Balancing in Design Defect Cases' 243.
[156] Miles Brundage and others, 'The malicious use of artificial intelligence: Forecasting, prevention, and mitigation' (2018) arXiv preprint arXiv:180207228.
[157] Cf. Latin, 'Good warnings, bad products, and cognitive limitations'.
[158] Thomas Verheyen, 'On Behavioural Asymmetry in Product Liability Law' (2021) 12 Journal of European Tort Law 40-64; Juan José Ganuza, Fernando Gomez and Marta Robles, 'Product liability versus reputation' (2016) 32 The Journal of Law, Economics, and Organization 213-241; Oswald A Mascarenhas, Ram Kesavan and Michael Bernacchi, 'Buyer–seller information asymmetry: Challenges to distributive and corrective justice' (2008) 28 Journal of Macromarketing 68-84.

### a) Article 3 AILD Proposal

A cornerstone of the AILD Proposal and its push for effective compensation is Article 3 AILD Proposal, which spells out specific disclosure duties for potential defendants. Member State courts may order a certain disclosure procedure, akin to yet significantly different from US pretrial discovery (see below), to enable potential defendants to ascertain if and whom to sue.[160] Note that, in contrast to the presumption of causality under Article 4, the disclosure mechanism is not restricted to fault-based liability claims and therefore applies to strict liability claims as well. However, it is limited to litigation involving high-risk AI systems.

The court order hinges on two prerequisites (Article 3(1) AILD Proposal): The potential claimant needs to provide facts and evidence sufficient to support the plausibility of the damages claim; and the potential defendant needs to have refused access to the necessary information despite having it at its disposal. Importantly, in a claim for damages, the potential claimant must have undertaken all proportionate attempts to obtain the evidence from the potential defendant (Article 3(2) AILD Proposal).

The disclosure order needs to be limited to what is necessary and proportionate, and include a general balancing of the legitimate interest of all parties (Article 3(4) AILD Proposal). In particular, the risk management system as well documentation, records and logs may be of interest for potential claimants.[161] However, trade secrets falling under the EU Trade Secrets Directive and other confidential information must be given due weight in this balancing exercise. As a result, specific confidentiality measures may be installed by the court, such as limiting access to documents, hearings or transcripts.[162] Third parties not originally involved in a litigation may only also be required to provide information under the same balancing conditions.[163]

The refusal of the potential defendant to provide information may not be considered by the court as a factor in assessing potential non-compliance.[164] This is a departure from the case law established in general non-discrimination law, where the CJEU ruled in *Meister* that potential victims of discrimination do not have a right to access information, but that the refusal of the potential perpetrator to provide necessary information may constitute one factor in the court's assessment concerning a prima facie case of discrimination.[165] In contrast, in the AILD Proposal, it seems justified not to factor a refusal into the non-compliance assessment as, in contrast to non-discrimination law, potential claimants may eventually obtain the information via a court order.

The consequences are different if the potential defendant fails to comply with the court order for disclosure: as discussed in detail below, this triggers a rebuttable presumption of non-compliance with the relevant duty of care (Article 3(5) AILD Proposal).

---

[159] Fn. 27; Rec. 30 PLD Proposal.
[160] Rec. 17 AILD Proposal.
[161] Cf. Rec. 26 AILD Proposal.
[162] Rec. 20 AILD Proposal.
[163] Rec. 19 AILD Proposal.
[164] Rec. 17 AILD Proposal.
[165] CJEU, Case C-415/10, Meister, ECLI:EU:C:2012:217 paras. 46-47.

### b) Article 8 PLD Proposal

The PLD Proposal contains an essentially similar mechanism for the disclosure of evidence in its Article 8. Its scope of application is significantly broader, however, than that of the equivalent AILD Proposal provision: Article 8 PLD Proposal may be invoked to obtain information regarding non-high-risk AI systems as well.

Again, claimants need to present facts and evidence for the plausibility of the claim; evidence must be at the disposal of the defendant; disclosure is limited to what is necessary and proportionate in consideration of the legitimate interests of all parties, particularly trade secrets and confidential information; and specific measures can be invoked to preserve the confidentiality of such information. Under these conditions, failure to comply with an order to disclose relevant evidence leads to a rebuttable presumption of defectiveness (Article 9(2)(a) PLD Proposal), just like the refusal to comply with an AILD Proposal disclosure order entails a presumption of non-compliance with the relevant duty of care (Article 3(5) AILD Proposal).

However, some minor differences between the two disclosure systems exist: the AILD Proposal is more detailed in its requirements in several cases. To start with, a claimant under the PLD need not previously ask the defendant to divulge the evidence (cf. Article 8(1) PLD Proposal), which is compulsory under Article 3(1) AILD Proposal. As a consequence, plaintiffs under the PLD need not have undertaken all proportionate attempts at gathering the relevant evidence from the defendant before a court order in a damages case (cf. Article 3(2) AILD Proposal). A specific provision empowering courts to order evidence preservation measures, articulated in Article 3(3) AILD Proposal, is also lacking in the PLD equivalent, and should be added.

Finally, and perhaps most importantly, the AILD Proposal disclosure mechanism can be invoked both during and before an actual trial (Article 3(1) AILD Proposal: "request of a potential claimant […] or a claimant"). In contrast, the PLD equivalent seems to apply only once a claim has been registered in court (Article 8(1) PLD Proposal: "claimant"; "defendant"). The difference can be substantial if, as in many countries and areas of the law, claimants have to pay court fees and pay a deposit upon filing their claim in court. These fees are likely to be much lower for a mere disclosure proceeding than for the main proceeding which includes the disclosure procedure. Hence, the PLD Proposal should include a pretrial discovery possibility as well to lower barriers for compensation.

### c) Assessment

The new (pre-)trial disclosure instruments constitute a welcome addition to the procedural toolbox available to potential and actual claimants.

#### i. Merits

First, the disclosure process contributes to effective enforcement as the potential claimant otherwise would likely be barred from gathering the necessary information to make an informed and rational decision about whether to sue the potential defendant. This again speaks in favour of extending the PLD discovery to pretrial settings. Second, in turn, effective enforcement strengthens incentives for compliance with the AI Act. Third, the process may prevent non-meritorious suits at an early stage, as potential claimants may refrain from litigation if they cannot gather enough evidence. This exonerates courts and saves social cost.[166] Fourth, under the AILD Proposal, the potentially requested information largely matches the documentation, record keeping and information obligations already established in the AI Act.[167] The additional burden for providers of high-risk AI systems will likely be low.

---
[166] Cf. Rec. 17 AILD Proposal.

### ii. Shortcomings and recommendations

On the flip-side, however, the alignment of the AILD Proposal disclosures with the AI Act may also constitute a significant shortcoming. The recipients of the evidence disclosure are, generally, strikingly different from those of the typical AI Act disclosures. The latter are addressed toward experts with significant technological knowledge, such as developers, tech consultants, AI watchdogs, or regulatory agencies.[168] In contrast, under both directives, evidence in liability proceedings will be primarily requested by claimants, such as consumers, or counsel without a strong technical background, and needs to be interpreted by judges trained in law, but not AI. While one could recur on expert witnesses to interpret technical disclosures, these are costly and therefore again raise the barriers for enforcement. Hence, specific information understandable by laypersons (specifically, counsel with experience in IT law) is needed to facilitate litigation. Such information requirements would come on top of the documentation required by the AI Act in its current form; it seems necessary, however, as the disclosure of evidence provisions otherwise risks becoming mere law in the books, devoid of practical effect.

This immediately points to another weakness of the proposals, however: the very requirement to provide facts and evidence to support the plausibility of the damages claim. Some threshold of this type is indeed necessary to shield companies and developers from vexatious suits or the weaponisation of the disclosure system to obtain information about trade secrets.[169] Nonetheless, it bears noting that a similar provision exists in antidiscrimination law with a limited success in facilitating effective private enforcement. According to the non-discrimination directives, claimants have to provide facts suggesting a prima facie case of discrimination.[170] As a result, the burden of proof is reversed, and the defendant needs to show that, in fact, the decision was not tainted by discrimination. However, even gathering evidence to support a prima facie case of discrimination often overwhelms potential claimants.[171] The AI disclosure rules should strive to avoid a similar fate. Hence, it should be sufficient for potential claimants to provide facts and evidence of damage and the involvement of an AI system for the disclosure of evidence rules to be triggered. This would balance the legitimate needs of companies to be shielded from vexatious suits with the individual and social desiderata of effective enforcement. If, however, the claimant is a potential or actual competitor of the defendant, the plausibility requirement should apply as foreseen under the current proposals. Additionally, the safeguards for trade secrets need to be particularly strictly enforced by the courts in such situations.

---

[167] While the wording of Article 3 AILD Proposal also covers evidence beyond the AI Act disclosures and records, litigation will likely focus on these for practical reasons.
[168] Philipp Hacker and Jan-Hendrik Passoth, Varieties of AI Explanations under the Law. From the GDPR to the AIA, and Beyond (Springer 2022) 361.
[169] See, e.g., James R McKown, 'Discovery of Trade Secrets' (1994) 10 Santa Clara Computer & High Tech LJ 35.
[170] See, e.g., Article 9 of Directive 2004/113/EC.
[171] Hacker, 'Teaching fairness to artificial intelligence: existing and novel strategies against algorithmic discrimination under EU law' 1143, 1168; Wachter, Mittelstadt and Russell, 'Why fairness cannot be automated: Bridging the gap between EU non-discrimination law and AI' 10; Zuiderveen Borgesius, 'Strengthening legal protection against discrimination by algorithms and artificial intelligence' 1572, 1577 et seq.

Third, both proposals establish the necessary condition, for disclosure of evidence, that information is at the defendant's disposal. This, in turn, can be difficult to gauge from the outside. While it is clear what type of information a provider or user *should* possess in observation of the AI Act's rules, it is not guaranteed that the evidence is indeed at the defendant's disposal. Furthermore, Article 8 PLD Proposal also applies to non-high-risk AI systems, for which the AI Act does not prescribe any specific information, logging or record-keeping requirements. Hence, it should be clarified that both provisions apply whenever the information is, *or should legally be* (e.g., under the AI Act), at its disposal.[172] This would cover cases in which the defendant, in violation of the AI Act or other statutory or contractual obligations, failed to produce or keep certain kinds of information, logs or records. Such breaches should not exclude the defendant's obligation to disclose pertaining evidence, even if this implies creating the information ex post for the sake of the damages case- For example, if the defendant failed to disclose metrics on accuracy and discriminatory impacts required under Article 11, Annex IV(2)(g) AI Act, it would have to produce these documents as part of the discovery procedure (cf. Rec. 31 PLD Proposal).

Fourth, in order to further lower barriers for claimants, the directives should ensure that the pretrial disclosure case can be brought at the same court as the case in the main proceedings, usually at least the court where the damage occurred.[173] If consumers or collective redress organizations have to sue at the place of the defendant's headquarters, this may entail prohibitive costs of legal fees (e.g., for foreign counsel).

Moreover, the general problems known from pretrial discovery persist,[174] particularly the threat of the tool being leveraged for obtaining trade secrets or pressuring financially more vulnerable entities. The US Federal Rules of Civil Procedure seek to mitigate the risk by allowing courts to grant protective orders "to protect a party or person from annoyance, embarrassment, oppression, or undue burden or expense," in particular the revelation of trade secrets. The protective order can allow a party not to reveal a trade secret or it can allow them to reveal it only in a specific way.[175] Years of US experience with these problems should be integrated into guidelines concerning the protection of trade secrets foreseen in the balancing test in both proposals.

In my view, however, the final and biggest problem with the AILD Proposal discovery procedure lies in its limitation to high-risk AI systems as per the AI Act. The idea is to spare non-high-risk providers the additional burden of drawing up information they are not required to keep or generate under the AI Act.[176] While generally understandable from a proportionality perspective, such a restriction is rendered futile by the applicability of the PLD disclosure mechanism to non-high-risk AI developers. Furthermore, it crucially depends on the correct classification of applications as high-risk. As mentioned, certain applications not considered high-risk under the AI Act (at the moment) certainly may significantly harm affected persons in material or immaterial ways, leaving them without effective redress options under the AILD Proposal. For example, AVs and emotion recognition systems are not, under the current AI Act draft, qualified as high-risk systems (and the fate of insurance pricing is uncertain).[177] Hence, these systems would have to be included either in the high-risk category of the AI Act or, failing that, in the list of systems to which the pretrial discovery procedure extends under the AILD Proposal. In these cases, given the risks posed by the models, the additional burden for providers seems justified even if they are (erroneously) spared this burden under the AI Act.

---

[172] See also Spindler, 'Die Vorschläge der EU-Kommission zu einer neuen Produkthaftung und zur Haftung von Herstellern und Betreibern Künstlicher Intelligenz', 696.
[173] Cf. Art. 7(2) of Regulation (EU) No 1215/2012.
[174] See, e.g., Stephen N Subrin, 'Discovery in Global Perspective: Are We Nuts' (2002) 52 DePaul L Rev 299; Hein Kötz, 'Civil justice systems in Europe and the United States' (2003) 13 Duke J Comp & Int'l L 61, 74; McKown, 'Discovery of Trade Secrets'.
[175] See F.R.C.P. Rule 26(c)(1)(G). Courts may appoint special masters to examine potential trade secrets subject to discovery, F.R.C.P. Rule 53(a); see also Peter F Daniel, 'Protecting Trade Secrets from Discovery' (1995) Tort & Insurance Law Journal 1033.
[176] Rec. 18 AILD Proposal.

### iii. Harmonization of mechanisms

Overall, it would make sense to harmonize the disclosure mechanisms by including the same prerequisites for disclosure, irrespective of whether the claim is based on the PLD or on national liability rules. Generally, the additional provisions in the AILD Proposal seem convincing and should be transplanted to the PLD framework as well. Requesting the information explicitly from the potential defendant before filing a claim for discovery in court makes economic and practical sense, for example.[178] Furthermore, it is crucial to offer potential claimants the possibility to start the discovery procedure independently of, and before, the main proceedings to lower economic barriers for effective access to courts and compensation. A single system for disclosure of evidence would also be easier to handle for all involved parties. There is no convincing reason to withhold evidence under the PLD which is required to be produced under the AILD Proposal, and vice versa.

## 4. Burden of proof

Next to the disclosure mechanism, the second key element of both the AILD Proposal and the PLD Proposal is to alleviate the claimants' burden to prove specific elements of the provisions underlying the damages claim. While the AILD Proposal addresses non-compliance with a duty of care, the PLD Proposal focuses on the corresponding product defectiveness. Moreover, both proposals reverse the burden of proof for the causal link between the breach of a duty of care ('fault') and the output of an AI system (AILD Proposal), or between defectiveness and damage (PLD) (see *Table 3*).[179] Such modifications of the burden of proof had been proposed by numerous scholars and expert groups before the Commission proposal concerning fault/duty of care,[180] defectiveness,[181] and causality.[182] In fact, other areas of EU law already include similar

---

[177] See the discussion around Fn. 85.
[178] Most Member State may offer defendants the opportunity to directly accept the claim and avoid paying legal fees for it if the claimant did not ask for the subject of litigation before going to court; but there might be exceptions to this rule, now or in the future.
[179] See now also Herbert Zech, 'Liability for AI: Complexity problems' in Sebastian Lohsse, Reiner Schulze and Dirk Staudenmayer (eds), *Liability for AI* (Nomos 2023 (forthcoming)).
[180] See, e.g., Herbert Zech, 'Entscheidungen digitaler autonomer Systeme: Empfehlen sich Regelungen zu Verantwortung und Haftung?' in Ständige Deputation des Deutschen Juristentages (ed), Verhandlungen des 73 Deutschen Juristentages • Hamburg 2020/Bonn 2022, Band 1: Gutachten: Ergänzungsband (C.H.Beck 2022), Gutachten A., 60; Spindler, 'User liability and strict liability in the Internet of Things and for robots' 125, 138 Martin Sommer, Haftung für autonome Systeme: Verteilung der Risiken selbstlernender und vernetzter Algorithmen im Vertrags-und Deliktsrecht, vol 4 (Nomos Verlag 2020) 375-379; Technologies, Report on Liability for Artificial Intelligence and Other Emerging Digital Technologies 48-49; Meik Thöne, Autonome Systeme und deliktische Haftung: Verschulden als Instrument adäquater Haftungsallokation?, vol 1 (Mohr Siebeck 2020) 257-261.
[181] See, e.g., Gerhard Wagner, 'Verantwortlichkeit im Zeichen digitaler Techniken' (2020) Versicherungsrecht 717, 735; provided, that it is proven that the system caused the damage: Technologies, Report on Liability for Artificial Intelligence and Other Emerging Digital Technologies 42 et seqq.; Wagner, 'Produkthaftung für autonome Systeme' 707, 747.
[182] See, e.g., Technologies, Report on Liability for Artificial Intelligence and Other Emerging Digital Technologies, 49 et seq.; Zech, 'Entscheidungen digitaler autonomer Systeme: Empfehlen sich Regelungen zu

alleviations of the burden of proof triggered by breaches of substantial law (see, e.g., Article 9(1) of the Market Abuse Regulation[183]).[184]

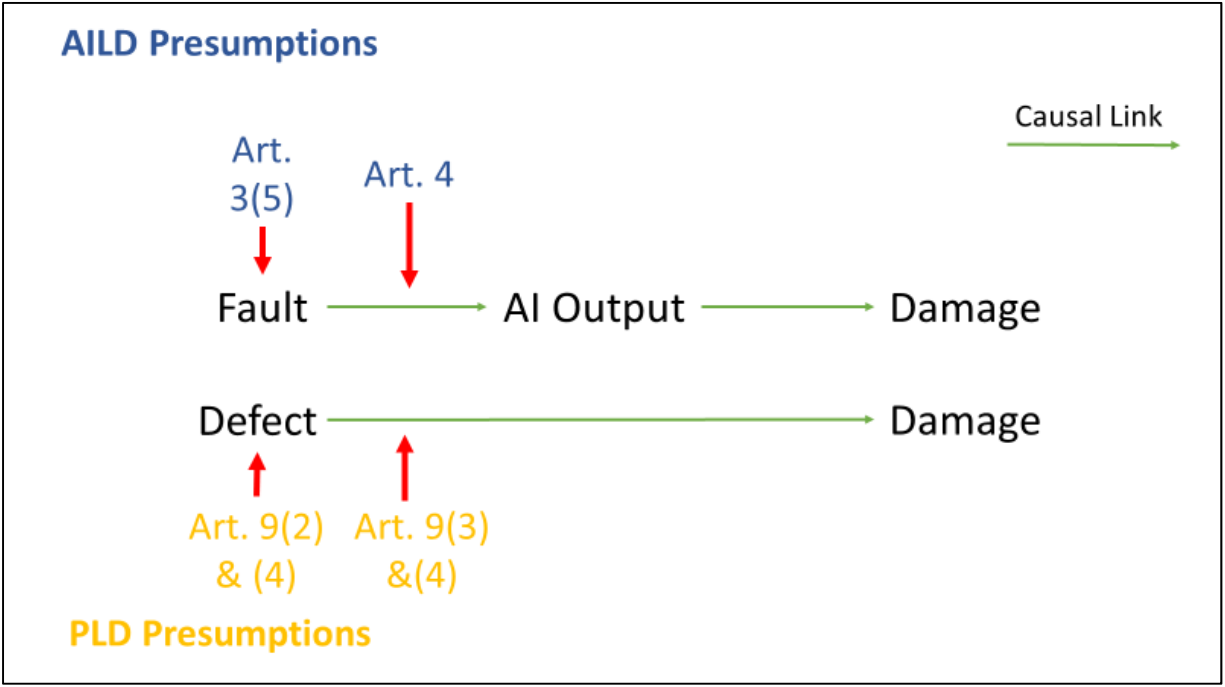

*Table 3: Overview of the presumptions of fault, defectiveness, and causality under the proposed AI liability framework*

### a) AILD Proposal – general elements

Regarding the AILD Proposal, it must first be stressed that the concept of causality itself is expressly not defined in the directive but left to the Member States.[185] Similarly, the proposal refrains from specifying what "fault" or "damage" means in order to facilitate the integration of the AILD Proposal into the liability systems of Member States.[186]

As mentioned above, it can be difficult in some types of AI models to determine which specific input features were most relevant for the output.[187] Based on this opacity risk, Article 4 AILD Proposal lays down a rebuttable presumption for the specific causal link between fault,

---

Verantwortung und Haftung?', A 1, A 100; Hacker, 'Europäische und nationale Regulierung von Künstlicher Intelligenz' 2142, 2145; Gerald Spindler, 'Neue Haftungsregelungen für autonome Systeme?' (2022) 77 JZ Juristen Zeitung 793, 797 et seq.; Spindler, 'User liability and strict liability in the Internet of Things and for robots' 125, 139; more reserved: Miquel Martín-Casals, 'Causation and Scope of Liability in the Internet of Things (IoT)' in Sebastian Lohsse/Reiner Schulze/Dirk Staudenmayer (ed), Liability for Artificial Intelligence and the Internet of Things (Nomos 2019) 201, 216; Thöne, Autonome Systeme und deliktische Haftung: Verschulden als Instrument adäquater Haftungsallokation? 257 - 261; Mario Martini, Blackbox Algorithmus (Grundfragen einer Regulierung künstlicher Intelligenz, Springer 2019) 286-287.

[183] European Parliament and the Council of the European Union, Regulation (EU) No 596/2014 of the European Parliament and of the council of 16 April 2014 on market abuse (market abuse regulation) and repealing Directive 2003/6/EC of the European Parliament and of the Council and Commission Directives 2003/124/EC, 2003/125/EC and 2004/72/EC (2014) [MAR]

[184] Hacker, 'A legal framework for AI training data—from first principles to the Artificial Intelligence Act'

[185] Commission, Proposal for a Directive of the European Parliament and of the Council on adapting non-contractual civil liability rules to artificial intelligence Rec. 10. [AILD Proposal] Rec. 10.

[186] Ibid, p. 11 and Rec. 10, 22.

[187] Fn. 28.

comprising at least the breach of a duty of care, on the one hand and the AI output (or the lack thereof) on the other. The presumption therefore does not cover the determination of fault as such, the existence or nonexistence of an AI output, the existence and scope of the damage, or the causal link between the AI output and the damage. Rather, the presumption linking fault and output is only triggered if the claimant has demonstrated all of these latter elements.

### b) PLD Proposal – general elements

In many respects, the provisions on the burden of proof enshrined in Article 9 PLD Proposal structurally mirror Article 4 AILD Proposal. The first paragraph of Article 9 states that the claimant generally must prove defectiveness, damage and the causal link between them. This reflects the current state under Article 4 PLD, which was nuanced in a number of important decisions in recent years, most recently the *Sanofi Pasteur* case.[188] They have not, however, introduced a reversal concerning the burden of proof as foreseen now under Article 9 PLD Proposal.

### c) A concrete example: the bank robbery

An example may further elucidate the intended function of Article 4 AILD Proposal and Article 9 PLD Proposal. Imagine that an unknown female attacker robs a bank. The security cameras capture video footage of the incident at a moment when the attacker does not wear a face mask. The bank subsequently runs its AI-based face recognition software on the footage. The AI system, which features a complex machine learning model, determines that one of the banks' own clients, Ms. Smith, was very likely the attacker.

The output triggers an automatic freeze of all of Mrs. Smith's assets at the bank. Ms. Smith is therefore unable to meet payment deadlines, is evicted from her home and faces significant late payment penalties on her credit line. After a few weeks, it turns out that the face recognition software misidentified Ms. Smith, who was not involved in the robbery. Only 25% of the images in the training data set for the face recognition system depicted women, which was known to the developers, SmartView Ltd., and the bank. Both are based in the EU. Ms. Smith now would like to recover damages both from the provider/manufacturer of the AI system (i.e., SmartView Ltd.) and the user (i.e., the bank).

Indeed, in such a situation, Ms. Smith faces several difficulties. First, without access to the training data set, she cannot establish that women were substantively underrepresented in it (fault/defectiveness). Second, even if she can prove substantive underrepresentation, she must show that the (negligent) choice of the particular training data set really led to her misidentification (causal link between fault and AI output (AILD Proposal); and between defectiveness and damage (PLD)). However, she might have been misidentified even by an AI system trained on a gender-balanced data set, as even high-performing models are never 100% accurate.[189] In such cases, Articles 3(5) and 4 AILD Proposal as well as Article 9 PLD Proposal seek to provide potential victims with effective redress options by reversing the burden of proof under certain circumstances.

---

[188] See, e.g., CJEU, Case C-621/15, Sanofi Pasteur, ECLI:EU:C:2017:484.
[189] Lindholm and others, Machine Learning, 65 et seq.

### d) AILD Proposal: non-compliance and causality

The AILD Proposal contains two presumptions reversing the burden of proof: Article 3(5) deals with non-compliance with a duty of care, and Article 4 with the causality between fault and AI output.

#### i. Additional requirements for the non-compliance presumption

Hidden away in Article 3(5), the AILD Proposal contains a presumption concerning the substantive element triggering liability under national law: non-compliance with the relevant duty of care. It builds on the disclosure mechanism contained in Article 3(1) and (2) AILD Proposal. If the defendant fails to comply with the respective court order to disclose or preserve evidence, a presumption of non-compliance with the relevant duty of care is activated. The defendant may rebut it by submitting evidence to the contrary (Article 3(5) and Rec. 21 AILD Proposal). The presumption is justified as it provides an effective incentive to comply with the disclosure order and not to engage in socially, and individually, costly delaying tactics.[190]

#### ii. Additional requirements for the causality presumption

While the conditions for the non-compliance presumption are straightforward, the AILD Proposal only mandates a rebuttable presumption of the causal link between the defendant's fault and the output of the AI system (or the lack thereof) if several additional elements are present. These can be divided into two sets of criteria: three necessary conditions which need to be fulfilled in all cases (Article 4 (1) (a)-(c) AILD Proposal); and another four individually sufficient case scenarios of which at least one needs to be fulfilled besides the three general, necessary conditions.

##### (1) Three necessary conditions

Concerning the three general, necessary elements, the claimant first needs to demonstrate the fault of the defendant (or of a person attributable to them), Article 4(a) AILD Proposal. The provision spells out that "fault" must at least consist in the non-compliance with a duty of care, in EU or national law, intended to protect against the damage that occurred. Arguably, in general, such a duty may arise from either statutory or case law. Article 2(9) AILD Proposal defines "duty of care" as "a required standard of conduct, set by national or Union law, in order to avoid damage to legal interests recognized international Union law level […]." Nothing in the wording of the definition or in the recitals suggests that only statutory law could establish a relevant duty. This reading is reinforced by a purposive interpretation of the relevant provisions of the AILD Proposal. What matters, in view of the presumption's purpose, is the difficulty of proving, in some situations, causality; and not the source of the legal obligation. This is not only important for countries in the common law tradition but also for continental jurisdictions with general clauses in tort law from which Member State courts derive more concrete duties of care.[191]

Since the concept of fault itself is not harmonized, Member States remain free to demand, besides a breach of duty, any subjective element concerning the defendant (e.g., negligence).

---

[190] Cf. Rec. 33 PLD Proposal.
[191] E.g., France and Germany.

Most often, the claimant will have to resort to the disclosure of evidence proceedings under Article 3 AILD Proposal to demonstrate this (negligent) breach of duty. In our robbery example, fault would consist in the violation of the relevant data governance provisions of Article 10(3) AI Act.

Second, the court must consider it reasonably likely that the fault has influenced the output (or the lack thereof), Article 4(1)(b) AILD Proposal. This is a typical requirement for factual presumptions in court proceedings.[192] The assessment is made on a case-by-case basis and may, incidentally, necessitate testimony of machine learning experts. The proposal, notably, does not require the claimant to demonstrate the likeliness of influence, so that Member States would arguably be free to require courts to investigate this matter *ex officio*. As an example of probable influence, the AILD Proposal suggests the violation of a defined perimeter of operation of an AI system; a counterexample would be the failure to register a system or to provide certain documentation, according to the Commission (Rec. 25 AILD Proposal). As we shall presently see, however, other problematic cases arise even within the ambit of application of Article 4(2) and (3) AILD Proposal.

The claimant does need to demonstrate, though, that the AI output (or the lack thereof) caused the damage, Article 4(1)(c) AILD Proposal. This third requirement will likely be easiest to prove for claimants as it directly affects them and their assets. Moreover, it will generally be just as simple (or difficult) to show that the output of an AI system caused some damage as it is to demonstrate that the action of a machine, or the output of a non-AI system, give rise to that damage. Hence, AI-specific difficulties of proof will likely not occur concerning this third and final general prerequisite for the causality presumption.

### (2) Four individually sufficient scenarios

As will be discussed below, however, causality between fault and output may not be the only element difficult to prove for Ms. Smith. At any rate, the alleviation mechanism proposed in Article 4 AILD Proposal, concerning this causal link, differentiates between four separate cases. They contain additional, individually sufficient requirements for the presumption to be triggered.

### (a) Case 1: claim against providers of high-risk system

The first case concerns a claim against the *provider* of a *high-risk* AI system (for prohibited AI systems, see below, Part e)). As mentioned, the provider is the entity that develops the AI system and places it on the market/puts it into service under its own name or trademark. In our example, this is SmartView Ltd. The presumption of Article 4(2) AILD Proposal then covers the causal link between the negligent use of the skewed training data set and the AI output, i.e., the misidentification. However, the presumption is only triggered if specific, exhaustively enumerated requirements of the AI Act are breached. These encompass:

- the data governance regime, including training data, of Article 10(2) to (4) AI Act;
- the transparency requirements of Article 13 AI Act;
- the human oversight provision of Article 14 AI Act;
- the performance, robustness and cybersecurity requirements of Article 15 AI Act;
- the corrective actions obligations, including withdrawal and recall, of Article 21 AI Act.

---
[192] Fn. 213.

In our case, the obligation to use training data sets that are representative and have the appropriate statistical properties concerning the expected affected persons (Article 10(3) AI Act) was arguably breached by the developing company. Hence, the presumption is triggered and SmartView Ltd. must rebut it (Article 4(7) AILD Proposal).

While this does benefit Ms. Smith, it seems obvious that it will still be difficult for her to prove the breach of the training data regime, or of any other article of the AI Act listed in Article 4(2) AILD Proposal, without access to information (data, models, logs) held by the developers. It is precisely this difficulty that the disclosure of evidence rules in Article 3 AILD Proposal target, as mentioned.

Article 4 AILD Proposal then exonerates the claimant from the necessity to provide, and pay for, expert witnesses detailing the exact causal impact of the breach of the AI Act on the output of the AI system. Such expert testimony must now be furnished by the developers to rebut the presumption. In this sense, Article 4 lowers the barriers, and costs, for potential victims of AI systems to plead their case in court. However, this cost improvement only extends so far: injured persons still have to hire AI experts to prove the violation of the AI Act in the first place.

### (b) Case 2: claim against users of high-risk system

The second case, specified in Article 4(3) AILD Proposal, concerns a claim against the *user* of a *high-risk* AI system, in our case the bank. The definition of "user" is again relegated to the AI Act, which in its current version specifies that it means "any natural or legal person, public authority, agency or other body using an AI system under its authority". What is noteworthy about this tautological definition is that it now includes consumers operating the AI system for nonprofessional purposes. Here, the presumption is triggered if the user breaches its obligations under Article 29 AI Act. Significantly, however, none of these obligations apply to consumers (Article 2(8) AI Act).[193] The requirements include:

- the obligation to use and monitor the AI system as specified in the instructions of use, Article 29(1), (4) AI Act;
- the requirement to assign human oversight only to natural persons with necessary competence, training and authority, Article 29(1a) AI Act;
- the obligation to provide only input data that is relevant regarding the intended purpose of the AI system, Article 29(3) AI Act.

Article 29, therefore, directly links to Article 14 AI Act: the latter requires developers to enable oversight during use, and the former obliges users to perform this oversight.[194] In our case, the bank did not implement any human oversight of the face recognition system. If the provider specified in the instructions of use that such oversight is required, which providers will likely do to potentially deflect liability or share it with the user, Article 29 AI Act was indeed breached. As a result, the causal link between this omission and the output of the AI system is presumed.

This, however, immediately points to a significant deficiency in the formulation of Article 4(1) AILD Proposal. Pursuant to this provision, only the link between fault and output is presumed. However, the breach of the user duty precisely concerns an action (human oversight) that only takes place *after* the AI output is produced. Therefore, the user can easily rebut the presumption:

---

[193] Rather, general contract or tort law obligations apply, see Recital 58 last sentence AI Act.
[194] See also Art. 29(1a)(i) and Art. 14(1)(2) AI Act EP Version.

irrespective of any oversight, the output of the AI system would have been the same.[195] What would be needed is a presumption that the act/omission of the user, i.e. the violation of Article 29's human oversight requirement, caused the damage. The crux of the case against the user, in terms of causality, lies precisely in proving that with human oversight, the AI output would have been corrected by the human in the loop (post-processing), *not* that the output would have been different.

Moreover, Recital 15 AILD Proposal suggests that in such cases, presumptions of causality are not even necessary. In my view, this is flawed for two reasons. First, proving the causal link between a lack of proper oversight and damage may be equally demanding, given that standards for proper oversight are lacking.[196] More specifically, second, the literal application of the reading enshrined in Recital 15 would render the reference to Article 29 AI Act largely pointless.

Hence, Article 4(3) AILD Proposal, as it stands, fails to achieve its stated objective. The same critique applies, *mutatis mutandis*, to the obligation of the provider to allow for effective oversight during use mentioned in Article 4(2)(c) AILD Proposal and originally contained in Article 14 AI Act.[197]

### (c) Case 3: claim concerning non-high-risk system

The third case concerns a claim that involves a *non-high-risk* AI system, irrespective of who the action is brought against. For example, a professional user could fail to monitor a generative AI system producing text that is automatically posted on a social network. As in the case of the Microsoft bot turned racist,[198] the language model may increasingly output derogative language and finally insults a specific person in a legally relevant way.

In this case, the presumption of causality between fault (failure to monitor) and AI output is not coupled to the violation of any specific duties of the AI Act (of which there are very few in these cases, cf. Art. 52 AI Act). Rather, the general provision in Article 4(1)(a) AILD Proposal governs, which only requires the breach of *any* duty of care laid down in Union or national law and intended to protect against the specific damage. However, this presumption only applies if the national court deems it excessively difficult for the claimant to meet the burden of proof.

---

[195] One way in which oversight may influence the output is if previous oversight would have detected the inclination of the AI system to underperform concerning women and the AI model would therefore have been corrected by the developers or other entities. In that case, the output in question would have been different because of the change in the design of the model prior to the application in question. This, however, seems quite far-fetched and not to be the scenario contemplated by Article 4 AILD Proposal. Similarly, one might argue that correct oversight would have detected an erroneous output, the system would have been redesigned on the spot, and a novel, correct output would have been generated. In this case, correct oversight would lead to a different, but only secondary output. However, an instantaneous redesign of the AI model is deeply implausible. Again, this scenario seems unconvincing.

[196] See Fn. 126.

[197] While Article 14 AI Act formally addresses the design stage, it obliges providers to build AI models such that human users may conduct effective oversight during use. The Commission Proposal reads: "High-risk AI systems shall be designed and developed in such a way [...] that they can be effectively overseen by natural persons during the period in which the AI system is in use." [emphasis added] This becomes even clearer in the EP Version of Article 14. See also Rebecca Crootof and others, 'Humans in the Loop' (2023) 76 Vand L Rev 429, 504;

[198] Oscar Schwartz, 'In 2016, Microsoft's Racist Chatbot Revealed the Dangers of Online Conversation' IEEE Spectrum <https://spectrum.ieee.org/in-2016-microsofts-racist-chatbot-revealed-the-dangers-of-online-conversation> accessed November 14, 2022

The assumed lower risk of the non-high-risk AI system therefore leads to higher demands on the claimant. However, such a rule is not justified where non-high-risk systems are as opaque and individually risky as high-risk systems, as specified in the AI Act, respectively. Again, this points to the necessity of including certain AI systems, like AVs, emotion recognition and insurance pricing models, in the category of high-risk systems. Failing that, the presumption of causality under Article 4(2) and (3) AILD Proposal should be specifically extended to cover such systems.

### (d) Case 4: claim against consumers

The final case concerns claims against persons who used the AI system for a personal, non-professional activity, i.e. consumers (Article 4(6) AILD Proposal). Note that the AI Act, in its current version, does not spell out specific requirements for consumers vis-à-vis AI systems beyond of the generic transparency requirements in Article 52 AI Act (concerning biometric identification, emotion recognition and content generation/manipulation systems).

Under national tort law, however, users will often be required to monitor the AI system they put into action to the extent that they can control its behaviour and prevent harm.[199] In such cases, Article 4(6) AILD Proposal now provides that the causality presumption between fault and AI output only applies in either of two cases. First, it is triggered if the consumer materially interfered with the conditions of the AI systems operation. This could be a case where an AI-based lawnmower was placed on a gravel path by the consumer and therefore kicked up a pebble that hit a pedestrian.[200] Second, the presumption equally applies if the consumer was required and able to determine the conditions of operation and failed to do so. For example, the consumer might know and be contractually obliged to use an AI system only in certain weather conditions but decided to use it in adverse weather conditions anyway.[201] This may, for example, be relevant for AVs operated in AI-assisted modes under difficult driving conditions.

Here, the output would indeed often have been different had the consumer not interfered with or changed the operational conditions. One might argue that, under the concrete circumstances, the outcome would again have been the same (in any deterministic AI system) irrespective of whether the consumer action constituted an interference with or failure to determine the operational conditions – the AV may just not work under certain conditions. However, I would argue that the outcome would have been different in the absence of the fault of the consumer in the sense that the AV would not have been placed in these conditions in the first place. Hence, the outcome would simply not have been generated. Only this interpretation seems to leave any scope of application for the presumption. Otherwise, the same problem as in the case of the monitoring by users would arise. Thus, the second interpretation should be preferred.

### e) Strengths, weaknesses and limitations of the AILD Proposal presumptions

The causality presumption, and the more limited non-compliance presumption, present an important step into the right direction. They seek to plug a crucial gap in the effective private enforcement of the AI Act and the concomitant compensation of affected persons for damages.

---

[199] See, e.g., Kirsty Horsey and Erika Rackley, Tort Law (6th edn, Oxford University Press 2019) 64, 73; Deakin and Adams, Markesinis and Deakin's Tort Law 91.
[200] Cf. Rec. 29 AILD Proposal ("choosing the area of operation").
[201] Cf. Rec. 29 AILD Proposal ("setting performance conditions of the AI system").

However, this aim is only partially achieved as the causality presumption still suffers from several shortcomings.

First, as shown, the presumption fails to capture obligations referring to actions that need to be undertaken *after* the AI output is generated. Such actions can, by definition, not change the previous AI output (unless one assumes that other post-processing actions following even earlier outputs would have led to a revision of the model and hence to a different output in the case in question).[202] This is particularly virulent in the case of Article 14 AI Act (human oversight).

Second, a similar reasoning applies to the violation of Article 13 AI Act, the transparency requirement. According to this provision, providers must design the system in a sufficiently transparent way. Transparency must ensure that both users and the provider may comply with their obligations, and that users are enabled to understand and use the system. As will be explored in depth below (Part V.3.b)), some of these explanations may also be provided after the output was generated. This gives rise to the same post-processing issue, unless it can be assumed that, with the right type of explanations, the developers themselves would have noted the problem and redesigned the AI model accordingly. The answer depends on the circumstances of the case, but it does clearly not include straightforward line of argumentation for claimants. It fails entirely if it is highly likely that only the user would have spotted the damage-causing error after the output was obtained. Ironically, and wrongly, Article 11 AI Act, which forces *pre-processing* transparency, is not even mentioned in Article 4(2) AILD Proposal.[203]

Third, in cases 1 and 2 (claim against provider or user of high-risk system), Article 4(4) AILD Proposal additionally holds that the presumption does not apply if the defendant shows that sufficient evidence and expertise is reasonably accessible for the claimant to prove the causal link between fault AI output. This immediately raises the question whether reasonable access to evidence and expertise includes hiring (potentially expensive) AI experts as forensic witnesses. While such a reading cannot be entirely ruled out, Recital 21 AILD Proposal clearly shows that the directive aims to minimize unnecessary expenses for claimants in order to lower the barriers of private enforcement of the AI Act. This reading is further supported by Recital 27 AILD Proposal, which mentions documentation and logs pursuant to the AI Act as examples of available access to evidence, but is silent on external sources.

Fourth,[204] the causality presumption extends to cases against consumers. It might seem odd that, while consumers will still face significant barriers to gain compensation from tortfeasors, they are simultaneously themselves the object of alleviated enforcement. However, it must be acknowledged that the presumption against consumers does not apply when the consumer simply launches and uses the AI system as described in the instructions for use and according to the intended purpose.[205] Any routine operation of the AI system by consumers therefore excludes the causality presumption. It is only triggered in specific cases (material interference with operation conditions; failed determination of operation conditions despite requirement and ability) in which the consumer knew or should have known that the AI system will not work under the given circumstances. In such a situation, it seems defensible to hold the consumer accountable for harm done to third parties if those third parties can prove the consumer's fault and causal damage. Third parties affected by consumers wrongly operating AI systems are in an equally difficult situation concerning the proof of the causal link between fault and AI output as other victims – including consumers – are vis-à-vis developers or professional users. The evident mishandling of the AI system by the consumer provides a sufficient foundation to justify the alleviated burden of proof in favour of wronged third parties.

---

[202] See Fn. 195.
[203] Cf. Rec. 22 AILD Proposal.
[204] See, e.g., the critical remarks by BEUC's Deputy Director General Ursula Pachl in Bertuzzi, The new liability rules for AI.
[205] Rec. 29 AILD Proposal.

Fifth, relatedly, the AILD Proposal does not contain a comprehensive system for a reversal of the burden of proving fault, except in the narrow case of Article 3(5) AILD Proposal. This implies that injured parties will generally have to demonstrate, and prove, violations of the AI Act, for example (Article 4(1) and (2) AILD Proposal). While the necessary information may be obtained under Article 3 AILD Proposal, it remains a tremendous challenge for end users to show fault in the training of highly complex AI systems.[206] They will necessarily have to hire expert AI modelling witnesses, driving up legal costs with uncertain outcomes. For example, even an expert will only be able to say after digging through heaps of data whether the training data set was relevant, free of errors, representative, complete, and had the appropriate statistical properties, as required under Article 10 (3) and (4) AI Act. The well-known fundamental paradox of information[207] resurfaces here to the claimant's detriment: one can only judge the content of information once one has paid for access or expert analysis of it. This will be taken up in Part h) below.

Finally, the proposal must be updated to cover harm inflicted by the use of a prohibited AI system (Article 5 AI Act).[208] Ideally, in this case, truly strict liability requiring only a causal link between the use of the prohibited system and the damage would apply (see below, Part V.1.b)i.). As a logically second-best, but practically identical solution, fault and the causal link between fault and AI output must be irrefutably presumed. Even if perfectly well-designed, prohibited AI systems must not be used, and developers putting them onto the market, as well as users deploying them, must compensate any damage caused by their operation.

### f) PLD: defectiveness and causality

The PLD Proposal goes slightly further than the AILD Proposal in reversing the burden of proof. It contains three sets of individually *sufficient* conditions for the presumption of either defectiveness or the causal link between the defect and the damage to be triggered. At this point, as mentioned, the PLD Proposal diverges from the current PLD and pertaining case law.[209]

#### i. Sufficient condition for defectiveness presumption

The first set of sufficient conditions for a presumption of the product's defectiveness is contained in Article 9(2) PLD Proposal. It is triggered by either of the following three criteria:

- failure of the defendant to comply with a disclosure obligation under Article 8(1) PLD Proposal (Article 9(2)(a) PLD Proposal);
- violation of individually protective mandatory safety requirements, in EU or national law (Article 9(2)(b) PLD Proposal);
- damage caused by obvious malfunction of the product (Article 9(2)(c) PLD Proposal).

---

[206] Cf. Samar Abbas Nawaz, The Proposed EU AI Liability Rules: Ease or Burden? (2022).
[207] Kenneth Arrow, 'Economic welfare and the allocation of resources for invention' in Harold M. Groves (ed), The rate and direction of inventive activity (Princeton University Press 1962) 615.
[208] Cf. also Wagner, 'Liability Rules for the Digital Age - Aiming for the Brussels Effect', 41.
[209] See Fn. 188.

The first criterion obviously mirrors Article 3(5) AILD Proposal. While under the AILD Proposal regime, however, non-compliance with a duty of care is only presumed in the case of non-compliance with the disclosure order, the presumption of defectiveness under the PLD Proposal may be triggered by two further elements, each of which activates the presumption. This provides the claimant with more options to succeed in court. The fulfilment of these two additional criteria must, however, be established by the claimant. Significantly, the breach of mandatory safety requirements now explicitly prompts the presumption of defectiveness, further aligning the two pillars of product regulation in the EU.[210] Concerning the other element, obvious malfunction, the proposal mentions an exploding glass bottle as a prime example (Rec. 33), taking up (perhaps involuntarily) a string of famous product liability cases from Germany.[211]

In our bank robbery case, Ms. Smith would again request disclosure of evidence concerning the modelling of the AI system (training data, hyperparameters, model type etc.). Since she was misidentified, she will pass the plausibility threshold under Article 8(1) PLD Proposal. If the manufacturer (SmartView Ltd.) fails to provide the information, defectiveness is presumed according to Article 9(2)(a) PLD Proposal. If, on the other hand, they do provide the evidence, Ms. Smith will have to hire an AI expert to sift through the data, who will likely find the gender imbalance in the training data set. This finding should be enough to establish a violation of Article 10(3) AI Act and, hence, defectiveness under Article 6 PLD Proposal (see above, Part IV.2.b)).

To save time and costs, Ms. Smith could alternatively attempt to avail herself of the defectiveness presumption under Article 9(2)(c) PLD Proposal. Arguably, the misidentification of a person by a face recognition system constitutes an obvious malfunction. The error is as evident as in the case of an exploding glass bottle mentioned in Recital 33 PLD Proposal. Since it can be assumed that the malfunction occurred during normal use and under ordinary circumstances, the presumption is indeed triggered. The developers, arguably, cannot rebut the presumption given the imbalance in the training data.

### ii. Sufficient condition for causality presumption

However, Ms. Smith will additionally have to establish the causal link between the defect and her damage. Enter Article 9(3) PLD Proposal. Causality is presumed if defectiveness is established and the damage is typically consistent with the defect in question. In light of the purpose of Article 9 PLD Proposal, the presumption in Article 9(2) must suffice for the establishment of defectiveness. Otherwise, injured persons will again find themselves without effective redress if they need to prove defectiveness without recourse to the presumption in order to avail themselves of the causality presumption.[212] The proposal is silent on what types of damage may be considered typically consistent with the defect. A convincing interpretation, derived from the threshold for factual presumptions in civil proceedings, could hold that typical consistency requires that the ordinary course of events may lead to such damages.[213] It would arguably be excluded if an average person with knowledge of the defect would not expect the kind of damage in question to occur.

---

[210] Cf. Rec. 33 PLD Proposal.
[211] Daily Wuyts, 'The product liability directive–more than two decades of defective products in Europe' (2014) 5 Journal of European Tort Law 1 30; BGH, Case VI ZR 91/87, Soda Bottle, BGHZ 104, 323; BGH, Case VI ZR 158/94, Mineral Water Bottle II, BGHZ 129, 353.
[212] For a more restrictive reading of "established", see Spindler, 'Die Vorschläge der EU-Kommission zu einer neuen Produkthaftung und zur Haftung von Herstellern und Betreibern Künstlicher Intelligenz', 698.
[213] Cf., e.g., BGH, Case V ZR 437/99, Actual Presumption, NJW 2001, 1127, 1128 et seq.

In the bank robbery case, all the damage positions (eviction costs and late payment penalties because of the asset freeze) can arguably be expected as a result of the misidentification of a person in the case of a bank robbery. Hence, causality is presumed as well.

### iii. Additional sufficient conditions for both presumptions

Article 9(4) PLD Proposal contains an additional presumption of defectiveness, causality, or both, which is triggered if the following three criteria are fulfilled (concerning the respective object of the presumption):

First, in the view of the court, it must be excessively difficult for the claimant, due to technical or scientific complexity, to prove the respective object of the presumption (defectiveness; causality; or both). Recital 34 lists several factors for complexity: "the complex nature of the product, such as an innovative medical device; the complex nature of the technology used, such as machine learning; the complex nature of the information and data to be analyzed by the claimant; and the complex nature of the causal link, such as a link between a pharmaceutical or food product and the onset of a health condition, or a link that, in order to be proven, would require the claimant to explain the inner workings of an AI system." The list is fairly comprehensive and convincing, even though machine learning as such need not necessarily be complex (e.g., linear regression).[214] The assessment therefore must be conducted on a case-by-case basis (Rec. 34 PLD Proposal).

Furthermore, the claimant needs to establish, on the basis of sufficiently relevant evidence, that the product contributed to the damage (second criterion) and that the factual object of the presumption (defectiveness; causality; or both) is likely (third criterion). In total, the provision therefore lowers the probability threshold of evidence the claimant needs to pass to obtain a favourable judgment: from the regular standard under national law (e.g., beyond reasonable doubt) to mere likeliness. With the third criterion, Article 9(4) PLD Proposal again mirrors the AILD Proposal in a striking, yet imperfect way. As seen, Article 4(1)(b) AILD Proposal equally establishes a likelihood condition, but with a slightly different wording ("reasonably likely" versus "likely"; "influenced the output" versus "cause of the damage"). In the legislative process starting now, these differences should be ironed out (unless there is a good, hitherto hidden reason for them).

Returning once more to the robbery example, Ms. Smith could successfully claim under Article 9(4) PLD Proposal that she does face excessive difficulties to prove both defectiveness and the causal link due to the innovative nature of the face recognition system and the complex nature of the deep learning system (e.g., a convolutional deep neural network). To prove the causal link, she would have to explain the inner workings of the AI system to demonstrate that a differently trained system would not have misidentified her. In addition, as in most cases, it is easy to establish that the product contributed to, and actually caused, the damage (as opposed to the *defectiveness* of the product causing the damage). Finally, given the gender-imbalanced training data, the system was arguably defective and at least likely caused the damage. Overall, Ms. Smith thus disposes of several highly promising routes to recover damages in court proceedings against the manufacturer of the face recognition software, i.e., the developers (SmartView Ltd.).

---

[214] See, e.g., Lipton, 'The mythos of model interpretability: In machine learning, the concept of interpretability is both important and slippery'; Arrieta and others, 'Explainable Artificial Intelligence (XAI): Concepts, taxonomies, opportunities and challenges toward responsible AI'.

### g) Strengths, weaknesses and limitations of the PLD presumptions

As the robbery example showed, the PLD Proposal does contain a burden of proof regime that, together with the disclosure of evidence mechanism, opens avenues for successful private enforcement of the AI Act or other relevant obligations. Just like the equivalent provision in the AILD Proposal, it presents a step in the right direction by lowering barriers for enforcement and therefore raising incentives for compliance. Again, however, several weaknesses bear noting.

First, under Article 9(2)(a) PLD Proposal, the presumption of defectiveness should also be triggered, as under Article 3(5) AILD Proposal, by the failure to comply with an order to preserve, not only to disclose, evidence.

More importantly, second, the fulfilment of the last two criteria of Article 9(2) PLD Proposal with the presumption of defectiveness – violation of mandatory safety requirements or obvious malfunction – must be proven by the claimant. Under some circumstances, this may be highly onerous, or even impossible. Defendants will usually comply in one way or another with the evidence disclosure obligation, as – at least under civil liability – some disclosure is better than an inverse presumption following a refusal to disclose. Defendants therefore have a significant incentive to overwhelm claimants with vast amounts of (inconclusive) data. As a result, claimants will likely have to hire expensive AI experts to analyze the data.

To avoid these additional upfront costs, claimants must establish either non-compliance with safety requirements or an obvious malfunction. Both, however, may turn out hard to prove. Safety requirements again concern the internal production process of manufacturers. Without access to specific documents and data, claimants will face the same difficulties as under the current regime. Therefore, it must be ensured that the disclosure rules under Article 8(1) PLD Proposal apply equally to the proof of mandatory safety requirements, to the extent that it is necessary to fulfil the criterion for a reversal of the burden of proof under Article 9(2)(b) PLD Proposal.

Whether an obvious malfunction can effectively be established will depend chiefly, in my view, on the AI system used. Very broadly speaking, machine learning systems fulfil two distinct tasks (in supervised learning settings): classification and regression.[215] The former denotes the allocation of input data to a specific category (e.g., image recognition), while the latter concerns the output of a more or less continuous score (e.g., credit scoring). An obvious malfunction will be significantly easier to establish in cases of erroneous classification, as in the robbery example. Here, it suffices to point out that the model chose the wrong category, which can often be externally verified (dog instead of cat; person X instead of person Y).[216] If, however, the AI system predicts continuous scores (regression) as in credit scoring, insurance pricing or job performance prediction, any error, and a fortiori an "obvious malfunction", will be hard to prove. Generally, easily accessible ground truth to compare the prediction with (the "true" credit score, insurance price or job performance index of a person) will not exist – which is why the AI system is used in the first place. Even with access to the data and the model, claimants will then face an uphill legal battle, and need to hire AI experts.

---

[215] Lindholm and others, Machine Learning.
[216] This may be different if outcomes are not logged or are difficult to access, for example in autonomous vehicles.

In sum, the alleviations are effective for cases with obvious malfunctions of the AI system, occurring more frequently in classification tasks. In all other cases, particularly concerning scoring, claimants need to be prepared to hire expensive AI. Therefore, the proposals need to be accompanied by effective strategies to facilitate access to litigation funding for claimants unable to bear the costs of hiring AI experts out of their own pockets.

### h) Harmonization of the burden of proof regime

Overall, for the same reasons mentioned concerning the disclosure mechanism, it seems preferable to fully harmonize the burden of proof regime between the PLD and the AILD Proposal. This holds even more if one accepts that "defectiveness" under the PLD usually implies as much a breach of a duty of care as "fault" under the AILD Proposal.

More specifically, short of uniting the PLD and the AILD Proposal into one regulation or directive (see above and below), harmonization would at least require the following. First, concerning the presumption of non-compliance, the two additional sufficient conditions of Article 9(2) PLD Proposal (violation of mandatory safety requirements; obvious malfunction) should be added to Article 3(5) AILD Proposal. There is no reason why these two criteria should not imply, at least presumptively, the breach of a duty of care just as much as the defect of the product. This would mitigate the difficulties of proving fault under the AILD Proposal regime outside of the narrow case of a refusal to comply with a court order. Second, concerning also the causality presumption, the likelihood conditions in Article 4(1)(b) AILD Proposal and Article 9(4)(b) PLD Proposal must be aligned in terms of wording and spirit.

### i) Rebuttal of the presumptions

Defendants may rebut any of the presumptions (Articles 3(5)(2), 4(7) AILD Proposal; Article 9(5) PLD Proposal). Importantly, concerning the causality presumptions, it should not be considered sufficient for a rebuttal to (correctly) state that almost any machine learning model inevitably commits some errors and that, therefore, even a close-to-perfect system could theoretically have caused the specific damage (by accidentally making the exact error that caused the damage). To hold otherwise would render the causality presumption obsolete in all cases of machine learning. In our robbery case, this would mean that the hypothetical possibility of alternatively trained models equally misidentifying Ms. Smith does not qualify as a rebuttal of the causal link between defect and damage.

### 5. Elements exclusive to the PLD Proposal

The PLD Proposal finally suggests the removal and attenuation of several liability limitations and exceptions currently enshrined in the PLD. At least three of these are relevant to AI systems: the (still insufficient) expansion of eligible damages; the later-defect defence; and the development-defect defence.

### a) Expansion of eligible damages

To start with, the €500 threshold applicable to property damage under Article 9(b) PLD is abolished (Article 4(6)(b) PLD Proposal). This is convincing as it sets proper incentives for economic operators to avoid defects even if they lead to small amounts of damage. Such harm can nevertheless be dispersed among a multitude of injured persons, causing significant total harm and concomitant social welfare losses without compensation under the current framework. Importantly, as Gerhard Wagner has convincingly argued, the restriction of eligible damage to *privately* used property should be abandoned, reflecting an outdated and formalistic understanding of products liability law as a part of consumer protection law (only).[217] Commercially used property must be included as well, particularly from an incentives perspective.

Furthermore, Article 4(6)(c) PLD Proposal convincingly adds the loss or corruption of data not exclusively used for professional purposes to the damage the economic operator can be held liable for. Such an amendment is necessary as the question of data ownership remains unresolved.[218]

Another constraint unique to both the existing and the proposed PLD regime is that only damage to property, life or health is covered (Art. 4(6) PLD Proposal).[219] The directive, in this sense, seeks to limit lawsuits brought to compensate for "pure economic loss".[220] However, this restriction is entirely at odds with the main types of high-risk AI systems operative in economic applications: employment, credit scoring, and insurance pricing. In all of these areas, injury will typically *not* involve damage to property or health – but cause direct financial losses. Hence, under the proposed regime, victims are relegated to the AILD Proposal regime – where they have to prove fault without the same types of presumptions as under the PLD for defectiveness.[221]

In sum, ironically, the liability regime likely remains dysfunctional precisely for high-risk applications. Hence, at least for these cases of high-risk AI systems, the eligible damage under the PLD should include pure economic loss as well,[222] lest the liability regime remain a paper tiger in exactly the domain in which it was supposed to offer relief (see below, Part V.1.b)i.(3)).

### b) Later-defect defence

An important tenet of product liability law is that manufacturers are only responsible for defects stemming from their sphere of influence.[223] This is epitomized by the so-called later-defect defence in Article 7(b) PLD: manufacturers are exempt from liability if "it is probable that the defect which caused the damage did not exist at the time when the product was put into circulation by him or that this defect came into being afterwards."

---

[217] Wagner, 'Liability Rules for the Digital Age - Aiming for the Brussels Effect', 33-34.
[218] Francesco Banterle, 'Data ownership in the data economy: a European dilemma', EU Internet Law in the digital era (Springer 2020).
[219] Recovery of non-material damage can be foreseen in Member State law, Recital 18 PLD Proposal.
[220] Deakin and Adams, Markesinis and Deakin's Tort Law 91, 126.
[221] Nawaz, The Proposed EU AI Liability Rules: Ease or Burden?.
[222] For a nuanced assessment of allowing specific cases of pure economic loss under tort law, see Deakin and Adams, Markesinis and Deakin's Tort Law, 127; see also Wendehorst, AI liability in Europe: anticipating the EU AI Liability Directive, 13; against coverage: Wagner, 'Liability Rules for the Digital Age - Aiming for the Brussels Effect', 53-55.
[223] Karin Alheit, 'The applicability of the EU product liability directive to software' (2001) 34 Comparative and International Law Journal of Southern Africa 188, 198; Tiago Sérgio Cabral, 'Liability and artificial intelligence in the EU: Assessing the adequacy of the current Product Liability Directive' (2020) 27 Maastricht Journal of European and Comparative Law 615-635; Joasia Luzak, 'A Broken Notion: Impact of Modern Technologies on Product Liability' (2020) 11 European Journal of Risk Regulation 630, 647; Gabriele Mazzini, 'A system of governance for artificial intelligence through the lens of emerging intersections between AI and EU Law' in A. De Franceschi/ R. Schulze (ed), Digital Revolution – New challenges for Law (2019) 22.

Scholars have long argued that this defence is at odds with the potential of AI systems to evolve after being put on the market and to develop defects.[224] From a technical perspective, however, it must be reiterated that ML models do not generally continue to learn after deployment, but only if they are explicitly programmed to do so (e.g., online learning).[225] Hence, even under the current framework, a defect that, at first glance, arises only after the system was put onto the market can be traced back to the initial decision of the manufacturer, before deployment, to enable continued adaptation. As a consequence, the later-defect defence should not apply to those scenarios even under the current framework.

The PLD Proposal makes this finding explicit. It reiterates the later-defect defence in Article 10(1)(c) PLD Proposal, but provides for exemptions from this defence in Article 10(2). Economic operators continue to be liable if the product is within the manufacturer's control and the defect concerns a related service; the software including updates or upgrades; or the lack of such updates or upgrades necessary to maintain safety. Such a clarification is needed to endow the aforementioned interpretation with a strong foothold in statutory law. From an economic and a technical point of view, it is entirely convincing to couple residual control over the product and its risks to continued liability. This is also acknowledged in the new Article 6(c) PLD, which adds a continued learning ability to the list of factors relevant for defectiveness (see also above, Part IV.2.b)).

Updates and upgrades, in particular, can be delivered online and are a routine staple of software maintenance nowadays. Significantly, only critical updates (maintenance of safety) are required to deflect liability under the later-defect defence. Quite strikingly, the PLD Proposal thereby establishes a non-contractual duty to provide safety updates. Currently, Art. 8(2) of Directive 2019/770 on digital content and services only provides for a contractual, waivable right to updates, including security updates, against the seller. The PLD proposal, in turn, strikes a sensible balance between incentives for continued software development on the one hand and regulatory overreach to compel the development of functionally superior products on the other. Generally, and convincingly, incentives for product development are delivered by market pressure, and developers should not be forced to add functional – as opposed to safety – features to their products by way of product liability law (cf. also Article 6(2) PLD Proposal).

### c) Development-defect defence

Another important defence for high-tech products is enshrined in Article 7(e) PLD. The manufacturer is exempt from liability if "the state of scientific and technical knowledge at the time when he put the product into circulation was not such as to enable the existence of the defect to be discovered." This so-called development-defect defence has been controversial ever since the PLD introduced it,[226] and Member States may currently derogate from it.[227] The CJEU has rightly held that the state of knowledge refers to an objective assessment and not to the subjective knowledge or capacities of the individual manufacturer.[228]

---

[224] See e.g., Cabral, 'Liability and artificial intelligence in the EU: Assessing the adequacy of the current Product Liability Directive', 629.
[225] See Fn. 134.
[226] Charlotte de Meeus, 'The Product Liability Directive at the Age of the Digital Industrial Revolution: Fit for Innovation?' (2019) 8 Journal of European Consumer and Market Law 152; Wuyts, 'The product liability directive– more than two decades of defective products in Europe' 30.
[227] Article 15(1)(b) PLD.

Article 10(1)(e) PLD Proposal perpetuates the development-risk defence and removes the derogation option for Member States to create a level playing field.[229] Next to hard-coding the mentioned CJEU judgment, it also aligns the period during which the defect was non-discoverable with the extended control obligation of manufacturers now enshrined in Article 6(1)(e) PLD Proposal: if the manufacturer retains control after the product was placed on the market or put into service, and the defect becomes discoverable during this time, the manufacturer is still liable.

Nonetheless, the development-defect defence remains debatable. The aim of the defence is to erect a liability shield for manufacturers of advanced technology goods and to thus spur innovation.[230] On the other hand, it exempts manufacturers from internalizing risks stemming from their production sphere and shifts losses to users.[231] However, the latter are arguably in no better position than the manufacturers to know about or protect themselves against technically or scientifically complex defects.

In my view, a compromise should be sought concerning development defects that balances the interests of spurring innovation with the rights of injured persons to compensation. In tort law theory, compensation by the state or a dedicated fund is recognized as a third way beyond the dichotomy of liability of manufacturers/injurers on the one hand and non-compensation of the injured party on the other.[232] Arguably, such a solution would strike a convincing compromise for development defects.[233] A fund dedicated to victims of technology risk could reimburse injured persons if they can show that, barring the development-risk defence, they would have had a claim against economic operators. This ensures effective compensation for harms not originating in the sphere of end users. It would be unjust to make end users of products the residual loss bearers for development defects. Contributory negligence would, of course, have to be factored in to maintain incentives for product users to act diligently.[234] Simultaneously, the fund solution exonerates developers, thus reducing possible insurance premiums. Importantly, it does not detract from incentives to prevent avoidable harm:[235] by definition, development defects cannot be recognized by economic operators even if state-of-the-art knowledge and resources are employed. Hence, a provision establishing an EU-wide compensation fund for persons injured by development defects should be added to the PLD Proposal.

## V. The way forward

The preceding discussion of the framework for liability of AI systems in the EU has demonstrated that the proposals contain steps into the right direction, but still exhibit significant shortcomings. From a policy perspective, the final part of the paper concentrates on four key areas for improving the currently proposed framework. In doing so, it develops a roadmap for the future of AI liability not just in the EU, but also in the US and other countries aspiring to adapt their legal systems to the peculiarities of AI. The four major proposals are:

- facilitating effective compensation via strict liability;
- boosting legal certainty and innovation;
- reining in unforeseeable effects, opacity and discrimination to build trustworthy AI;
- and addressing the increasing and urgent challenge of rendering AI sustainable.

---

[228] CJEU, C-300/95: Commission v. United Kingdom, ECLI:EU:C:1997:255 para. 29.
[229] PLD Proposal, 12.
[230] Wuyts, 'The product liability directive–more than two decades of defective products in Europe' 30.
[231] Ibid.
[232] Calabresi, The cost of accidents: a legal and economic analysis; Steven Shavell, 'On liability and insurance' (1982) 13 Bell Journal of Economics 120; Giuseppe Dari-Mattiacci, 'Tort law and economics' (2003) Working Paper, https://ssrncom/abstract=347801 31.
[233] Cf. also Bertolini, Artificial intelligence and civil liability, 87.
[234] Shavell, Foundations of economic analysis of law, 184, 188; Dari-Mattiacci, 'Tort law and economics' 16 et seq.
[235] Cf. Shavell, 'On liability and insurance' 127.

Within the scope of this article, I can only sketch possible scenarios and solutions; a more detailed exposition must be left to follow-up papers.

### 1. Effective compensation: strict liability versus negligence revisited

The entire approach of the Commission proposals rests on the assumption that the PLD installs a system of strict liability, while the AILD Proposal addresses fault-based liability regimes in the Member States. This dichotomy, however, collapses upon closer scrutiny.

#### a) Foundations: fault under the AILD Proposal and the PLD

The AILD Proposal indeed applies to fault-based liability only.[236] Even though the concept of fault is not fully specified in the AILD Proposal,[237] it does contain an element of the breach of a duty of care (Article 4(1)(a) AILD Proposal). By contrast, the Commission[238] and part of the legal commentary maintain that the PLD provides for a strict liability framework, which dispenses of the need to prove fault.[239]

In reality, however, fault requirements resurface in the determination of defectiveness in many scenarios, not in the sense of intent or negligence, but of an objective breach of duty.[240] The key take-away from the discussion of the concept of defectiveness in Part IV.2. is that, both under the current and the proposed PLD framework, the finding of a design or warning defect[241] is tantamount to showing that the defendant violated certain standards of conduct – in other words, that they breached a duty of care (cf. Article 2(9) AILD Proposal).[242] Under the traditional understanding of design defects, the manufacturer chose a specific design despite available alternatives. As a result, irrespective of the theoretical vantage point one assumes, claimants always need to establish a breach of duty, and hence fault, under the PLD in these cases. It is a fiction that liability under the PLD is truly strict; rather, fault resurfaces in the guise of product defectiveness.[243]

---

[236] See also Rec. 10 AILD Proposal.
[237] Rec. 10, 22 AILD Proposal.
[238] Rec. 2 and 3 PLD: 'liability without fault'; PLD Proposal, 2; AILD Proposal, 2.
[239] See e.g. for manufacturing and design defects: Hans Claudius Taschner, 'Produkthaftung-Noch einmal: Verschuldenshaftung oder vom Verschulden unabhängige Haftung?' (2012) ZEuP: Zeitschrift für europäisches Privatrecht 560, 563; Nawaz, The Proposed EU AI Liability Rules: Ease or Burden?.
[240] David G. Owen, Products Liability Law (3rd edn, Thomson West 2015), 315 et seqq.; Zech, 'Liability for AI: public policy considerations', 147, 154; Wagner, 'Robot Liability', 27, 34.
[241] In the case of a manufacturing defect, liability is indeed strict as it, at least generally, covers even outliers which cannot be avoided even by adhering to a maximalist duty of care, see Wagner, 'Produkthaftung für autonome Systeme', 712.
[242] Zech, 'Liability for AI: public policy considerations', 147, 150 et seqq.; Wagner, 'Robot Liability', 25, 31 et seq.; Gerhard Wagner, 'Einleitung zum Gesetz über die Haftung für fehlerhafte Produkte (Produkthaftungsgesetz – ProdHaftG)', Münchner Kommentar BGB, vol 7 (8 edn, 2020) para. 18-23.

This is not merely an academic or terminological question. Rather, it is crucial to understand and calibrate the dynamics and risks of private enforcement. To bring a successful claim under the PLD framework, consumers and other affected persons have to prove fault just as much as under any other fault-based liability system if the case turns on design or warning defects – which it will in most instances concerning AI liability.[244] Hence, claimants need exactly the same kind of support, concerning evidence and burden of proof, under a PLD (design or warning defect) case as under a fault-based case rooted in Member State liability law.

This suggests that the artificial distinction between the supposedly strict liability under the PLD and the fault-based framework under the AILD Proposal is, to a large extent, artificial and should be given up. Rather, a coherent and unified framework for AI liability is needed which addresses the main risks, facilitates enforcement, and prevents confusion on the part of consumers and other affected persons concerning their rights to compensation. This is clearly *not* what the current proposals achieve.

### b) Proposal for a comprehensive AI liability framework

Hence, alternatives have to be developed. To the extent necessary and possible, these alternatives should take foreseeable reactions, and circumvention strategies, by affected stakeholders into account. For example, scholars have repeatedly argued for strict liability for AI developers (producers) in the past.[245] In my view, such an approach must be nuanced. A revision of the AI liability framework may draw inspiration from the resolution of the European Parliament concerning civil liability for AI systems,[246] but should broaden the scope to include, next to professional users, manufacturers, too.[247] Simultaneously, the rules must be implementable for AI developers large and small. Hence, liability caps, subsidized insurance for SMEs, restrictions to specific cases of illegitimate harm, and other measures must be combined to make strict liability affordable for "innovative" companies, as will be explained below.

In a nutshell: truly strict liability should be reserved for economic operators and professional users of high-risk AI systems. Concerning Foundation Models, strict liability must be tailored to their high-risk deployments. SMEs and non-high-risk AI systems should only be subjected to rebuttable presumptions of fault and causality in order to prevent over-deterrence.[248] Consumers using AI should face fault-based liability with a limited presumption of causality.

Finally, in all these scenarios, defendants must be able to invoke contributory negligence by injured persons. This is generally the case under tort law and corresponds to basic insights from tort law and economics about maintaining incentives for potential victims to act diligently in the presence of AI systems.[249]

---

[243] This holds at least for design and instruction defects, see Reimann, 'Product liability', 238.
[244] Geistfeld, 'A roadmap for autonomous vehicles: State tort liability, automobile insurance, and federal safety regulation', 1619.
[245] Wendehorst, 'Strict liability for AI and other emerging technologies' 179; Zech, Liability for AI: public policy considerations 155; cf. also Wagner, 'Robot Liability', 47; but see also Yavar Bathaee, 'The artificial intelligence black box and the failure of intent and causation' (2018) 31 Harv JL & Tech 889, 932 et seq.
[246] European Parliament, Resolution of 20 October 2020 with recommendations to the Commission on a civil liability regime for artificial intelligence (2020).
[247] The latter were neglected under the EP Resolution, see Zech, Liability for AI: public policy considerations.
[248] Spindler, 'User liability and strict liability in the Internet of Things and for robots'

### i. Illegitimate-harm, high-risk (and prohibited) AI models: truly strict liability

Economic operators and professional users of high-risk AI systems should be subject to truly strict liability in the sense that neither negligence nor, in general, a breach of duty, a defect, or non-compliance need be shown by claimants.[250] Truly strict liability is thus essentially based on causation only, usually (and rightly) with a defence for *force majeure*.[251]

As postulated throughout this paper, the category of high-risk systems cannot be simply transplanted from the AI Act, though, unless the latter adopts an individualized risk perspective.[252] AVs, emotion recognition systems and insurance pricing models need to be among those considered high-risk for the sake of AI liability (and, preferably, for the AI Act as well). Significantly, truly strict liability must also apply to developers and all users of prohibited AI systems to dis-incentivize their deployment from the outset (see also above, Part IV.4.e).

#### (1) General reasons for truly strict liability

The reason for the establishment of truly strict liability is threefold. First, as is well known from basic tort law and economics, strict liability, in contrast to rules based on negligence or duty of care, regulates the activity level of those facing liability, too.[253] This seems particularly important for high-risk AI systems, where the socially optimal amount of activity – for example, of products sold – should be directly influenced by the liability system. Second, strict liability dispenses of the need to investigate a potential breach of the duty of care or a defect, which is costly, time-consuming and prone to judicial error in technically complex environments.[254] By implication, a strict liability regime need not wait for the development and adaptation of technical standards for defects or duties of care, which often take considerable time to be formulated or updated.[255] It thus incorporates automatically the state of the art, and auto-adapts to technological change. Third, strict liability corresponds to the old adage that those who reap the benefits of a particular product should also bear the burden, i.e., fully internalize the costs of the product to society (*qui habet commoda ferre debet onera*).[256] Only strict liability makes the injurer fully internalize the harm the system causes.[257]

#### (2) Restriction to illegitimate-harm models

The European Parliament was therefore justified, in my view, in proposing a truly strict system of liability for high-risk AI systems under which certain operators would have to compensate any eligible losses caused by the model.[258] To properly define the ambit of truly strict liability for high-risk AI systems, however, we need to introduce a novel, key distinction between two types of models: illegitimate-harm and legitimate-harm models. Illegitimate-harm models are AI systems that, from a social perspective, should not cause harm during their correct operation; legitimate-harm models, conversely, are meant to cause harm if functioning properly. More specifically, in illegitimate harm models, a considerable fraction (e.g., > 20%) of the specific output in a given use case will cause damage to the affected persons that could not have been avoided if a perfect prediction had been made by the AI system. In other words, even perfect decisions will often lead to damage.

---

[249] See Fn. 234.
[250] Cf. Wendehorst, 'Strict liability for AI and other emerging technologies', 159; Zech, Liability for AI: public policy considerations.
[251] Wendehorst, 'Strict liability for AI and other emerging technologies', 159.
[252] See text accompanying Fn. 88.
[253] Shavell, Foundations of economic analysis of law, 196.
[254] Ibid, 188 et seq.
[255] Spindler, 'User liability and strict liability in the Internet of Things and for robots'.
[256] Reinhard Zimmermann, The law of obligations: Roman foundations of the civilian tradition (Juta and Company Ltd 1990) 201 and 209.
[257] Shavell, Foundations of economic analysis of law 260.
[258] Parliament, Resolution of 20 October 2020 with recommendations to the Commission on a civil liability regime for artificial intelligence, Art. 4.

AI liability is a complex issue, and hence, we cannot expect easy solutions. Some demanding distinctions will have to be made. Under the current system, the main criteria are defectiveness and fault, which are difficult to operationalize concerning AI systems, as seen. The dividing line proposed here – between illegitimate- and legitimate-harm models – arguably will often, but not always, be more straightforward regarding specific use cases. The general justification is: if a model, operating properly, should and will not cause harm, the economic and distributive justice reasons just exposed speak in favour of strict liability. This concerns, for example, machine learning used in the context of autonomous driving or medical AI. In cases of truly strict liability, to facilitate insurability,[259] a liability cap for claims based on truly strict liability should be introduced.[260] By contrast, if the model is *meant* to cause specific harm to some persons, for example by ranking and rejecting some candidates, operators must be granted the possibility to demonstrate that the model functioned properly, harm was justified, and liability thus not warranted. Hence, in these cases, a presumption of fault is in order. Notable examples include credit scoring, insurance, and recruitment.

### (3) Combination with pure economic loss

Admittedly, there is a certain risk in coupling truly strict liability with the compensation of pure economic loss: scholars are rightly worried about uncontrollably broad liability.[261] This risk, however, is sufficiently mitigated in the proposal made here, in my view. First, truly strict liability only applies to prohibited or high-risk AI systems which tend to cause illegitimate harm. Many machine learning use cases in the digital economy (e.g., insurance; credit scoring; recruitment) will not fall under these categories. Second, the combination of truly strict liability with the recovery of pure economic loss should be limited to prohibited AI systems and those high-risk AI systems which, typically, primarily cause financial harm. Otherwise, the AI liability system effectively and unjustifiably shields these high-risk systems from liability (see above, Part IV.6.a)). This second condition, however, excludes medical AI and autonomous vehicles. Ultimately, the dreaded combination will therefore affect none of the major high-risk categories in the economy.

Third, liability caps should be introduced to facilitate insurance, as seen. In addition, SMEs[262] should be able to apply for subsidies concerning insurance premia. This would help them field the costs for strict liability and incentivize SME innovation even in high-risk fields with illegitimate harms. The typical insurance exemption in cases of gross negligence counteracts moral hazard,[263] which might otherwise be introduced by subsidizing insurance. Overall, with these safeguards in place, truly strict liability may safely be established alongside the compensation of pure economic loss.

---

[259] But see also Michael Denga, 'Deliktische Haftung für künstliche Intelligenz' (2018) 34 Computer und Recht 69, 75.
[260] Spindler, 'User liability and strict liability in the Internet of Things and for robots' 137; see also Bertolini, Artificial intelligence and civil liability, 41, 93.
[261] Wagner, 'Liability Rules for the Digital Age - Aiming for the Brussels Effect', 56.
[262] For the definition of SMEs, see Article 3(3a) AI Act.
[263] Shavell, Foundations of economic analysis of law, 128 and 443.

### ii. Legitimate-harm, high-risk AI models: presumption of defectiveness/fault and causality

Legitimate-harm models, in turn, will in the normal course of operation primarily cause justified harm. In other words, from a social perspective, some harm should occur if the model performs well because the system differentiates between affected persons, some of whom will be denied favours they initially applied for. As mentioned, examples include many scoring models, such as credit scoring, résumé screening and insurance pricing, but also some use cases of face recognition (harm as a result of legitimately denying entry; of justified arrest etc.).

A decisive weakness of strict liability attached merely to causality is its excessive scope. Under truly strict liability, even perfect legitimate-harm models will trigger liability if they cause harm in their operation: for example, when rejecting candidates for job offers who are truly not qualified.[264] This, however, sets socially suboptimal incentives and is perceptibly unjust.

Therefore, concerning legitimate-harm models, the law cannot avoid determining defectiveness, or lack thereof, to properly limit the scope of covered scenarios. For design and instruction defects, fault thus resurfaces within such a strict liability system, as under the current PLD framework. As a consequence, it will be of utmost importance to develop clear thresholds for AI defectiveness (see below, Part V.2.b)).

To ensure effective compensation, defectiveness and breach of the duty of care should be presumed, as well as the causal link between these liability triggers and the damage. Nevertheless, it must be ensured that the presumption can be effectively and without undue burden rebutted by operators and professional users. For example, consider a non-biased, high-performing job application scoring model. A candidate might be rejected and now seek compensation because she believes that she was unfairly discriminated against. Since the job application scoring system constitutes a legitimate-harm model, defectiveness and breach of the duty of care (in essence: unjustified discrimination) are legally presumed. They could be rebutted, however, by disclosing information about training data, general fairness metrics and their application to the individual case. The disclosure of such information is already a (future) obligation under Article 11, Annex IV(2)(g) and (3) AI Act[265] and Article 15(1)(h) GDPR,[266] at least in the case of automated decision-making.[267]

Hence, the rebuttal option does not add any additional burden for developers or users and constitutes a proportionate balance between effective compensation and a legal environment enabling AI innovations. By functioning as an infor<mation-forcing liability rule, the presumption incentivizes the party with an information advantage, i.e., the developers and users, to document and disclose the relevant parameters.[268] Incidentally, such a system also dispenses with lengthy, costly and potentially opportunistically used disclosure of evidence mechanisms.

---

[264] Wendehorst, 'Strict liability for AI and other emerging technologies', 167.
[265] Information about training data and general fairness metrics needs to be disclosed by the AI system provider, i.e., the developer, right after training the model.
[266] Specific information about score distributions or feature relevance, in individual cases, must be provided by the data controller after the individual decision, see next fn. for references.
[267] Cf. Andrew D Selbst and Julia Powles, 'Meaningful information and the right to explanation' (2017) 7 International Data Privacy Law 233, 241 et seq.; Hacker and Passoth, Varieties of AI Explanations under the Law. From the GDPR to the AIA, and Beyond, 343, 349 et seq.
[268] Cf. the concept of information-forcing penalty defaults in Ian Ayres and Robert Gertner, 'Filling gaps in incomplete contracts: An economic theory of default rules' (1989) 99 Yale LJ 87, 91.

### iii. Foundation Models

Foundation Models also merit a specific liability regime. Not only do they constitute the most advanced, and currently perhaps most socially relevant, type of AI systems (e.g., ChatGPT, GPT-4, DALL·E 2, Stable Diffusion),[269] but they also present specific liability challenges due to the way they are deployed. As discussed above, first and foremost, the AI Act should be updated concerning the rules governing such widely adaptable and potent systems (along roughly the lines suggested by the European Parliament). A general rule of truly strict liability for Foundation Models, however, does not seem advisable from an incentive perspective. The problem of any strict liability regime is to limit liability such that it can be controlled, and insured against, by the manufacturer.[270] One Foundation Model, however, might be used in 1,000 AI applications, only one of them being a high-risk application.

Under such circumstances, it does not seem justified to label the entire Foundation Model as high-risk concerning liability (the same reasoning applies to the AI Act classification[271]). Rather, if a Foundation Model is used in a non-high-risk application, the liability rules for non-high-risk AI must apply. To hold differently would unduly disadvantage Foundation Models vis-à-vis Non-Foundation Models (which, if used in non-high-risk settings, only entail non-high-risk liability) and trigger potentially prohibitive liability costs for Foundation Model providers. Only concerning specific high-risk deployments, truly strict liability is justified and may set in. In such a situation, however, liability of the operators of the high-risk AI system – for example, a company using the Foundation Model for recruitment – and of the Foundation Model developer(s), including those fine-tuning the system,[272] must be coordinated.

In my view, Article 26 GDPR might provide a model here.[273] Under this provision, two or more controllers who are jointly responsible for data processing (e.g., Instagram and the operator of a professional Instagram account[274]) are jointly and severally liable to data subjects. Internally, they must reach an agreement in which they can efficiently distribute data protection obligations and compliance mechanisms, which must be disclosed in its most relevant parts.

This mechanism should be adapted for Foundation Model settings, which equally concern at least two AI-developing/deploying parties as well as potentially injured persons. To ensure effective compensation, both the Foundation Model provider(s) and the high-risk system operator should be jointly and severally liable under the strict liability rules sketched above. However, they should be encouraged to draw up an internal agreement in which they clarify who handles which obligations of the AI Act. In this way, compliance can be efficiently achieved by the party best suited to fulfil the obligation. Depending on the source of the error that caused damage, the party internally not responsible for compliance with the affected AI Act provision may then seek indemnification from the other party(ies). For example, if the training data was skewed and the Foundation Model provider was responsible internally for training data quality, the high-risk system operator could still be sued by an injured person.

---

[269] See note 1.
[270] Cf. Deakin and Adams, Markesinis and Deakin's Tort Law, 126.
[271] See, in greater detail, Philipp Hacker, Andreas Engel and Marco Mauer, 'Regulating ChatGPT and other Large Generative AI Models' (2023) ACM Conference on Fairness, Accountability, and Transparency (FAccT '23) 1112, 1115-1117.
[272] See ibid.
[273] See also ibid, 1117; Edwards, 'Regulating AI in Europe: four problems and four solutions'.
[274] Cf. CJEU, Case C-40/17, Fashion ID, ECLI:EU:C:2019:629.

However, they could then turn around and seek compensation from the Foundation Model provider initially responsible for the damage.

### iv. SMEs and non-high-risk AI: presumption of defectiveness/fault and causality

As a third exception to the general rule of truly strict liability, SMEs as well as operators and users of non-high-risk AI applications should only be covered by a presumption of defectiveness, breach of duty and causality, as explored in Part IV.4. Regarding SMEs, this rule could apply for the first three years after their product has been placed on the market, irrespective of whether it concerns a high-risk application or not.

Such a rule would arguably strike a reasonable balance of interest. On the one hand, it extends a certain liability shield to SMEs and operators of non-high-risk AI applications by enabling them to demonstrate a lack of fault. This addresses the fear that strict liability would stifle innovation.[275] Economically speaking, it amounts to a subsidization of AI development and uptake for smaller companies, which typically have fewer resources available for legal counsel, as well as for less critical scenarios. On the other hand, it again acknowledges the difficulty of injured persons to prove fault and causality. In addition, SMEs should be able to apply for direct liability insurance subsidies, as discussed above (Part: Combination with pure economic loss).

### v. Consumers: fault-based liability with limited presumption of causality

The final exception from the general rule of truly strict liability ought to be established for cases against consumers using AI products.[276] Here, the law should provide for fault-based liability without a presumption of fault, but with a presumption of causality as in Article 4(6) AILD Proposal. Such a rule should spur the uptake of AI by consumers, while a presumption of fault or even strict liability would likely achieve over-deterrence. Furthermore, consumers are generally not in a better position, with respect information about the system, than injured persons. Moreover, questions of consumers' fault will rarely turn on technical issues but rather on questions of the practical mishandling of AI devices. Such scenarios are not different from those involving non-AI products, and hence need not be covered by specific presumptions (except for the rare cases of Article 4(6) AILD Proposal).

## 2. Innovation

A balanced approach to AI liability should not only ease compensation for injured persons, but also provide a framework to foster innovation, development and uptake in the AI sector. In my view, the following three aspects are crucial for this: an alignment of liability law with the AI Act; and legal certainty via safe harbours.

### a) Alignment with the AI Act

The first desideratum is a general alignment of the duties under the PLD and the AILD Proposal framework with the obligations under the AI Act. Ideally, a violation of the AI Act would function as a necessary and sufficient condition for liability under either framework. This would significantly lower compliance, administration and enforcement costs as all stakeholders (affected persons, courts, developers) would only have to internalize and comply with, or analyse and adjudicate, one set of rules. Every deviation causes legal and compliance costs as well as administrative friction, which may quickly become prohibitive, particularly for start-ups and SMEs. The result would be a further concentration of the digital economy in the hands of companies with significant market power and pockets deep enough to pay for legal counsel, compliance tools, and related insurance. This seems undesirable from the perspective of workable competition.[277]

---
[275] See Schneeberger, Stöger and Holzinger, The European legal framework for medical AI 128.
[276] Cf. Wendehorst, 'Strict liability for AI and other emerging technologies', 157.

A full alignment of liability with the AI Act would also imply that compliance with the latter shields companies entirely from civil liability. However, this deference of the civil liability framework is only defensible if the AI Act fully and adequately addresses all risks presented by AI with respect to individual harm eligible to be compensated under the civil liability framework. Even though the AI Act is still under discussion, it does not seem premature to predict that such a halcyon state will not be reached in the final version. Importantly, as the last section of this paper will show in detail (see below, Part V.4.), environmental and sustainability risk of AI are not adequately addressed in the AI Act (even though the EP version contains some rules to this effect).[278] Hence, as a second best, non-compliance with the AI Act can only constitute a sufficient, but not a necessary condition for finding defectiveness and breach of a duty of care under the PLD and the AILD Proposals, respectively.

### b) Safe harbours and legal certainty

Since a full alignment with the AI Act is not desirable at the moment, it is all the more important to foster innovation by providing legal certainty. For AI developers, legal uncertainty can have true chilling effects; legal certainty, conversely, helps companies to be and stay compliant. They can then decide whether they would like to fulfil the criteria themselves or outsource compliance to specialized companies, many of which are currently springing up in the AI ecosystem. This frees resources for technical and economic development and removes potentially over-deterrent fears of liability with software engineers and computer scientists not trained in law.

To achieve legal certainty, ample and quick use should be made of harmonized standards and, specifications foreseen in Articles 40 and 41 AI Act, respectively. These provisions should be used to define safe harbours for developers using, to the best extent possible, state-of-the-art measures and metrics with quantitative thresholds.[279] For example, vague references to training data being "relevant, representative, and to the best extent possible, free of errors and complete" (Article 10(2) AI Act) and having "appropriate statistical properties" (Article 10(3) AI Act), to "sufficiently transparent" systems (Article 13(1) AI Act) or to an "appropriate level of accuracy, robustness and cybersecurity" (Article 15(1) AI Act) should be translated into use-case-specific quantitative measures. For example, performance measure thresholds could be specified for AI recruitment tools. Tools with an accuracy of over 80% and a normalized F1 score of over 0.8[280] could then be presumed to meet the requirements of an appropriate level of "accuracy" (which should read "performance", to be technically correct[281]).

---

[277] See, e.g., Rec. 3 DMA.
[278] Moreover, the AI Act contains virtually no obligations for non-high-risk AI systems (Article 52 AI Act) and for non-professional users (Article 2(8) AI Act). Even obligations of professional users are sparse and not meant to be exhaustive (Recital 58 AI Act).
[279] Cf. Hacker, 'A legal framework for AI training data—from first principles to the Artificial Intelligence Act' 299.

In this way, safe harbours would offer legal certainty for AI developers and users. Those who choose to deploy AI not reaching the safe harbours would do their "at their own risk". Conversely, quantitative thresholds could be specified to mark a presumption of non-compliance as well. In this way, standards and specifications would designate quantitative red zones (non-compliance), green zones (compliance) and grey zones (unclear compliance status). Developers could adapt their models and strategies accordingly.

This is all the more important as the latest versions of the AI Act have increasingly included references to fundamental rights not only in the recitals, but also as explicit elements of articles and obligations. While fundamental rights are obviously the bedrock of the rule of law in the EU, their infamous conceptual vagueness makes them uneasy bedfellows for inclusion in secondary law AI regulation. More specifically, rules referencing fundamental rights are hardly operationalizable with a sufficient degree of certainty in specific scenarios. Here, safe harbours are particularly important here.

### 3.    Trustworthy AI

Effective compensation and innovation cannot, however, be the only goals of AI liability. Trustworthy AI has emerged as a key concept in AI ethics[282] and constitutes a cornerstone of the EU approach to the regulation of AI.[283] While its understanding by the EU Commission has been criticized,[284] the AI liability package explicitly endorses "the objective […] to promote the rollout of trustworthy AI to harvest its full benefits for the internal market."[285] Key aspects of trustworthiness in AI,[286] such as the hedging of unforeseeable behaviour, the reduction of opacity and mitigation of discrimination AI systems, continue to be of major importance not only from an ethical but also from a regulatory perspective. Working toward these goals implies effectively addressing the main risks of AI mentioned in Part II.

However, as the following section shows, some of these crucial dimensions are not properly addressed in the new liability framework and should urgently be bolstered. Other aspects of trustworthy AI, such as non-manipulation[287] and privacy or data protection,[288] transcend the scope of this paper.

---

[280] The F1 score takes both false negative and false positive ratios into account, see Lindholm and others, Machine Learning, 88.
[281] Accuracy is only one of many performance measures and quite unhelpful in many scenarios, for example medical AI (imbalanced problem); see, e.g., ibid, 88.
[282] Taddeo, 'Modelling trust in artificial agents, a first step toward the analysis of e-trust'; Pieters, 'Explanation and trust: what to tell the user in security and AI?'; Grodzinsky, Miller and Wolf, 'Developing artificial agents worthy of trust:"Would you buy a used car from this artificial agent?"' 21; Ferrario, Loi and Viganò, 'In AI we trust incrementally: A multi-layer model of trust to analyze human-artificial intelligence interactions'.
[283] See Fn. 53.
[284] Laux, Wachter and Mittelstadt, 'Trustworthy Artificial Intelligence and the European Union AI Act: On the Conflation of Trustworthiness and the Acceptability of Risk'.
[285] AILD Proposal, 2.
[286] See Fn. 282.

### a) Autonomy and unforeseeability

An important component of trustworthy AI is predictability.[289] This feature is obviously challenged by the unforeseeability of some types of AI output, which results from the models' autonomy.

The current proposals arguably seek to mitigate the effects of unforeseeability by lowering barriers to compensation, via (supposedly) strict negligence, presumption of fault and causality. More importantly even, the AI Act aims to render unforeseeable damage less likely by imposing several boundary conditions on the development and deployment of high-risk AI. For example, the data governance regime in Article 10 AI Act seeks to ensure that training, validation and test data are sufficiently diverse and representative of the target population. By adhering to the regime, providers significantly raise the likelihood that their model will generalize well to unseen cases in the field.[290] Conversely, the output requirements in Article 15 AI Act are specifically designed to minimize error rates and prevent negative feedback loops in online learning scenarios.[291] These are indeed important conditions for the use of any AI system, not only high-risk applications.

One specific problem regarding unforeseeability, however, is that the concept of causality is not harmonized in the proposals.[292] Some national tort systems maintain a normatively enriched understanding of causality under which injurers are exempted from compensating damages that, at the moment of the act or omission, were objectively unforeseeable.[293] With AI, however, unforeseeable output, and damage caused by it, is foreseeable in the abstract.[294] Therefore, the EU liability regime should specify in both directives that damage eligible for compensation includes harm caused by unforeseeable acts or omissions of an AI system.

---

[287] For an analysis in this direction, see, e.g., Helberger and others, 'EU Consumer Protection 2.0'; Hacker, 'Manipulation by algorithms. Exploring the triangle of unfair commercial practice, data protection, and privacy law'; N Helberger and others, 'Choice Architectures in the Digital Economy: Towards a New Understanding of Digital Vulnerability' (2022) 45 Journal of Consumer Policy 175; Wendehorst, The Proposal for an Artificial Intelligence Act COM (2021) 206 from a Consumer Policy Perspective 64 et seqq.; Claire Boine, 'AI-enabled manipulation and EU law' (2021) Available at SSRN 4042321; Hubert Thomas, 'Opinion 3/2018 EDPS Opinion-on online manipulation and personal data' (2018); Frederik Zuiderveen Borgesius and others, 'Online political microtargeting: promises and threats for democracy' (2018) 14 Utrecht Law Review 82-96.
[288] Wachter, Mittelstadt and Russell, 'Counterfactual explanations without opening the black box: Automated decisions and the GDPR'; Selbst and Powles, 'Meaningful information and the right to explanation'; Mittelstadt, Russell and Wachter, Explaining explanations in AI; Margot E Kaminski and Gianclaudio Malgieri, Multi-layered explanations from algorithmic impact assessments in the GDPR (2020); on generative AI, see Hacker, Engel and Mauer, 'Regulating ChatGPT and other Large Generative AI Models', Technical Report.
[289] Cf. D Harrison McKnight and Norman L Chervany, 'The meanings of trust' (1996) 34.
[290] Lindholm and others, Machine Learning, 296 et seqq.
[291] See Fn. 134 for details on online learning.
[292] See, e.g., Zech, 'Liability for AI: Complexity problems' (forthcoming).
[293] Selbst, 'Negligence and AI's human users' et seqq.; BGH, Case I ZR 31/51, Sluice Accident, BGHZ 3, 261 para. 15: An adequate causal connection exists "if a fact was in general, and not only under particularly peculiar, quite improbable circumstances […] suitable to bring about the outcome."
[294] Denga, 'Deliktische Haftung für künstliche Intelligenz', 72.

### b) Complexity and opacity: explainable AI

A second crucial dimension of trustworthiness highlighted in AI ethics is explainability,[295] i.e., the degree to which an observer may understand the causes of the system's output.[296] Such reasons for an AI outcome may be obscured by two distinct and independent versions of opacity: institutional and technical opacity.[297] The former denotes strategic withholding of information by AI developers or users; the latter refers to the difficulty of pinpointing the causes of a model's output due to its technical complexity.

The disclosure of evidence regime of the AILD and PLD Proposal clearly seek to directly overcome institutional opacity, while the burden of proof regime tackles this issue indirectly by burdening operators, including developers, if and when they do not provide necessary information.[298]

The proposals are much less explicit concerning the challenge of technical opacity, however. From a technical perspective, many models satisfy explainability at least in the sense that one may swiftly determine what features are most relevant for an individual decision (local explanation) or for the entire model (global explanation).[299] For example, linear regression or simple decision trees are interpretable ex ante, delivering quite easily the weights for individual features in the model.[300] For more complex models, though, such as deep neural networks, ensemble models or support vector machines, only post-hoc explanations approximating feature relevance can be provided for individual decisions.[301] Global explanations are even more difficult.[302] Moreover, the accuracy of these explanations continues to be debated.[303]

EU law has, in recent years, enshrined several provisions to force explanations (in the sense of feature relevance) for AI models.[304] The AI Act rather timidly adds a generic transparency

---

[295] Pieters, 'Explanation and trust: what to tell the user in security and AI?'; see also Brent Daniel Mittelstadt and others, 'The ethics of algorithms: Mapping the debate' (2016) 3 Big Data & Society 2053951716679679, 6 et seq.
[296] Definition from: Or Biran and Courtenay Cotton, Explanation and justification in machine learning: A survey (2017); see also Zachary C Lipton, 'The mythos of model interpretability: In machine learning, the concept of interpretability is both important and slippery' (2018) 16 Queue 31.
[297] See, e.g., Burrell, 'How the machine 'thinks': Understanding opacity in machine learning algorithms'; Andrew D Selbst and Solon Barocas, 'The intuitive appeal of explainable machines' (2018) 87 Fordham L Rev 1085.
[298] See also Rec. 3 AILD Proposal; Commission, Questions and answers on the revision of the Product Liability Directive, under 9.
[299] Arrieta and others, 'Explainable Artificial Intelligence (XAI): Concepts, taxonomies, opportunities and challenges toward responsible AI', 90; Marco Tulio Ribeiro, Sameer Singh and Carlos Guestrin, '"Why should i trust you?" Explaining the predictions of any classifier' (2016) Proceedings of the 22nd ACM SIGKDD International Conference on Knowledge Discovery and Data Mining 1135Lipton, 'The mythos of model interpretability: In machine learning, the concept of interpretability is both important and slippery'.
[300] See, e.g., Cynthia Rudin, 'Stop explaining black box machine learning models for high stakes decisions and use interpretable models instead' (2019) 1 Nature Machine Intelligence 206.
[301] Marco Tulio Ribeiro, Sameer Singh and Carlos Guestrin, " Why should i trust you?" Explaining the predictions of any classifier (2016); Arrieta and others, 'Explainable Artificial Intelligence (XAI): Concepts, taxonomies, opportunities and challenges toward responsible AI' 93.
[302] Industry may use a global average of local SHAP values, cf. Scott M Lundberg and Su-In Lee, 'A unified approach to interpreting model predictions' (2017) 30 Advances in Neural Information Processing Systems 4765; see also Sebastian Lapuschkin and others, 'Unmasking Clever Hans predictors and assessing what machines really learn' (2019) 10 Nature Communications 1.
[303] Anna Hedström and others, 'The Meta-Evaluation Problem in Explainable AI: Identifying Reliable Estimators with MetaQuantus' (2023) arXiv preprint arXiv:230207265.
[304] See the overview in Adrien Bibal and others, 'Legal requirements on explainability in machine learning' (2021) 29 Artificial Intelligence and Law 149; Mateusz Grochowski and others, 'Algorithmic transparency and explainability for EU consumer protection: unwrapping the regulatory premises' (2021) 8 Critical Analysis of Law

regime in its Article 13 as well as an obligation to provide feature relevance as part of the technical documentation required under its Article 11(1) and Annex IV(2)(b).[305]

### i. Black-box models: disclosure and presumptions of fault or defectiveness

The proposals implicitly tackle technical opacity in two ways: with the disclosure and the presumption regime. As seen, under both the AILD Proposal (Article 3(1)) and the PLD Proposal (Article 8(1)), defendants have to provide relevant evidence that is *at their disposal*. This invites the question: what if the information is not at the defendant's disposal because a black-box AI system is used? For example, it is currently practically impossible to obtain feature relevance scores for the entire model when support vector machines (SVM), a type of machine learning model, is used.[306] This information may be crucial, however, to determine if a credit scoring or other SVM-based decision was driven by bias. Similar difficulties may surface regarding deep neural networks.[307]

Hence, at first glance, the defendant need not disclose this information since it is not at their disposal (i.e., impossible to obtain). However, the choice of using a black-box model cannot, in my view, lead to a greater burden on injured persons. Rather, the defendant should bear the consequences of her choice of an inscrutable model. This follows from a purposive interpretation of the disclosure norms, which are meant to alleviate, not perpetuate, difficulties of proof for victims. Here again, it is essential that the disclosure is also due if the evidence *should* be, but due to the block-box character of the model *is* not, at the defendant's disposal.[308]

### ii. Explainable AI: Presumptions of causality

As seen, Article 4(2) AILD Proposal mentions the transparency regime of Article 13 AI Act among those provisions whose violation triggers a rebuttable presumption of causality. The analysis above[309] has already suggested, however, that this framework is insufficient for any obligations compelling post-processing activities, i.e., an action after the output of the AI system is delivered. This problem resurfaces specifically regarding explainability. If Article 13 AI Act prescribes post-hoc explanations, which is a reasonable understanding in some circumstances,[310] the presumption of the causal link between fault (i.e., violation of Article 13 AI Act) and AI output is immediately rebuttable: as the explanation would have occurred *after* the output, it cannot have caused it.[311] However, injured parties will face an equally uphill legal battle to prove a causal link between an incorrect or non-existent post-hoc explanation and damage: they would have to show that the explanation would indeed have had an effect on subsequent events and would not have been ignored or misunderstood,[312] as most statutorily

---

43; Thierry Kirat and others, 'Fairness and Explainability in Automatic Decision-Making Systems. A challenge for computer science and law' (2022) arXiv preprint arXiv:220603226, 30 et seqq.
[305] Hacker and Passoth, Varieties of AI Explanations under the Law. From the GDPR to the AIA, and Beyond.
[306] Arrieta and others, 'Explainable Artificial Intelligence (XAI): Concepts, taxonomies, opportunities and challenges toward responsible AI' 90.
[307] Ibid.
[308] See also text accompanying Fn. 172.
[309] Part IV.5.d)ii.(2)(b).
[310] Cf. Hacker and Passoth, Varieties of AI Explanations under the Law. From the GDPR to the AIA, and Beyond 359.
[311] See Fn. 195 for a more detailed argument.
[312] Ashraf Abdul and others, COGAM: measuring and moderating cognitive load in machine learning model explanations (2020).

provided information generally is.[313] Hence, to make any sense for post-hoc explanations, the presumption must cover the causal link between the violation of Article 13 AI Act and the damage, not only between the violation and the output.

Moreover, a reference to Article 11 AI Act is missing in Art. 4(2) AILD Proposal; Recital 22 AILD Proposal suggests that the provision is missing because the Commission considers that it is not intended to protect against damage of affected persons. However, this assessment should be questioned particularly regarding the transparency requirement of Article 11 in conjunction with Annex IV(2)(b) AI Act. While it may indeed be debated if feature relevance should be the required explainability metric in all situations,[314] one cannot seriously doubt that such explanations should, at least by way of an analysis by providers and professional users, ultimately protect affected persons, too.[315] Therefore, Article 11 in conjunction with Annex IV(2)(b) AI Act should be added to the list of requirements whose violation prompts the presumption of causality under Article 4(2) AILD Proposal.

Beyond the AILD Proposal, it should be noted that failure to use explainable AI may constitute a product defect under the PLD framework as well, particularly if the explanation would significantly enhance the safety of the product.[316] Nevertheless, it remains unclear what specific explanatory metric may be used to avoid design defects in such situations. Hence, implementing acts should spell out specifically what type of explanation (feature relevance; counterfactual etc.) is needed in what use case under the AI Act and the AI liability regime. This would increase both legal certainty and the safety of affected persons.

### c) Discrimination: algorithmic fairness

The third crucial dimension of trustworthiness in AI is competence, i.e., the capacity of models to reach the goals set by their human users or developers.[317] The fulfilment of this criterion is not only threatened by suboptimal operational performance but also and specifically by discrimination exhibited by AI systems. A wealth of scholarship has shown shortcomings of the current EU non-discrimination framework concerning the mitigation of AI-driven discrimination,[318] also in the context of high-risk AI applications such as credit scoring,[319] insurance,[320] or employment.[321]

---

[313] Cf. Omri Ben-Shahar and Carl E Schneider, More than you wanted to know. The Failure of Mandated Disclosure (Princeton University Press 2014).

[314] See, e.g., Wachter, Mittelstadt and Russell, 'Counterfactual explanations without opening the black box: Automated decisions and the GDPR'; Hacker and Passoth, Varieties of AI Explanations under the Law. From the GDPR to the AIA, and Beyond, 363-365.

[315] See, e.g., Ribeiro, Singh and Guestrin, " Why should i trust you?" Explaining the predictions of any classifier; Hacker and Passoth, Varieties of AI Explanations under the Law. From the GDPR to the AIA, and Beyond et seqq..

[316] Hacker and Passoth, Varieties of AI Explanations under the Law. From the GDPR to the AIA, and Beyond et seqq.; Hacker and others, 'Explainable AI under contract and tort law: legal incentives and technical challenges' et seqq.; Froomkin, Kerr and Pineau, 'When AIs outperform doctors: confronting the challenges of a tort-induced over-reliance on machine learning' 97.

[317] See, e.g., Mark Coeckelbergh, 'Can we trust robots?' (2012) 14 Ethics and information technology 53, 54-57; Ferrario, Loi and Viganò, 'In AI we trust incrementally: A multi-layer model of trust to analyze human-artificial intelligence interactions', 531-532.

However, the proposed liability framework gives this troubling gap fairly short shrift. The AI Act comprises sparse provisions on the examination of AI modelling data regarding prohibited biases and requirements for the representativeness and statistical appropriateness of training data;[322] an exception to process sensitive data for algorithmic fairness purposes;[323] information duties concerning potential discriminatory impact;[324] and an obligation to prevent, as far as possible, biased feedback loops in online learning settings.[325] The liability framework proposed by the Commission, however, only incompletely maps onto these provisions.

### i. Discrimination and the AILD Proposal

The causality presumption in Article 4(2) AILD Proposal refers to Article 10 (data governance) and 15 AI Act (feedback loops), but omits Article 11 AI Act (information duties). Here again, the same arguments as in the explainability section speak for the addition of Article 11 AI Act to the norms triggering the causality presumption.[326] Another, potentially even more pernicious issue arises regarding the applicability of the AILD Proposal. The two proposals establish a radical dichotomy between strict and fault-based liability, with the causality presumption under Article 4 AILD Proposal only covering the link between fault and output. However, liability under the non-discrimination directives is strict, not fault-based, as the CJEU has repeatedly stressed.[327] Hence, under a narrow reading, the causality presumption would not even apply to non-discrimination cases. The Commission, however, seems to view discrimination as fault.[328] Under such a reading, by contrast, the provision might also apply to Article 82 GDPR, where liability is even more clearly fault-based.[329]

Returning to liability for discriminatory AI systems, the AILD Proposal should explicitly extend its disclosure of evidence and burden of proof rules to non-discrimination law and liability under the non-discrimination directives. As mentioned, a key problem in non-discrimination litigation is enforcement, which often turns on the (unavailability of) access to training data and the model.[330] Article 3 AILD Proposal could theoretically provide such access, statutorily overruling the *Meister* case.[331]

---

[318] See Fn. 4.
[319] See, e.g., Hacker, 'Teaching fairness to artificial intelligence: existing and novel strategies against algorithmic discrimination under EU law'; Katja Langenbucher, 'Responsible AI-based credit scoring–a legal framework' (2020) 31 European Business Law Review 527; Zehlike, Hacker and Wiedemann, 'Matching code and law: achieving algorithmic fairness with optimal transport'; Nikita Aggarwal, 'The norms of algorithmic credit scoring' (2021) 80 The Cambridge Law Journal 42; Katja Langenbucher, 'Consumer Credit in The Age of AI' (2022) LawFin Working Paper No 42.
[320] See, e.g., Angelo Borselli, 'Insurance by algorithm' (2018) Eur Ins L Rev 35; Hacker, 'Teaching fairness to artificial intelligence: existing and novel strategies against algorithmic discrimination under EU law'.
[321] See, e.g., Hacker, 'Teaching fairness to artificial intelligence: existing and novel strategies against algorithmic discrimination under EU law'; Jeremias Adams-Prassl, 'What if your boss was an algorithm? Economic Incentives, Legal Challenges, and the Rise of Artificial Intelligence at Work' (2019) 41 Comp Lab L & Pol'y J 123; Valerio De Stefano, '"Negotiating the Algorithm": Automation, Artificial Intelligence, and Labor Protection' (2019) 41 Comp Lab L & Pol'y J 15; Adams-Prassl, Binns and Kelly-Lyth, 'Directly Discriminatory Algorithms'; see also Eva Kocher, Digital Work Platforms at the Interface of Labour Law (Bloomsbury Publishing 2022).
[322] Article 10(2)(f) and (3) AI Act.
[323] Article 10(5) AI Act.
[324] Article 11(1) in conjunction with Annex IV(2)(g) and (3) AI Act.
[325] Article 15(3) AI Act.
[326] See text accompanying Fn. 315.
[327] CJEU, Case C-177/88, Dekker, ECLI:EU:C:1990:383; CJEU, Case C-180/95, Draehmpaehl, ECLI:EU:C:1997:208.
[328] Commission, Questions and answers on the revision of the Product Liability Directive, under 5; Spindler, 'Die Vorschläge der EU-Kommission zu einer neuen Produkthaftung und zur Haftung von Herstellern und Betreibern Künstlicher Intelligenz', 704.
[329] See Art. 82(3) GDPR.

In contrast to Article 4, disclosure of evidence is not restricted to fault-based liability in the first place. However, facts and evidence supporting the plausibility of the claim must be provided. As shown, this mirrors the, often prohibitive, requirement in current non-discrimination law of providing facts suggesting a prima facie case of discrimination.[332] Without clear guidelines concerning the threshold of plausibility, the disclosure of evidence mechanism risks to be mere law in the books in the field of non-discrimination (and beyond).

### ii. Discrimination and the PLD Proposal

With respect to the PLD framework, it is not a priori ruled out that discriminatory output by an AI system constitutes a product defect. However, the PLD Proposal does not explicitly reference any non-discrimination directives, while it does expressly link to product safety and cybersecurity requirements (Article 6(1)(f) PLD Proposal). This does provide an argument *e contrario* against importing requirements from the EU non-discrimination law acquis into the concept of defectiveness. The Commission is also of the opinion that discrimination does not trigger a claim under the new PLD framework.[333]

On the other hand, one may argue that non-discrimination constitutes a particularly strong expectation by the overall public and most likely also the specific expectation of many intended end users and affected persons of AI systems (Article 6(1)(h) PLD Proposal). While the consideration of non-discrimination requirements for the sake of defectiveness can therefore not be automatic, as in the case of cybersecurity requirements, it is submitted that they should nevertheless be taken into account on a case-by-case basis. Otherwise, the AI liability framework risks to become a dead letter precisely in the area where bolstered enforcement is arguably most needed.

More generally, to provide legal certainty, implementing acts should, as in the case of explainable AI, seek to delineate what kind of fairness metrics are admissible to prevent illegal discrimination in specific use cases.[334] The computer science literature has developed an astounding array of fairness metrics in previous years,[335] some of which are incompatible with one another or can be fulfilled only partially at a given moment.[336] The integration of these results into EU non-discrimination remains a distinctive challenge at the intersection of law and computer science.

---

[330] See text accompanying Fn. 171.
[331] See text accompanying Fn. 165.
[332] See text accompanying Fn. 171.
[333] Commission, Questions and answers on the revision of the Product Liability Directive, under 5.
[334] For a scholarly account, see, e.g., Kirat and others, 'Fairness and Explainability in Automatic Decision-Making Systems. A challenge for computer science and law', 14 et seqq.; Wachter, Mittelstadt and Russell, 'Bias preservation in machine learning: the legality of fairness metrics under EU non-discrimination law'; Dino Pedreschi, Salvatore Ruggieri and Franco Turini, 'A study of top-k measures for discrimination discovery' (2012) Proceedings of the 27th Annual ACM Symposium on Applied Computing 126.
[335] See, e.g., Meike Zehlike, Ke Yang and Julia Stoyanovich, 'Fairness in Ranking, Part I: Score-based Ranking' (2022) ACM Computing Surveys (CSUR); Dana Pessach and Erez Shmueli, 'Algorithmic fairness' (2020) arXiv preprint arXiv:200109784.
[336] See, e.g., Jon Kleinberg, Sendhil Mullainathan and Manish Raghavan, Inherent Trade-Offs in the Fair Determination of Risk Scores (Schloss Dagstuhl-Leibniz-Zentrum für Informatik 2017); Sorelle A Friedler, Carlos Scheidegger and Suresh Venkatasubramanian, 'The (im) possibility of fairness: Different value systems require different mechanisms for fair decision making' (2021) 64 Communications of the ACM 136; Zehlike, Hacker and Wiedemann, 'Matching code and law: achieving algorithmic fairness with optimal transport'; Corinna Hertweck

### 4. Sustainability

The final development perspective for the future of AI regulation is the one with undoubtedly the greatest social and economic significance: ensuring that AI, and technology more generally, do not exacerbate the climate crisis, but are put on a track toward greater environmental sustainability.[337]

#### a) A blind spot in AI regulation

In computer science, awareness for the sustainability risks of AI is rapidly growing in the machine learning community.[338] And with good reason: the estimated contribution of the ICT sector to global greenhouse gas emissions lies between 1.8 and 3.9%,[339] potentially surpassing global air traffic (ca. 2.5%).[340] The carbon footprint of machine learning more specifically has increased dramatically over the last years.[341] Given the current pace of AI development and uptake, the contribution of machine learning to global warming is real, noticeable, and rising.[342] Hence, clear and convincing incentives should be introduced to make AI more sustainable.

In legal scholarship, the implications of law for climate change, and vice versa, are no longer restricted to environmental law,[343] but increasingly spill over into the general legal literature.[344] However, when it comes to AI regulation more specifically, questions of climate change and

---

and Tim Räz, Gradual (In) Compatibility of Fairness Criteria (2022); Zehlike and others, 'Beyond incompatibility: Trade-offs between mutually exclusive algorithmic fairness criteria in machine learning and law'.
[337] See, e.g., Philipp Hacker, 'Sustainable AI Regulation' (2023) Working Paper, presented at the Privacy Law Scholars Conference 2023, https://arxivorg/abs/230600292. There are, of course, other sustainability concerns arising in the context of AI as well, such as economic or social sustainability, which are primarily addressed here via trustworthy AI; cf. Friederike Rohde and others, Nachhaltigkeitskriterien für künstliche Intelligenz (2021), 28.
[338] See, e.g., Freitag and others, 'The real climate and transformative impact of ICT: A critique of estimates, trends, and regulations'; Cowls and others, 'The AI gambit: leveraging artificial intelligence to combat climate change—opportunities, challenges, and recommendations'; Mariarosaria Taddeo and others, 'Artificial intelligence and the climate emergency: Opportunities, challenges, and recommendations' (2021) 4 One Earth 776; see also Fn. 41.
[339] Freitag and others, 'The real climate and transformative impact of ICT: A critique of estimates, trends, and regulations'; see also OECD, Measuring the Environmental Impacts of AI Compute and Applications: The AI Footprint (2022), 25-26.
[340] Council, ACM TechBrief: Computing and Climate Change, 1.
[341] Ibid: potentially by a factor of up to 300,000.
[342] OECD, Measuring the Environmental Impacts of AI Compute and Applications: The AI Footprint, 5; but see also, for a decoupling of the growth of performance and energy consumption, Radosvet Desislavov, Fernando Martínez-Plumed and José Hernández-Orallo, 'Compute and energy consumption trends in deep learning inference' (2021) arXiv preprint arXiv:210905472
[343] See, e.g., Stepan Wood and Benjamin J Richardson (eds), Environmental Law for Sustainability (Bloomsbury Publishing 2006).
[344] See, e.g., W. Neil Adger and Andrew Jordan (eds), Governing Sustainability (Cambridge University Press 2009); Klaus Bosselmann, The principle of sustainability: transforming law and governance (Routledge 2016); Beate Sjåfjell and Benjamin J. Richardson (eds), Company Law and Sustainability (Cambridge University Press 2015); Anne-Christin Mittwoch, Nachhaltigkeit und Unternehmensrecht (Mohr Siebeck 2022); Jan-Erik Schirmer, Nachhaltiges Privatrecht (Mohr Siebeck, forthcoming).

sustainability still constitute a significant blind spot.³⁴⁵ Only the European Parliament added several provisions to this effect in its position adopted on June 14, 2023.³⁴⁶

The original Commission proposal of the AI Act remained largely silent on questions of environmental sustainability.³⁴⁷ This evasiveness (mere encouragement, in three years, voluntary codes of conduct, see last fn.) is, however, emblematic: the legal framework for sustainable AI is pushed toward an uncertain future, while climate change has already become a definite, transformative and destructive force in the present.

The Commission maintains that the AILD Proposal connects with and fosters the European Green Deal.³⁴⁸ However, this interaction is narrowly focused on facilitating the deployment of AI for furthering sustainability goals. "The uptake of AI applications is beneficial for the environment," the Commission concludes.³⁴⁹ While AI does have a role to play here,³⁵⁰ this analysis is dangerously one-sided. It ignores the many ways in which AI can, and actually does, contribute to climate change, which will almost certainly significantly increase in the coming years.³⁵¹

### b) Regulating for sustainable AI

A full-blown proposal for a regulatory framework supporting sustainable AI development and deployment transcends the scope of this paper. However, two important strategies can be sketched that should be incorporated into the proposals discussed at the EU level. Simultaneously, they may serve as a blueprint for sustainable technology regulation more generally.³⁵²

#### i. Sustainability impact assessment

First, the AI Act should mandate a "sustainability impact assessment" for AI systems.³⁵³ Such an assessment could draw on the manifold proposals concerning impact assessments for AI systems in general.³⁵⁴ To this end, a provision structurally mirroring Article 9 AI Act (risk

---

³⁴⁵ But see Mario Martini and Hannah Ruschemeier, 'Künstliche Intelligenz als Instrument des Umweltschutzes', (2022) Zeitschrift für Umweltrecht 515; 521 et seqq.; generally on the regulation of the digital transformation and digitization, Herbert Zech, 'Nachhaltigkeit und Digitalisierung im Recht ' (2022) 2022 Zeitschrift für Digitalisierung und Recht 123; see also: Natasa Perucica and Katarina Andjelkovic, 'Is the future of AI sustainable? A case study of the European Union' (2022) 16 Transforming Government: People, Process and Policy 347; Ugo Pagallo, Jacopo Ciani Sciolla and Massimo Durante, 'The environmental challenges of AI in EU law: lessons learned from the Artificial Intelligence Act (AIA) with its drawbacks' (2022) 16 Transforming Government: People, Process and Policy 359; Scott Robbins and Aimee van Wynsberghe, 'Our New Artificial Intelligence Infrastructure: Becoming Locked into an Unsustainable Future' (2022) 14 Sustainability 4829.
³⁴⁶ For an overview and critique, see Hacker, 'Sustainable AI Regulation', Part III.
³⁴⁷ See Article 69(2) AI Act (voluntary codes of conduct); Article 84(4) AI Act (encouragement of the Commission to include environmental sustainability of AI systems into voluntary codes of conduct in its first review three years after AI Act's date of application.
³⁴⁸ AILD Proposal, p. 4.
³⁴⁹ AILD Proposal, p. 4.
³⁵⁰ See, e.g., Josh Cowls and others, 'The AI gambit: leveraging artificial intelligence to combat climate change—opportunities, challenges, and recommendations' (2021) AI & Society 1.
³⁵¹ See Council, ACM TechBrief: Computing and Climate Change.
³⁵² On that desideratum, see Nat. Acad. of Sciences, Fostering Responsible Computing Research: Foundations and Practices, 2022, 55.
³⁵³ See now also Hacker, 'Sustainable AI Regulation', Part IV.
³⁵⁴ Andrew D Selbst, 'An Institutional View of Algorithmic Impact Assessments' (2021) 35 Harvard Journal of Law & Technology; Kaminski and Malgieri, Multi-layered explanations from algorithmic impact assessments in

management system) should be added to the AI Act. It would apply to developers of high-risk and non-high-risk AI systems alike as the carbon footprint of AI systems is unrelated to their level of risk for health, safety or fundamental rights. Crucially, during the modelling phase, developers should compare different model types (e.g., linear regression versus deep learning) not only with respect to their performance, but also their estimated climate footprint.[355] Already, there are tools available to measure the carbon impact of models.[356] Simply put, if two model types exhibit similar performance, the developers would be obliged under the new provision to choose the more sustainable model for further development and deployment. In this way, the current fixation on performance measures can be complemented by even greater environmental awareness and concrete, low maintenance steps to include sustainability in the wider target function of ML development.

In fact, in many scenarios, sustainability and performance may even go hand in hand. A current trend in machine learning is the use of pre-trained models, which have been trained on some more general data for a certain task class (for example, image[357] or speech recognition[358]).[359] They are then fully trained by developers working on a concrete problem with domain-specific data. Such pre-trained (Foundation) models are not only often more performative and have become the state-of-the-art architecture in numerous tasks,[360] but they may also be less energy consuming overall as the pre-training only has to be done once for many different model deployments.[361] However, ironically, regulation might thwart these efforts. The most potent pre-trained models are Foundation Models – precisely the ones whose development the Council version of the AI Act[362] and the AI liability directives significantly dis-incentivise. Here, the wheel comes full circle, but not in an efficient or sustainable way. This again points to the importance of changing the rules for Foundation Models, as contemplated above (Part V.1.a)i.(2)(b)).

### ii. Sustainable design defect

A second policy option could consist of qualifying the deployment of a model with suboptimal sustainability parameters as a design defect under Article 6 PLD Proposal. Again, a risk-utility analysis would have to show that a reasonable alternative, i.e., a feasible, comparably performing model with a smaller carbon footprint, was not realized. Ideally, the test for this specific "sustainable design defect" would be the same as the one for the sustainability impact

---

the GDPR; Margot E Kaminski and Gianclaudio Malgieri, 'Algorithmic impact assessments under the GDPR: producing multi-layered explanations' (2020) International Data Privacy Law 19.

[355] The exact impact is not easy to measure. An index including Scope 1, 2 and 3 Emissions for necessary compute resources (e.g., energy; carbon emissions) for training and retraining could be used as a proxy. For a more comprehensive impact measure (including production, transport, and end-of-life, as well as water consumption), see OECD, Measuring the Environmental Impacts of AI Compute and Applications: The AI Footprint, Annex A; on Scope 1, 2 and 3 Emissions, see IPCC, Working Group III Contribution to the Fifth Assessment Report of the Intergovernmental Panel on Climate Change (2014), 122.

[356] Overview in OECD, Measuring the Environmental Impacts of AI Compute and Applications: The AI Footprint, 28; see also https://codecarbon.io/.

[357] See, e.g., Gustavo Carneiro, Jacinto Nascimento and Andrew P Bradley, Unregistered multiview mammogram analysis with pre-trained deep learning models (Springer 2015).

[358] Juliette Millet and others, 'Toward a realistic model of speech processing in the brain with self-supervised learning' (2022) NeurIPS https://arxiv.org/abs/2206.01685.

[359] Xu Han and others, 'Pre-trained models: Past, present and future' (2021) 2 AI Open 225.

[360] Ibid.

[361] David Patterson and others, 'Carbon emissions and large neural network training' (2021) arXiv preprint arXiv:210410350, 15.

[362] On this, see, e.g., Hacker, Engel and Mauer, 'Regulating ChatGPT and other Large Generative AI Models', 1114; Helberger and Diakopoulos, 'ChatGPT and the AI Act'.

assessment required under the AI Act. This would again link the regulatory with the liability approach, as advocated throughout this paper.

Via these dual, but interrelated tracks, it is submitted, the law can help shift the transition from the performance orientation of AI models[363] not only to ethically responsible AI in terms of algorithmic fairness and explainable AI (XAI), but also to what may be termed SAI – Sustainable AI.[364] Such a strategy, elaborated in the domain of SAI, could then be adapted and scaled to incentivize, and mandate, sustainable computing and technology more broadly.

## VI. Conclusion

As AI systems become ever more potent and mainstream, at the very latest with the arrival of large generative models such as ChatGPT, GPT-4, DALL·E 2 and Stable Diffusion, a clear, coherent and balanced system for AI must be established. The EU is taking the lead in this endeavor, as in many other fields of digital regulation.

The dual proposals of the Commission on liability for AI take steps into the right direction, but they do not go far enough. Overall, the two half-hearted directives do not add up to one convincing whole. They fail to provide a uniform framework for AI liability in the EU which would balance ease of compensation with sufficient legal certainty for AI development and deployment. EU lawmakers should have the courage to enact *one* coherent *regulation* – not two disparate directives –, thereby truly harmonizing tort liability for products, including software and AI, at the EU level.

This Article has analyzed in detail the various provisions the two directives propose. Important take-aways include that the dichotomy between the fault-based AILD Proposal and the supposedly strict liability PLD Proposal is fictional and should be abandoned; that an EU framework for AI liability should comprise one fully harmonizing regulation instead of two insufficiently coordinated directives; and that the current proposals unjustifiably collapse fundamental distinctions between social and individual risk by equating high-risk AI systems in the AI Act with those under the liability framework.

Finally, the paper has mapped out four key elements for the future of AI regulation in Europe and beyond. In the EU, they should be included in amendments of the proposed directives and of the AI Act; simultaneously, they establish a roadmap for the future of AI liability in the US and other countries aspiring to adapt their legal systems to the peculiarities of AI. Such a framework should contain:

- a staggered approach to liability, ranging from strict liability for certain high-risk AI applications and specific high-risk Foundation Model (≈ large generative models) scenarios, to presumptions of fault and causality for SMEs and non-high-risk AI applications, and simple fault-based liability for consumers using AI;
- a real framework for innovation, including safe harbours and support for SME developers by way of subsidized insurance for liability;
- clear provisions for fostering trustworthy AI, particularly in the fields of explainable AI and non-discrimination, which need to be covered by AI liability alleviations;

---

[363] Cf. Andrew A Chien, 'Good, better, best: how sustainable should computing be?' (2021) 64 Communications of the ACM 6.
[364] For signs of such a shift occurring, see, e.g., Patterson and others, 'Carbon emissions and large neural network training', 2.

- a roadmap for addressing sustainability questions in AI and technology more broadly, through sustainability impact assessments, tailored rules for Foundation Models, and sustainable design defects.

Given the current state of climate change, the move towards sustainable AI and information technology more broadly is arguably the most urgent, but also the most undertheorized, part of the future roadmap for AI liability. If the EU acts first on this matter, it might trigger a "Green Brussels Effect" and further spur the race for green AI and other technology, potentially benefiting research, industry, and users alike.

# Annex: List of Recommended Amendments

## 1. General considerations

1) AI liability should be implemented in the EU in the form of a regulation, not of directives.
2) Whoever in public statements claims to be using AI may be sued under the AI liability regime (even if the claim is factually incorrect).
3) EU lawmakers should acknowledge that the PLD is, as far as design defects are concerned, fault-based, just like many national liability systems. Consumers therefore have to prove fault (= design defects) in most cases concerning AI under the PLD as well.
4) The dichotomy between the AILD Proposal and the PLD Proposal is therefore artificial and should be abandoned.

## 2. Applicability

5) While an alignment with the AI Act concerning the definition of AI (or software) is paramount, the AILD Proposal must be decoupled from the AI Act concerning risk classification.
6) Preferably, autonomous vehicles, emotion recognition systems and insurance pricing models should be included in the high-risk category of the AI Act; failing that, the rules of the AILD Proposal must be extended to these systems.
7) The open-source exception contained in Recital 13 PLD Proposal should be extended to the AILD Proposal or, with the appropriate provisos, to the AI Act, particularly concerning general-purpose AI systems.
8) The PLD (and the AILD Proposal/AI Act) should also apply to commercial software which integrates open source software.
9) The Commission should use the implementing acts foreseen under Article 4(b)(1) AI Act to tailor the allocation of responsibilities between Foundation Model developers and the users of these systems for high-risk purposes to specific use cases

## 3. Defectiveness and breach of duty

10) For the purpose of AI systems, technical standards should, to the extent they exist, presumptively define defectiveness within their scope of application.
11) The presumption of non-compliance contained in Article 3(5) AILD Proposal should apply equally if a human agent took, or failed to take, the final decision leading to the damage caused by the AI output.
12) The ill-fated reference to automated decision-making in Recital 15 AILD Proposal should be deleted.

### 4. Disclosure of evidence

13) Disclosure of evidence rules must apply whenever the information is, *or should legally be* (e.g., under the AI Act), at the defendant's disposal.
14) Potential claimants only need to provide facts and evidence of damage and the involvement of an AI system for the disclosure of evidence rules to be triggered; the plausibility standard should only apply to competitors as claimants.
15) The directives should ensure that the pretrial disclosure case can be brought at the same court as the case in the main proceedings
16) A specific provision empowering courts to order evidence preservation measures should be added to Article 8 PLD Proposal, with an ensuing presumption of defectiveness if the defendant fails to comply.
17) Article 8 PLD Proposal should contain a pretrial discovery possibility.
18) Disclosure of evidence rules should specify that information must be given, to the extent possible, in a format understandable by average claimants' counsel.
19) The PLD and the AILD Proposal disclosure of evidence mechanism should be entirely harmonized.

### 5. Burden of proof

20) The disclosure rules under Article 8(1) PLD Proposal must apply equally to the proof of mandatory safety requirements, to the extent that such facts are necessary to trigger the reversal of the burden of proof under Article 9(2)(b) PLD Proposal.
21) Claimants must be able to invoke simultaneously the presumptions of defectiveness and causality under Article 9(2) and (3) PLD Proposal. This should be clarified.
22) Proof of defectiveness under Article 9(2) PLD Proposal will often turn on whether claimants may establish an obvious malfunction. This is often significantly easier in classification tasks (e.g., face recognition) than in regression tasks (e.g., scoring). The proposals therefore need to be accompanied by effective strategies to facilitate access to litigation funding for claimants unable to bear the costs of hiring AI experts to analyze disclosed data or show obvious malfunctions.
23) Concerning the presumption of non-compliance, Article 3(5) AILD Proposal should mirror the more detailed provision in Article 9(2) PLD Proposal. This implies adding two sufficient conditions (violation of mandatory safety requirements; obvious malfunction).
24) Concerning (also) the causality presumption, the likelihood conditions in Article 4(1)(b) AILD Proposal and Article 9(4)(b) PLD Proposal must be aligned in terms of wording and spirit.
25) A presumption should be added to Article 4(2) and (3) AILD Proposal that a post-processing violation of the AI Act by the provider or user caused the damage.
26) A recital should be added to both proposals concerning the causality presumptions: It should not be considered sufficient for a rebuttal to (correctly) state that almost any machine learning model inevitably commits some errors and that, therefore, even a close-to-perfect system could theoretically have caused the specific damage (by accidentally making the exact error that caused the damage).

## 6. Representative actions

27) Article 6 AILD Proposal should be aligned with the more open provisions of Article 5(2)(b) PLD Proposal to include, among injured persons eligible to be represented, natural persons acting in a professional capacity.

## 7. Defences

28) A provision establishing an EU-wide compensation fund for persons injured by development defects should be added to the PLD Proposal.

## 8. Pure economic loss and eligible damage

29) At least for cases of high-risk AI systems, the eligible damage under the PLD should include pure economic loss as well, lest the liability regime remain a paper tiger in precisely the domain in which it was supposed to offer relief, i.e., high-risk AI applications in the economy.
30) The restriction of eligible damage to privately used property should be abandoned; commercially used property must be included as well.

## 9. The way forward

### a) Strict liability versus fault-based liability

31) Truly strict liability (= based on causation only) should be reserved primarily for economic operators and professional users of high-risk AI systems.
32) In addition, truly strict liability must apply to developers and all users of prohibited AI systems to dis-incentivize their deployment from the outset.
33) A presumption of fault/defectiveness and causality should govern if the high-risk AI system, in the course of its normal operation, causes legitimate harm, for example by rejecting unqualified candidates in job interviews. Truly strict liability is not justified in this context.
34) SMEs and non-high-risk AI systems should only be subjected to a presumption of fault and causality. In addition, SMEs should be eligible for liability insurance subsidies.
35) If a Foundation Model is used in a non-high-risk application, the liability rules for non-high-risk AI must apply. Only concerning a specific high-risk use case, truly strict liability may be triggered for Foundation Models. For such situations, the mechanism in Article 26 GDPR should be adapted to Foundation Model settings.
36) Consumers using AI should face fault-based liability only with a limited presumption of causality.
37) Injurers must be able to invoke contributory negligence of injured persons.

### b) Innovation

38) Ideally, violation of the AI Act should be a necessary and sufficient condition for defectiveness and breach of a duty of care under the PLD and the AILD Proposals, respectively. However, given the current deficiencies of the AI Act, non-compliance with the AI Act can only constitute a sufficient, but not a necessary condition for finding defectiveness and breach of a duty of care.
39) Technical standards and specifications should designate quantitatively defined red zones (non-compliance), green zones (compliance) and grey zones (unclear compliance status) to operationalize vague concepts of the AI Act, for example in the domain of explainability, non-discrimination, performance, and training data quality.

### c) Trustworthy AI

#### i. Unforeseeability

40) The EU liability regime should specify that damage eligible for compensation includes harm caused by unforeseeable acts or omissions of an AI system.

#### ii. Explainability

41) AI developers and users, not injured persons, should bear the consequences of using black-box AI models. Hence, the presumptions of fault and defectiveness should apply, respectively, even if evidence cannot be produced by the defendant due to the black-box character of the model (Article 3(5) AILD and Article 9(2) PLD Proposal).
42) To make sense for post-hoc explanations, the presumption in Article 4(2) AILD Proposal must cover the causal link between the violation of Article 13 AI Act and the damage, not only between the violation and the output.
43) Article 11 in conjunction with Annex IV(2)(b) AI Act should be added to the list of requirements whose violation prompts the presumption of causality under Article 4(2) AILD Proposal.
44) Implementing acts should spell out specifically what type of explanation (feature relevance; counterfactual etc.) is needed in what use case under the AI Act and the AI liability regime.

#### iii. Non-discrimination

45) It should be explicitly clarified that Articles 3 and 4 AILD Proposal apply to claims under non-discrimination law.
46) Article 11 in conjunction with Annex IV(2)(g) and (3) AI Act should be added to the list of requirements whose violation prompts the presumption of causality under Article 4(2) AILD Proposal.
47) Discriminatory output of AI systems should be considered on a case-by-case basis in the defectiveness assessment under Article 6 PLD Proposal.
48) Implementing acts should spell out specifically what type of fairness metric is needed in what use case under the AI Act and the AI liability regime.

### d) Sustainable AI

49) The AI Act should mandate a "sustainability impact assessment" for AI systems.
50) The development of a model that would have failed the sustainability impact assessment should qualify as a design defect under Article 6 PLD Proposal.
51) Ethical AI design should include, next to algorithmic fairness and XAI, also SAI – Sustainable AI.